\documentclass[preprint2]{aastex}
\input epsf.tex
\begin{document}
\title{HST Observations of the White Dwarf Cooling Sequence of M4\altaffilmark{1}}
\author{Brad M. S. Hansen\altaffilmark{2}, Harvey B. Richer\altaffilmark{3}, Greg. G. Fahlman\altaffilmark{3,4},  
Peter B. Stetson\altaffilmark{5}, James Brewer\altaffilmark{3}, Thayne Currie\altaffilmark{3}, Brad K. Gibson\altaffilmark{6}, Rodrigo Ibata\altaffilmark{7},
R. Michael Rich\altaffilmark{2}, Michael M. Shara\altaffilmark{8}}
\altaffiltext{1} {Based on observations with the NASA/ESA Hubble Space Telescope, obtained at the Space Telescope Science
Institute, which is operated by the Association of Universities for Research in Astronomy, Inc., under NASA contract NAS 5-26555.
These observations are associated with proposal GO-8679}
\altaffiltext{2}{Department of Astronomy, University of California Los Angeles, Los Angeles, CA 90095, hansen@astro.ucla.edu, currie@astro.ucla.edu, rmr@astro.ucla.edu}
\altaffiltext{3}{ Department of Physics \& Astronomy, 6224 Agricultural Road, University of British Columbia, Vancouver, BC, V6T 1Z4, Canada, richer@astro.ubc.ca }
\altaffiltext{4}{ Canada-France-Hawaii Telescope Corporation, P.O. Box 1597, Kamuela, HA, 96743, fahlman@cfht.hawaii.edu}
\altaffiltext{5}{ National Research Council, Herzberg Institute of Astrophysics,
5071 West Saanich Road, RR5, Victoria, BC, V9E 2E7, Canada, Peter.Stetson@nrc.gc.ca}
\altaffiltext{6}{ Centre for Astrophysics Supercomputing, Mail No. 31, Swinburne University, P.O.Box 218, Hawthorn, Victoria 3122, Australia, bgibson@astro.swin.edu.au}
\altaffiltext{7}{ Observatoire de Strasbourg, 11, rue de l'Universite, F-67000 Strasbourg, France, ibata@newb6.u-strasbg.fr}
\altaffiltext{8}{ Department of Astrophysics, Division of Physical Sciences, American Museum of Natural History, Central Park West at
79th St, New York, NY, 10024-5192, mshara@amnh.org}

\slugcomment{\it 
}

\lefthead{Hansen {\it et al.}}
\righthead{M4 White Dwarf Sequence}

\begin{abstract}

We investigate in detail the white dwarf cooling sequence of the globular cluster Messier~4. In particular
we study the influence of various systematic uncertainties, both observational and theoretical, on the
determination of the cluster age from the white dwarf cooling sequence. These include uncertainties in
the distance to the cluster and the extinction along the line of sight, as well as the white dwarf mass,
envelope and core compositions and the white dwarf--main sequence mass relation. 
We find that fitting to the full two-dimensional colour-magnitude diagram offers a more robust method
for age determination than the traditional method of fitting the one-dimensional white dwarf luminosity function.
After taking into
account the various uncertainties, we find a best fit age of 12.1~Gyr, with a 95\% lower limit of 10.3~Gyr.
We also perform fits using two other sets of cooling models from the literature. The models of Chabrier et al
(2000) yield an encouragingly similar result, although the models of Salaris et al (2000) do not provide as good a fit.
Our results support our previous determination  of a delay between the formation of the Galactic halo and the onset of star formation
in the Galactic disk.

\end{abstract}


\section{Introduction}

The use of globular clusters as a laboratory to study stellar evolution has a long and venerable
tradition. In particular, a comparison of theoretical models with the observed properties of cluster stars
allows one to determine an age for the cluster and the use of these ages as a lower limit to the
age of the universe has become a standard cosmological test. The age determination rests on fitting
models to the observed location of the main-sequence turnoff in the cluster colour magnitude diagram.
Significant effort from many groups (reviewed in VandenBerg, Bolte \& Stetson 1996; Krauss \& Chaboyer 2003)
 has been invested in
examining the detailed dependence of this method on both observational and theoretical systematics.
The current state of the art is summarised in Krauss \& Chaboyer (2003), and yields a 95\% range for
the age of the Galactic cluster system of $12.6^{+3.4}_{-2.2}$~Gyr. The principal observational limitation
is the uncertainty in the determination of the distance to a given cluster (e.g. Bolte \& Hogan 1995)
while there are also several theoretical uncertainties, which include the treatment of convection,
the treatment of helium diffusion and the conversion from effective temperature to observed colour.

A direct consequence of the large age of the globular clusters is that a significant fraction of the
stars that were originally born in the cluster have completed their stellar evolution, ending their
days as white dwarf stars. A detailed study of the white dwarf population of a cluster allows us
the opportunity to probe the upper main sequence of a population~II system as never before, as well
as providing the opportunity to determine a cluster age by an entirely different method. As with
the main-sequence turnoff method, there are both observational and theoretical systematics to be
considered, but they are {\em different} systematics, so that we may hope to learn independent 
information by comparing the results of the two methods.

The problem with studying white dwarfs is that they are faint and difficult to detect in the
crowded environment of a globular cluster. While the cluster main sequence has been studied
since the 1950s (Arp, Baum \& Sandage 1952; Sandage 1953), the detection of cluster white dwarfs
had to await the advent of the Hubble Space Telescope (Elson et al 1995; Paresce, de Marchi \&
 Romaniello 1995; Richer et al 1995). To obtain useful age information requires extending the
observed cooling sequence to very faint magnitudes, requiring a significant investment of 
HST time. In Richer et al (2002) \& Hansen et al (2002, hereafter paper~I)  we reported the first investigation
of a large number of cluster white dwarfs in the nearby globular cluster Messier~4. Comparison
of the observed luminosity function with a single predetermined set of cooling models yielded an age for
the cluster $12.7 \pm 0.7$~Gyr ($2\sigma$ random error).

This paper represents an extension of the analysis performed in paper~I. The 
purpose is twofold. As this is the first data set to reach these magnitude levels in a population~II
system, we investigate some general features of the observed cooling sequence and the physics
that underlies the distribution of white dwarfs in the colour-magnitude diagram. The second
goal is to investigate in detail the various systematic effects, both observational and theoretical,
that influence the age we infer for the cluster from the cooling sequence. 

The organisation is as follows. Section~\ref{WDobs} describes our definition of the white
dwarf cooling sequence and the derivation of physical properties. Of particular importance
is our treatment of the extinction along the line of sight. Section~\ref{Theory} then describes
the details of the theoretical models which we use to fit the cooling sequence. In section~\ref{Features}
we discuss some general features of the cooling sequence, noting the location of important physical
regimes, before we proceed to the detailed model fitting in Section~\ref{Fits}. The reader impatient
for the bottom line may proceed to Section~\ref{Discuss}.

\section{The White Dwarf Sequence}
\label{WDobs}

The data are based on two HST programs, GO~5461 and GO~8679, taken in cycles 5 and 9 respectively.
The data consist of $15\times 2600$s exposures in F555W and $9\times 800$s exposures in F814W
in the first epoch (Richer et al 1995, 1997; Ibata et al 1999) and $98\times 1300$s in F606W and
$148\times 1300$s in F814W in the second epoch (Richer et al 2002, 2004a; Hansen et al 2002).
Here we review  briefly the data reduction procedure -- more extensive descriptions can be found
in Richer et al (2004ab).

The data were preprocessed according to the procedure outlined in the ALLFRAME cookbook (Stetson 1994;
Turner 1997). For each individual frame, we used DAOFIND with a
 4-$\sigma$ finding threshold to generate a star list.
The registered frames were then combined into 16 images, corresponding
to the two filters in each of the two epochs for each of the four CCDs in
the WFPC2 camera.
  The registered frames were combined using IRAF's IMCOMBINE task to
  reject the brightest $n$ pixels, and averages (not medians) were
constructed of the remaining
  pixels.  This method of combining was preferred as it provided
  us with higher S/N images; the noise in the average of two
  frames will be similar to the noise in the median of three frames
(Stetson 1994). 
  Hence, provided we reject less than 1/3 of
  the pixels, the noise in the high-pixel rejected averaged frame will
  be less than that in a median of all the data. In practice this
technique
allowed us to use 93\% of the $F606W$ frames in the long exposures in
our field and
94\% of the $F814W$ images in constructing the final images. 
Finally, we generated a finding list for the {\it combined} frames in
the second epoch data.

  The M4 fields are dominated by cluster stars. As a result, 
  the transformations used to register the frames
  were dominated by cluster stars
  (indeed, as the matching radius was decreased, the non-cluster stars
  were preferentially rejected). 
  A consequence of this is that non
 cluster
  stars appear to move when frames from the two epochs are blinked.
  To improve transformations between individual frames, we generated
  a list of stars likely to be cluster members by 
  isolating those near the origin on the
  $(x_{new}-x_{old})$ vs.\ $(y_{new}-y_{old})$ diagram (membership is confirmed by
  the narrow locus such stars trace out in a CMD).
  The list
  of candidate cluster stars is on the coordinate system of the
  transformed frames.  We then performed the inverse transformation
  of this coordinate list back onto
  the system of each of the pre-registered frames. Using this new list
  the individual
  (pre-transformed) frames were re-reduced using the ALLSTAR program with
  the recenter option enabled.  These `cluster only' ALLSTAR files,
  which have only cluster candidates, were matched with DAOMASTER, using
  a 20 term transformation equation. By experimentation, we found that
  the 20 term transformation reduced systematic errors. Transformations with
  fewer terms lead to apparent streaming motions in plots of $dx$ vrs $dy$,
  indicative of poor fits -- often occurring near the edges of the chips.
  As prescribed by Stetson,
the DAOMASTER matching
  radius was reduced until it was a few times larger than the standard
 deviation of the differences in the transformed positions.
  The 20 term transformation between the frames
(based
  on cluster stars only) was used to produce new transformed images.
   Note however that this time we used the MONTAGE2
  program to expand the frames by a factor of 2 (pc CCD) or 3 (wf
  CCDs). The expanded, registered images were then combined again
  into four images, one for each filter in each epoch.

  By expanding the frames, we found that the measured radial dispersion
in the cluster
  proper motion was reduced from 3.9 to 2.5 mas/year. The intrinsic
dispersion amongst the M4 stars
  is estimated to be about 0.5 mas/year, so the measured dispersion is
representative
  of the errors associated with our technique.  In effect, by expanding
  the frames we implemented a form of `drizzling' (Fruchter \& Hook
1997).

\subsection{Photometry}

Photometry on the expanded frames was performed using ALLSTAR.
Stetson's library PSFs were not appropriate for the
  expanded frames, so new PSFs for these frames were built following
  standard techniques outlined in the DAOPHOT manual.
   All objects in the master coordinate list were visually
  classified (as to stellar or extended, isolated or possibly
contaminated).

Photometry requires some care in this field because the crowded field results in a high background.
To define the photometric zero-point, we define a set of fiducial stars in the least crowded parts of
the field and perform aperture photometry and use the synthetic zero-points from Holtzmann
et al (1995) to place them on the WFPC2 system. All other stars are measured relative to this fiducial
set of stars.

To perform a proper comparison, we furthermore need to convert the WFPC2 magnitudes into the ground
system, using the transformations of Holtzmann et al. This is necessary as we hope to later compare
to various other models in the literature, most of whom are only available in the ground-based system.
 The transformations are in terms of V-I, whereas
we really want them defined in terms of F606-F814. To that end, we determine the reverse transformations
\begin{eqnarray}
\rm V-I & = & -0.018 + 1.617 \left( F606W-F814W \right) \nonumber \\
 &&  - 0.066 \left( F606W-F814W \right)^2
\end{eqnarray}
and
\begin{eqnarray}
\rm V & = & F606W + 0.014 + 0.344 \left( F606W - F814W \right) \nonumber \\
 &&  + 0.037 \left( F606W - F814W \right)^2 
\end{eqnarray}
for $F606W-F814W<0.66$ and
\begin{eqnarray}
\rm V & = & F606W - 0.018 + 0.407 \left( F606W - F814W \right) \nonumber \\
 &&  + 0.015 \left( F606W - F814W \right)^2 
\end{eqnarray}
for $F606W-F814W>0.66$.

These transformations are important as our deep exposures were in the F606W bandpass, which is broader than the 
more usual F555W bandpass (and therefore attractive for such expensive projects as this one) but has the consequence that
there is  
a significant colour-dependence in the WFPC2/ground transformation. Note this also has an important interaction with
the extinction, as the colour transformations are determined using bright standard stars, whose spectral energy distributions
may imperfectly mimic that of a reddened white dwarf. We will return to this point in \S~\ref{Redden}.

\subsection{Proper motions }

At this stage, the proper motion of a star (relative to the cluster) can
be
determined by examining its shift in $x$ and $y$. Cluster stars can be
isolated in the data as they will have small proper motions.
An important point to note here is that we were
matching
long exposures in $F606W$ (127.4 ksec) and $F814W$ (192.4 ksec) with
much shorter
exposures in $F555W$ (31.5 ksec) and $F814W$ (7.2 ksec). The question
thus arises
as to how we could measure positions in the earlier epoch data for the
faintest stars where the S/N is obviously poor. For a star at $V = 29.0$
the expected
S/N (from the HST ETC) in our deep $F606W$ frames is 6.7 while a star at
this
magnitude on the shorter $F555W$ frames has S/N = 2.5. This is just
about
what we measured on our frames. The reason we were able to measure
positions
of such faint objects in the first epoch are severalfold.

First, we
applied the 
finding list from the deep frames to the shallower frames. 
If the background noise in an image is Gaussian, then a
$2.5\sigma$ positive deviation will occur in about 6 pixels out of every 1000.
If, in a $750 \times 750$ image (neglecting the vignetted areas at the low-x and low-y
sides of the WFPC2 CCDs), we were to mark all of the $2.5\sigma$ peaks as
detected astronomical objects, we would expect more than 3,000 false
detections.  However, if we consider only the area within 0.5 pixels radius
of an object confidently detected in the long second-epoch exposure, we
expect a probability $\sim \pi \times 0.5^2 \times {6\over1000} \approx
0.005$ of finding a positive 2.5$\sigma$ deviation which is purely the result
of random noise in the first-epoch image.  That is to say, we expect of
order 5 false cross-identifications for every 1000 correct re-detections.
Even at $1.8\sigma$, presumed re-identifications will be correct 19 times out of 20.
For this reason, the knowledge that an actual astronomical object is present
somewhere nearby based upon the long-exposure second-epoch images allows us
to be confident that most of the claimed re-detections on the short-exposure
first epoch images correspond to true re-detections.  The first-epoch
astrometric positions, then, while poorer than those of the second epoch,
are nevertheless good enough to distinguish stars that are moving with the
cluster from stars and galaxies that are not.

Secondly, the proper motions of the
M4 stars are
actually quite large (about 1 HST pixel with respect to an extragalactic background over the 
6 year time baseline (Kalirai et al. 2004)).
While the lower $S/N$ image would not produce good enough photometry,
the $S/N$ is sufficient to give the centroid, which is crucial for
astrometry.
In Figure~\ref{plot_pm} we illustrate the quality of the proper motion
separation between cluster and field from the deep images in the outer
field. Some of the brightest stars do not
exhibit clean separation due to their near saturation. However, it is
clear from this diagram that most
field objects can be easily eliminated down to quite faint magnitudes by
making the generous proper motion cut within a total value of 0.5 pixel
($\sim 8$ mas/yr) of that of the mean cluster motion over the 6 year
baseline
of the observations.

Isolating the cluster main sequence in this manner is easy, given the large cluster proper motion. To isolate
the white dwarfs, a little more caution is required, as clearly our ability to centroid is expected to
get worse as the stars get fainter and so the proper motion errors will be larger for the faintest stars
(which are of most interest to us).
 To determine the
appropriate proper motion cut for our purposes, we will perform the separation in a different way than is usually done. 
First we apply a cut in the colour-magnitude diagram, to isolate all the stars (both cluster and background) that
are of appropriate colour and magnitude to be
 candidate white dwarfs. We then examine the proper motion displacements of only these candidates,
in order to determine the appropriate separation into cluster and background objects, without being biased
by the excellent separation of the brighter main sequence stars. 

Figure~\ref{CMD} shows the colour-magnitude diagram for all the stars in our sample, along with the colour
cut we apply. Figure~\ref{PMS} shows the proper motion displacement (relative to the bright cluster
main sequence stars) of all the stars that lie blueward of
this criterion. Clearly we can distinguish cluster from background to very faint magnitudes. However, since
we hope to obtain age discrimination by examining the structure in the luminosity function, there is 
a concern that the exact location of the cut could influence the finer details of what we call our
observed luminosity function. Including a self-consistent treatment of the proper motion cuts in the
incompleteness estimates (\S~\ref{ArtiStar}) should take this into account but we will also further
 investigate this by imposing two different proper motion cuts to see if
there is any significant difference. The cut designated A will be the default value and we will examine
the influence of cut B on the results later. 

Figure~\ref{Red} shows the resulting white dwarf cooling sequence for proper motion cut A. 
The total number of white dwarfs detected in the three WF chips is 272 (103 in WF2, 78 in WF3 and 91 in WF4).
This is the data set we attempt to model in \S~\ref{Theory}. Note that henceforth we will operate in
V and I, rather than F606 and F814.

De Marchi et al (2003) have published an independent analysis of this data which reaches significantly
different conclusions from those presented here. In particular their luminosity function does not extend
as deep as ours, causing them to conclude that all one can assert is that the age of M4 is larger than 9~Gyr.
We consider their analysis to be flawed and disagree strongly with their result on several points, which we discuss in Appendix~\ref{deMarchi} and
Richer et al (2004b). 

\subsection{Incompleteness}
\label{ArtiStar}

An important part of our modelling procedure in later sections is the transformation between
theoretical models and observational parameters. In a cluster environment, our ability to accurately
measure the magnitude and colour of faint objects can be strongly affected by the presence of many
brighter stars in the field. 
To accurately model the resulting errors we have performed extensive
artificial star tests, adding stars of known colour and magnitude at specific locations
in the field and then rereducing the frames to measure the resulting dispersion in
colour, magnitude and position of the artificial stars.

The added stars were drawn from a fiducial model white dwarf sequence
 and 4 trials of 1250 stars each were added into each chips in 
the
field. The stars were added onto a grid which ensured that no
artificial stars
 overlapped
with each other. The position of the added star at the intersection
points of this grid was given a small random offset
to insure that stars were not all
added at the same positions in different trials, but the shift
was small compared with the grid and PSF size. This technique (first
used
by
Piotto \& Zoccali 1999) allows large numbers of stars to be added into
a single frame while ensuring that they do not interfere with each other
(i.e. do not change the crowding statistics in the images).
This guarantees that the corrections thus derived provide accurate
incompleteness statistics. We 
 kept track of all the stars,
irrespective
at which magnitude they were recovered. We also determined the position
at which the star is detected -- necessary to properly model the effects
of proper motion selection when the proper motion accuracy degrades with
magnitude.
 From this we derived a 
``correction matrix" with input magnitude along one axis and recovered
magnitude
along the other, as a function of proper motion cutoff. The diagonal elements of this matrix are the recovery
fractions for stars recovered
within the same magnitude bin as input while the off-diagonal ones are
the
recovery fractions for stars
recovered at magnitudes either brighter or fainter than the input value. 
Figure~\ref{F0} shows the total recovery fractions (within a proper motion cutoff of 0.5 pixels)
for the three WF chips. Of course, this is only the crudest measure of the selection effects.
For the purposes of modelling, we also required the distribution of recovered magnitudes (and positions)
as a function of input magnitude.
Figure~\ref{sigs} shows the $1\sigma$ (i.e. 68\% limits -- the distributions are not
strictly Gaussian) bounds of the distribution (for both stars recovered at magnitudes both brighter
and fainter)
 as a function of input magnitude
for the WF4 chip (the distributions for the other chips are similar). Note that this distribution
reflects only those artificial stars which were recovered within 0.5~pixels of their input position,
so this should represent the scatter in the stars we are actually trying to model.

\subsection{Distance and reddening}
\label{Redden}

The principal uncertainty in the age determinations by main sequence fitting is the uncertain knowledge
of the distance and extinction, which influence directly the estimate of the turnoff mass and thus the age. 
Similarly, the uncertainty in the distance and extinction translates into an uncertainty in the
masses of the white dwarfs, although the relation to the age is not quite as straightforward,
as we shall see.

The distance and extinction quoted in papers~I and II were determined by fitting low metallicity
subdwarfs to the observed lower main sequence (Richer et al 1997). The quoted result was a 
distance modulus in V of $12.51\pm0.09$, with E(V-I)=$0.51\pm 0.02$. The reddening parameter
for this line of sight is estimated to be a nonstandard $R_V=3.8$, which yields a true 
distance modulus 11.18, in good agreement with astrometric determinations (Peterson et al 1995)
and Baade-Wesselink measurements (Liu \& Janes 1990). These methods constrain the distance to
M4 to be $1.73\pm0.14$~kpc.

While appropriate for most applications, a distance and extinction determination using subdwarf fitting is 
a potential source of systematic errors in our case. For one thing, it depends 
 on the assumed
metallicity (and one of the attractions of the white dwarf age method is that it is supposedly not
particularly sensitive to the metallicity). Furthermore, in practice one has to slightly adjust the colours
of some of the local subdwarfs whose metallicities are close but not exactly the same as that of the
cluster stars, thereby potentially introducing errors from model atmospheres. Finally, there are some differences between
 the spectral energy distributions of main
sequence stars and white dwarfs, so the final extinction in a given bandpass can be slightly different
for the two different classes.
Errors in the extinction contribute to errors on the age determination, since
the reddening vector has a significant component parallel to the cooling sequence (Figure~\ref{Red}).
For these reasons, we would like to determine the extinction directly for the white dwarfs.
We do this as follows.

The apparent magnitude of white dwarfs (of fixed temperature) at the top of the cooling sequence is
determined by the distance, the extinction and the white dwarf radius (and hence mass). In
fitting a model to the observed cooling sequence, one can then trade off variations in these parameters 
while keeping the apparent magnitude fixed. As a consequence, different extinctions imply
different masses at the top of the cooling sequence if we insist that the model sequence overlap the observations.
The mass dependence of white dwarf cooling means that the choice of extinction (and hence mass) has implications
for the luminosity function at fainter magnitudes and thus affects the age determination.

Thus we wish to explore the sensitivity of the relation between the inferred white dwarf mass
and the extinction.
At the hot end, the theoretical cooling and atmosphere models yield
\begin{equation}
M_{I,0}=2.52 (V-I)_0 + 11.48-5 \log \left( \frac{R}{10^9 cm} \right)
\end{equation}
where we have used the subscript `0' to denote unreddened quantities and $R$ is
the radius of the white dwarf, which we shall henceforth denote $R_9$ (i.e. in terms
of $10^9$cm) to distinguish it from the reddening parameter R.
 This fit is performed in the range
$0.1 < (V-I)_0 < 0.4$, corresponding to effective temperatures $7500<{\rm T_{eff}}<10^4$~K.
This was chosen to be cool enough to avoid the dependence of the radius on the
surface hydrogen layer mass (not negligible for the hottest white dwarfs) and 
hot enough to avoid the non-monotonic effects of molecular hydrogen absorption on the colours.

We also determine
 an empirical fit to the observations (transforming from the flight system to ground-based system via
the prescriptions of Holtzman et al 1995)
\begin{equation}
I = 2.53 (V-I) + 22.27
\end{equation}
in the range $0.5<V-I<0.9$. This is not the true top of the cooling sequence but
rather the location of the observed cooling sequence that corresponds to the approximate
location of the temperature range defined by the preceding theoretical fit. Note also that
there is some subtlety involved in defining what we term the `cooling sequence', as we refer
here to those white dwarfs that evolve from single star evolution. A handful of outliers
may lie slightly above this sequence due to the presence of binaries -- either as a result of the 
truncation of stellar evolution
in binaries, which leads to undermassive white dwarfs or simply due to two unresolved white dwarfs of
similar age. Such undermassive white dwarfs have been observed in the field (Bergeron,
Saffer \& Liebert 1992), in clusters (Cool et al. 1998) and indeed in this cluster (Sigurdsson et al 2003). 
We have thus excluded a handful of objects that lie above the readily apparent cooling sequence.
An example of such a fit is shown in Figure~\ref{CMDfit}.

Now, $I=M_I+\mu+A_I$ and $V-I=(V-I)_0+A_V-A_I$. Setting $A_I=\alpha A_V$, we can eliminate $V-I$
and $I$ to get
\begin{equation}
A_V = 0.96 - 0.025 (V-I) - 12.26 \log R_9
\end{equation}
where we have used $\alpha=0.6$ (we have determined that this number is robust for $3.1<R<3.8$ by using model atmospheres,
extinction curves from Fitzpatrick(1999) and bandpass models from Holtzman et al (1995)). If we perform the fit in the
middle of the fitting range ($V-I = 0.6$), we have a relation between the extinction and
the white dwarf radius (and hence mass) which must be satisfied to satisfy the observed
location of the cooling sequence
\begin{equation}
A_V = 0.94 \pm 0.1  - 12.26 \log R_9 +  2.45 \delta \mu 
\end{equation}
where we have also added the effect of a distance uncertainty $\delta \mu$.
Thus, we observe a strong
covariance between $A_V$ and $R_9$.
Note also the weak dependence on colour, which supports
our assumption that the white dwarf mass is approximately constant over this colour range.
This method is similar to the 
method used by Renzini et al (1996) and Zoccali et al (2001) to measure the distances to 
the clusters NGC~6752 and 47~Tuc by comparing the cluster cooling sequence with a set of
nearby white dwarf standards. However, in our case the sequences are shifted parallel to the
reddening vector, rather than vertically in the CMD.

What is a reasonable mass? The models of Renzini \& Fusi-Pecci (1988) predict 
$0.53\pm 0.02 M_{\odot}$, which Renzini et al (1996) argue is a robust prediction.
Cool, Piotto \& King (1996) find a mass of $0.55 \pm 0.05 M_{\odot}$ for the top of the cooling
sequence in NGC~6397, assuming a standard distance and extinction. Similar numbers have been
derived for the standard values in M4 as well (Richer et al 1997). These numbers are slightly
smaller than the mean value
 in the solar neighbourhood (e.g. Bergeron, Saffer \& Liebert 1992), where the distribution is sharply
peaked near $M \sim 0.6 M_{\odot}$. This would require $A_V = 1.7$ for our nominal distance, although an
extinction $A_V=1.3$ can be obtained with $\mu= 11.03$, which is just within the observational errors.
We will return to the issue of distance/extinction in \S~\ref{DeltaD}.
In principle, the brightest white dwarfs in globular clusters are now accessible to spectroscopic analysis
with large telescopes from the ground (Moehler et al 2000; Moehler 2002), offering the prospect of
direct spectroscopic gravity (and hence mass) determinations. Initial results for the hottest white dwarf in NGC~6397 find a rather low
gravity ($\log g \sim 7.3$) suggesting an undermassive white dwarf ($\sim 0.36 M_{\odot}$), characteristic of
binary evolution (e.g. Kippenhahn, Kohl \& Weigart 1967). This illustrates that a sample of at least several gravity determinations per cluster are
necessary to properly isolate the characteristic mass emerging from single star evolution. At present the 
spectra for M4 are not good enough to measure gravities, but do show that the hottest
dwarfs have hydrogen atmospheres (Moehler 2002).

Thus, our procedure is as follows. For each given model, we choose a characteristic white dwarf
mass at the top of the cooling sequence. This then determines the appropriate extinction for that
particular model, given an assumed distance. In this manner we ensure that each of our models provides a consistent fit to
the top of the cooling sequence. However,  for the purposes of defining a default model, we start
with a value of 0.55~$M_{\odot}$, which implies $A_V=1.39$ for our nominal distance $\mu=11.18$.
Although this 
value is very similar to that determined by the subdwarf method, it has been determined directly from the
white dwarf sequence, so that the agreement reflects an encouraging concordance.

\section{Theoretical Inputs}
\label{Theory}

We have discussed above the characterisation of the observed Messier~4 white dwarf cooling sequence. 
In order to determine an age we must fit a theoretical model to the observed distribution of colours
and magnitudes. We have already described two essential elements of the model -- namely the model of
the observational errors and the determination of the extinction and reddening. In addition we now
describe the various elements of the cooling models.

\subsection{Atmosphere Models}

Atmospheric models are necessary to convert effective temperatures and luminosities into
observable quantities such as colours and magnitudes. This is a major issue in the determination
of ages using main sequence stars, as the influence of metals on the spectrum can lead to colour variations
depending on the exact metallicity of the cluster stars. White dwarfs, on the other hand, have primarily
hydrogen and helium atmospheres and so, in principle, are less susceptible to this kind of systematic
error. Even the few white dwarfs which do show metals in the atmosphere show only trace amounts which
are not enough to significantly influence broadband colours.

There is one atmospheric subtlety which needs to be considered in white dwarf atmospheres. For
white dwarfs at sufficiently cool effective temperatures ($\rm T_{eff}<5000$~K), atmospheric hydrogen
has a significant molecular component -- which results in collisionally induced absorption by
$H_2$ which can have a significant influence on the colours (Mould \& Liebert 1978; Bergeron, Saumon \& Wesemael 1995;
Hansen 1998; Saumon \& Jacobsen 1999). We will adopt as our default $\rm T_{eff}$--colour transformations
those of Bergeron, Wesemael \& Beauchamp (1995) (augmented where necessary by fits to the models
of Bergeron, Leggett \& Ruiz 2001). We also use a model we calculated ourselves (Hansen 1998; 1999)
but the Bergeron models provide a better fit (\S~\ref{AtmosComp}).

Figure~\ref{CMD_comp} shows the fit of an atmospheric model to the observed cooling sequence, assuming
an extinction $A_V=1.39$ and a white dwarf radius $R=8.9\times 10^9$cm. The model is clearly an excellent
fit to the main locus of points. The deviation at the upper end is because, at the very highest temperatures,
there is a small dependence of radius on temperature which leads to slightly lower magnitudes. The
exact shape of a true cooling sequence will depend on how the mass varies along the cooling sequence.

\subsection{Cooling Models}

In addition to the atmosphere models, which relate $\rm T_{eff}$ to the observed magnitudes,
we also require a set of cooling models, which determine the rate at which each white
dwarf cools. White dwarf cooling is, in principle, quite simple to calculate. The luminosity
is powered by residual thermal energy left over from the prior nuclear burning history.
It is even insensitive to the exact temperature at which the white dwarf forms (since
hot white dwarfs cool rapidly by neutrino emission -- which means that the slow cooling
begins where neutrino cooling ends, for all white dwarfs). Although cool white dwarfs
do possess convection zones on the surface, such zones contain negligible mass and the parameterisation
of convective efficiency has little effect on the cooling rate (unlike for main sequence stars).

However, white dwarf cooling models do have their own set of uncertainties.
 \begin{itemize}
\item White dwarfs possess a carbon/oxygen core surrounded by thin shells of helium and
hydrogen. The thickness of these surface layers has an effect on the cooling as they
regulate the rate at which heat is lost from the star. Stellar evolution models provide
some guidelines as to the expected size of these layers, and asteroseismology offers the
promise of one day determining the layer masses directly, but for now the mass of the
hydrogen and helium layers remain a free parameter which we will vary within the expected
range.  Figure~\ref{Cool} shows the influence of different layer masses on the cooling ages
for our default model defined below.
\item In addition to the surface layers, the internal chemical compositional profile in the
carbon/oxygen core can influence the cooling by virtue of its influence on the global heat capacity. Clearly
a white dwarf that can store more heat will cool more slowly. This C/O compositional profile
is a function of mass and is also affected by the behaviour of the main sequence models and
by the $\rm ^{12}C(\alpha,\gamma)^{16}O$  reaction cross section, which is poorly known. 
\item Cool white dwarfs crystallise in their cores, resulting in a release of latent heat which
provides an extra source of heat to slow the cooling. In addition, there is the possibility that
the equilibrium mixture of carbon and oxygen may be different in the solid and liquid phases. In
that case the resulting adjustment in the density structure can lead to an additional release of
energy termed chemical separation energy. This too will depend on the nature of the compositional
profile, as well as the phase diagram of the C/O mixture.
 \end{itemize}

In subsequent sections we will investigate the influence of these various effects, both
by including variations in our own set of models and by making comparisons with others in
the literature.

\subsection{The Main sequence lifetime}

In order to get a proper estimate of the true age, we need to associate a given white dwarf
with a main sequence progenitor, as the true age of the star is the sum of the white dwarf
cooling age and  the additional main sequence lifetime associated
with that object. The main sequence lifetime is affected by overall metallicity and so we
must be careful to choose an appropriate model and not simply appropriate such fits as
have been used in previous papers on determining ages from white dwarf cooling models.
 We will use the lifetime-mass relation from the models of Hurley
et al (2000) for $Z=0.1 Z_{\odot}$. Figure~\ref{tms} shows the relation between mass and
pre-white dwarf lifetime for models of solar metallicity, and models of lower metallicity.

\subsection{The Main Sequence -- White Dwarf Mass relation}
\label{IFMR1}

A related quantity is the relationship between main sequence mass and white dwarf mass.
In principle, this can be predicted from a set of stellar models as well. However, our
incomplete understanding of 
 mass loss in the advanced stages of stellar evolution makes this a less than
secure proposition, and we also make use of empirically determined prescriptions. 
This is traditionally done by measuring the mass of white dwarfs and main-sequence
turnoff stars in a variety of clusters of different age (Weidemann 1987, 2000; Koester 
\& Reimers 1996; Claver et al 2001). Figure~\ref{IFMR} shows the range of relations
we will explore in this paper, compared to the empirical relation
of Weidemann (2000). The solid curve denotes our default relation, based
on that of Wood (1992)
\begin{equation}
M_{wd} = 0.55 M_{\odot} {\rm exp} \left( 0.095 ( M_{ms} - M_{TO} ) \right) \label{WoodM}
\end{equation}
where $M_{TO}$ is the turnoff mass appropriate to a particular age. This has the same
form as the relation A used by Wood,  but we have adjusted it so that the turnoff mass
always gives $0.55 M_{\odot}$ for all ages. This is necessary to be consistent with our
method of determining the extinction in \S~\ref{Redden}. Other theoretical relations
are those due to Dominguez et al (2000) and Hurley et al (2000). Some scatter is
evident for main sequence masses $>2 M_{\odot}$. This is also approximately the
amount of scatter accomodated by the observations (Claver et al 2001). We should
note that most empirical measures are for solar or super-solar metallicity stars 
and that some difference is expected for lower metallicity stars. We will test this
by testing predictions for both solar and sub-solar metallicity models.

One encouraging point to note is that the scatter in the predictions is primarily
for main sequence masses $>2 M_{\odot}$, which means that cluster ages $>10$~Gyr are
not significantly affected. We will return to this point in \S~\ref{IFMR2}.

\section{Features of the  cooling sequence}
\label{Features}

Given that this dataset is the most extensive sample of population~II stellar remnants
known, it is of interest to discuss a few general features of the cooling sequence
before turning to detailed model fits. Along our cooling sequence we can identify
several important epochs in the life of a white dwarf.

\subsection{ Hot white dwarfs}
\label{HotStuff}
Young, hot white dwarfs cool very rapidly -- the hottest cool by neutrino emission.
The neutrino contribution to the cooling rate drops to less than 10\% within $3 \times 10^7$~years.
The effective temperature at this point is $\sim 25,000$~K, which corresponds to 
approximately $V-I \sim 0.29$ and $I \sim 22.9$ for this cluster. Our cooling sequence has
3 stars that lie above this point, suggesting the {\em current} birthrate of white dwarfs
in our cluster field is $\sim 10^{-7}$~yr$^{-1}$. Rather than rely on just three stars, we can
also estimate this rate by determining the cooling age of the tenth faintest white dwarf on
our sequence. This occurs at $I =24$ and corresponds to $\rm T_{eff} \sim 9700$~K for our default
parameters. The cooling age for a 0.55~$M_{\odot}$ dwarf at this temperature is $5 \times 10^8$~years,
yielding a rate estimate $\sim 10/5\times 10^8 \sim 2 \times 10^{-7} yr^{-1}$. If the current
white dwarf birthrate is this high, we expect $\sim 10^8 \times 2 \times 10^{-7} \sim 20$ Horizontal Branch
stars in our field.  Richer et al (1997) estimate only $\sim 2$,  based on scaling
the relative numbers of horizontal branch and giant stars. However, the cluster shows evidence for
significant mass segregation. We have fit multi-mass Michie-King models to the full M4 data set (Richer et al 2004a),
including data from fields at smaller radii. These fields are also proper motion selected, using first epoch
data from the program GO-5461 and second epoch archival data from program GO-8679. On the basis of these
models we can relate the mass fraction in our field to the global value
and can thus convert the birthrate in our
field to a global white dwarf birthrate $\sim 1.5 \times 10^{-6}$~year. This now agrees well with
the expected birthrate given the total number of observed Horizontal Branch stars in the cluster
($\sim 150/10^8 yrs \sim 1.5 \times 10^{-6} yr^{-1}$). We will return to the issue of mass segregation
in \S~\ref{numbers}.
\subsection{DA versus DB white dwarfs} In the Galactic disk, some fraction of white dwarfs
show helium atmospheres (DB) versus hydrogen atmospheres (DA). We will discuss this in the context
of the cool models later, but for the purposes of the upper part of the cooling sequence, it
is important to note that the choice of atmospheric composition can result in a change
$\sim 0.2$~magnitudes at fixed colour. This can explain much of the dispersion observed
in the upper part of the cooling sequence but is too small to account for the `overluminous'
white dwarfs which lie above the main sequence. We will identify these as potential binary
members below. For hotter white dwarfs, hydrogen atmospheres are much more common than helium
atmospheres (e.g. Sion 1984), so we use hydrogen atmosphere models to determine the extinction in \S~\ref{Redden}.
This is further justified by the detection of hydrogen in the first spectra of hot white dwarfs in M4 (Moehler 2002).
\subsection{ The ZZ Ceti strip}
As the white dwarf cools, the effective temperature begins to drop below the point where
the atmospheric constituents remain ionized. As the white dwarfs pass through the
range $\rm T_{eff} = 11,500$--$12,500$~K, hydrogen atmospheres acquire a thin convection zone
near the surface. It is also in this range of effective temperatures that we find the
ZZ~Ceti stars. These are stars which display photometric variability due to the presence
of pulsations.
The nature of the extinction uncertainty means that the blue edge could lie in the colour range
$V-I = 0.35$--0.53, and the red edge $V-I = $0.42--0.60. This range encompasses about 7 white dwarfs
on our CMD, although only one or two would actually be ZZ Ceti stars (depending on where it is
truly located). A study of the variability in this field (Ferdman et al 2003) revealed no 
statistically significant photometric variations among these stars, but this does not pose
a strong constraint as the integration times (1300s) are of order the variability timescale
for ZZ~Ceti stars, so that much of the variation would be washed out.
\subsection{The Luminosity Function Jump}
\label{Jump}
The most striking feature of the cluster white dwarf luminosity function is the sharp jump in
the numbers for $I>26$. This is a direct consequence of the behaviour of the white dwarf cooling
curve. As the white dwarf continues to cool beyond the ZZ~Ceti strip, the opacity of the now
neutral atmospheric hydrogen decreases. The photospheric pressure $P \propto g/\kappa$ and 
the resulting
 increase in the pressure affects the outer boundary condition of the cooling models, causing
the cooling to slow down. This results in a pile-up of stars at this point - as shown in 
Figure~\ref{Lt} (for a $0.6 M_{\odot}$ model). The approximate location of this jump is at $T_{wd} \sim 5$~Gyr and
$\log L/L_{\odot} \sim -4.1$. It is important to note that this feature is primarily a result
of the effective temperature -- thus it does not depend strongly on the mass or core composition of
the white dwarf. This in turn implies that it should be a universal feature of all models, so that
there is little age information in the location of the jump.
\subsection{Core crystallisation}
Crystallisation of the core material occurs at roughly the same time as the change in the outer
boundary condition. Thus,  the release of latent heat can
also contribute to a temporary slowing of the cooling. However, there is an important difference between
these two potential contributors. The point at which crystallisation begins depends on the central
density and thus the mass of the white dwarf -- as a consequence, white dwarfs of different mass receive the
release of latent heat at different epochs in their cooling curves. However, the change in the atmospheric
boundary condition is a direct consequence of the effective temperature and thus occurs at the same
colour (except for changes in the reddening) for all white dwarfs. 
Crystallisation is very important, however, as it makes the cooling mass dependent (Figure~\ref{Mcool}).
This mass-dependent behaviour contributes significantly to differences in shape between the various
model luminosity functions and which leads to much of the age discrimination in the methods we
describe in \S~\ref{Fits}.
\subsection{ Colour Evolution and  the Luminosity Function cutoff}
At the faint end of the cooling sequence there are a couple of features one might hope to see.
One is the change in the white dwarf colours due to the absorption of infra-red light by molecular
hydrogen collisionally induced absorption (Mould \& Liebert 1978; Bergeron, Saumon \& Wesemael 1995).
 This causes white dwarf colours (at least in the optical
and near-IR) to get bluer as they cool, instead of redder (Hansen 1998; Saumon \& Jacobsen 1999).\footnote{
This has been referred to by some as the `blue hook', which could lead to confusion in the
globular cluster context, as this term is already used to describe a phenomenon associated with anomalous
Horizontal Branch stars!} 
As shown in Figure~\ref{CMD_comp}, we cannot claim evidence of
this distinctive feature, although such a statement is dependent on the assumed extinction value. We also
do not claim to have measured a real truncation in the white dwarf luminosity function -- our
model fits make use of the full distribution in colour and luminosity at the faint end to get constraints
on the age. This point will be discussed at greater length in \S~\ref{Fits}.
\subsection{White Dwarf Binaries}
The binary fraction amongst white dwarfs in clusters is of interest for several reasons -- not least
of which is the suggestion that they might be efficient producers of type~Ia supernovae (Shara \& Hurley 
 2002). Binary white dwarfs are identified as being overluminous -- either as a result of
the light of two unresolved white dwarfs being interpreted as a single object or else because
the white dwarf is of lower than average mass (and hence larger radius) due to the truncation
of stellar evolution in binaries (Kippenhahn, Kohl \& Weigert 1967). This is a phenomenon frequently encountered in globular
clusters (Hansen, Kalogera \& Rasio 2003), including this one (Sigurdsson et al 2003). Note that there is an
important difference between these two cases. For binaries containing two normal white dwarfs, only those of
comparable cooling age will appear overluminous -- our estimate will only be a lower limit to the overall
binary frequency. For binaries containing an undermassive white dwarf, no such correction applies as all
binaries of this type are readily identifiable.

We can get a rough idea by measuring the overluminous fraction of white dwarfs in our field.
Restricting our sample to $I<25$ (to avoid thorny issues of incompleteness and photometric
scatter) we find 5 overluminous white dwarfs out of a total of 45 stars in Figure~\ref{Red}. There is also one
anomalously massive white dwarf -- which lies significantly below and to the left of the cooling sequence. 
The identification of any single star as a cluster member is always open to question since
some background objects can have fortuitous motions which fall within the proper motion cut.
However, this star has a proper motion well within the cluster cut and all measures are
consistent with a cluster white dwarf. If it is a true cluster member then the location $\sim 1.4$~magnitudes 
below the cooling sequence implies a mass $>1 M_{\odot}$ and an age $\sim 10^8$~years.
If this object is the result of a merger of two $0.5 M_{\odot}$ dwarfs, then it suggests a minimum rate of white dwarf
mergers $\sim 10^{-8}$ per year in this field alone. To convert this into a global rate we can again make
use of the Michie-King models of Richer et al (2004a), which suggest a global rate $\sim$ 10 times larger
than in our field alone, 
i.e. $\sim 10^{-7}~yr^{-1}$.\footnote{If the M4 rate is characteristic
of Galactic globular clusters, then this implies a rate $\sim 10^{-5} yr^{-1}$ in the Galaxy as a whole. 
Such a number is of interest for the question of whether globular clusters are a significant source of
Type~Ia supernovae (e.g. Shara \& Hurley 2002) . If this number is characteristic, then globular clusters contribute roughly 1\% of
the estimated Galactic rate. Of course, the rate could be higher in clusters more centrally condensed than M4 or
indeed in the center of M4 itself (binary hardening due to external perturbations proceeds faster in the
denser central regions). Also
we note that we can only count white dwarf mergers that {\bf fail} to become supernovae. However, given the
characteristic distribution of white dwarf masses, our number is probably an upper limit.}

The binary fraction of white dwarfs ($\sim 0.11 \pm 0.05$) we observe is 
 interestingly similar 
 to the one inferred for field white dwarfs (Holberg et al 2003)
if we restrict the sample to binaries that would remain hard in M4 (The disk also contains a roughly
similar fraction of wider white dwarf binaries that would be disrupted by three-body encounters in M4).
This ratio is also consistent with our estimates for the binary merger rate relative to the total birthrate
of white dwarfs in our field.
Our binary fraction is slightly larger than the $1\%$ inferred for the lower main sequence (Richer et al 2004a),
but not necessarily a problem as the statistical errors are relatively large and the two numbers
reflect very different parts of the stellar population. 

Binaries are more massive than the average star, so our binary fraction estimate may not be 
appropriate for the cluster as a whole. The first epoch data (Richer et al 1995) for this cluster covers a much
wider area of the cluster than our second epoch data. Furthermore, a shallow second epoch taken by 
Bedin et al (2001) allows us to separate cluster white dwarfs by proper motion in the inner fields in the same manner
as for our default field, albeit to a much brighter magnitude limit. However, the depth achieved in these
inner fields is sufficient to address the question of binary membership. Figure~\ref{binfrac} shows the
upper white dwarf cooling sequence in 3 fields closer to the core than our field. The data were divided
into 3 radial bins so that roughly equal numbers of white dwarfs were found in each. We consider only the
magnitude range $23<F814<24.5$, as for fainter white dwarfs the photometric scatter is too large and for
hotter white dwarfs the temperature dependence of the white dwarf radius leads to an uncertainty.
In all three fields, the numbers are consistent with a $10\%$ binary fraction, although this is based on
small numbers. In the outer two fields
 we see fractions 2/23 and 1/20. The inner field, however, shows a larger potential binary
fraction of $0.31\pm 0.15$ (5/16). However, the photometric scatter in the inner field is larger, so that
3 of the binary candidates are uncertain, i.e. all fields are consistent with a $10\%$ binary fraction. 

\section{Modelling the Luminosity Function}
\label{Fits}

We have discussed above the various elements required to model the distribution of white dwarfs
in the cluster.
 The model should match not only the distribution in luminosity but in colour
as well. Since one can always keep a V or I band magnitude fixed while trading off mass/radius
versus $\rm T_{eff}$, it is clear we need to include colour information to get the best possible fit.
 Traditionally, white dwarf ages are determined by fitting models to the luminosity function.
In principle then, one could fit to both the V-band and I-band luminosity functions simultaneously.
Alternatively, one could fit to the two dimensional distribution of stars in the CMD directly (this is referred
to as fitting the Hess diagram in stellar population studies). The luminosity function fits have
the advantage that the one dimensional nature allows us to use smaller bins while still retaining
enough stars in each bin to be statistically meaningful. However, only a two-dimensional fit can
take proper self-consistent account of the colour information. 

To begin our model, we
 choose a (main sequence) mass function of given slope $x$ (where $dN/dM_{ms} \propto M_{ms}^{-(1+x)}$) 
in the white dwarf progenitor mass range
and sample it in a Monte Carlo fashion. We convert all main sequence stars to
white dwarfs based on the initial--final mass relation (our default model will assume relation~\ref{WoodM})
 and assign a magnitude
and colour based on the white dwarf cooling age once the main sequence age has been subtracted.
This is then fed through the appropriate extinction and distance. We also Monte Carlo sample
the results of our artificial star tests and assign an observed magnitude
based on the probability distribution inferred for the observational photometric scatter. 

\begin{itemize}
\item {\bf Fitting to the Luminosity Function}:

 We then form V and I band luminosity functions from our model. We marginalise over the total
normalisation to get the best fit (Avni 1976). The scaling is performed relative to the 
number of stars in the
ranges $27.3<V<29.1$ and $25.8<I<27.3$. The $\chi^2$ fits to the observed luminosity functions are
applied over the entire magnitude range brighter than the 50\% completeness limit
(29.1 in V and 27.3 in I). The bins are 0.3 magnitudes large in both luminosity functions except for
the top of the two cooling sequences, where the regions $V<27.3$ and $I<25.8$ are counted as one bin
because of the low numbers of stars at these bright magnitudes.

\item {\bf Fitting the Hess diagram}:

A more quantitative method for taking account of colour information is to bin the white dwarfs in
the Hess diagram directly. Our grid for doing this is shown in Figure~\ref{Grid2}. For $I<26$, we
use a single bin, as the white dwarfs are relatively sparse and there is little age information
in the fine structure of this part of the cooling sequence.
 After the luminosity function jump
we use a finer grid, but use only those regions of the diagram that lie above the 50\% completeness limits in
both V and I. Once again, the overall normalisation is a free parameter over which we marginalise to
get the best fit. Normalisation is scaled relative to the total
 count in the bins shown
in Figure~\ref{Grid2}. 

\end{itemize}

\subsection{The Default Model}

 We begin with a fit to our default model. The cooling models are based on the cooling models
of Hansen (1999), with a hydrogen layer mass fraction $q_H=10^{-4}$ and a helium layer mass
fraction $q_{He}=10^{-2}$. We assume the white dwarfs at the top of the cooling sequence are
$0.55 M_{\odot}$ and the initial-final mass relation as defined in equation~(\ref{WoodM}). This
choice of mass, along with the best fit distance ($\mu_0=11.18$) implies an extinction
$A_V=1.39$. Colours and magnitudes are calculated using the transformations from Bergeron, Wesemael
\& Beauchamp (1996). The resulting fit is shown in Figure~\ref{Tx3}.

The best fit to the Hess diagram is an excellent $\chi^2=8.7$ for 11-2=9 degrees of freedom (i.e.
$\chi^2_{dof}=0.97$),
located at $(x,T)$=(-0.85,12.4~Gyr). The 2$\sigma$ age limit is $T=10.9$--14.2~Gyr (where
again, we interpret `$2\sigma$' as representing the range encompassed by the 95\% confidence
interval -- the distribution is {\bf not} gaussian).
 Strikingly,
the luminosity function fits yield somewhat different limits, with the V~LF favouring older
ages and the I~LF favouring younger ages, although both fits overlap the $2\sigma$ region
of the Hess~fit (albeit in different regions). This is an illustration of why we need to
examine the full colour information. An examination of the 50\% completeness limits in
Figure~\ref{Grid2} shows that the V and I~LF reflect different samples (since we
exclude different stars on the basis of the completeness limits). Henceforth we will
base our estimates on the Hess~fits as this is the only  way to treat
the colour information self-consistently.

At this point it is instructive to examine what distinguishes a good fit from a bad fit.
Figure~\ref{Hess_Tx3} shows the Hess diagram for our best fit solution $(x,T)=$(-0.85,12.4~Gyr).
Clearly both the colour and magnitude distributions are very well fit by the model. Figure~\ref{LF_Tx3}
shows the V and I~LF for the same solution. In both cases the shape of the luminosity function
rise is well-modelled and  the degradation in the $\chi^2$ is driven by the last few bins, which
is just where we expect problems to occur when we try to model a two-dimensional distribution
with a one-dimensional model.

We also examine two cases of what we classify as a bad fit. Figure~\ref{Hess_bad1} shows the
Hess~fit if we retain the same mass function slope as our best fit model, but reduce the
age to 10~Gyr. The $\chi^2_{dof}=5.7$ for this model and is driven up by two problems. The first
is that the model fails to reproduce the correct ratio of bright to faint white dwarfs, while
the second problem is that the colour distribution of the model is too blue at the faint
end. This is due to the fact that the lower age means the model white dwarfs are more massive, hence
fainter at late times, and
so enter the faintest magnitude bins while still hotter and bluer than their less massive counterparts.
Our second case of a bad fit is shown in Figure~\ref{Hess_bad2}. In this case we examine 
what happens when we keep the same age as our best fit model, but increase the mass function
slope $x$ to $x=1.35$ (Salpeter's value). In this case we again find the problem that the model cannot fit the
ratio of bright to faint white dwarfs (although the problem is in the opposite direction in
this case!) and furthermore, although the shapes of the colour distributions are roughly
correct, the relative numbers of white dwarfs in the three different fainter magnitude ranges 
are no longer well fit. The $\chi^2_{dof}=2.5$ in this case.

 The models used here are those that correspond
most closely to that favoured by our prior prejudices -- the extinction is approximately
that inferred from the subdwarf fits to the main sequence, the white dwarf mass at the
upper end corresponds to that expected from stellar evolution and the initial-final
mass relation is similar to that used in other contexts. The cooling models also
assume traditional hydrogen and helium layer thicknesses.
We now wish to explore the sensitivity of our results to a variety of model uncertainties. 

\subsection{Number Counts}
\label{numbers}

Before we investigate other parameters, we note that there is a piece of information we
have not yet used. The mass function slope $x$ that we quote is used only to determine the
probability distribution of progenitor masses. At no point in the above analysis have we
made any attempt to connect this to the observed number counts of white dwarfs and main
sequence stars. In fact, we will argue below that the relation between 
the quantity $x$ and the true mass function is not trivial. However, to begin with
we will treat $x$ as the true mass function slope. In that event, and
 assuming continuity of the cluster mass function, we can obtain  additional
information as follows.

 Each solution we obtain predicts a total number
of white dwarfs with $V<29$. To proceed further, we need to make an assumption that connects the mass function
of the white dwarf progenitors to that of the observed main sequence stars. The main sequence is successfully
modelled (Richer et al. 2004a) as a power law mass function with $x\sim -0.9$ over the range
$0.095$--$0.56 M_{\odot}$. To connect to the power law at
the upper end, we assume continuity of $dN/dM$ at some break mass~$M_b$. Thus, integrating these two power laws,
the ratio of white dwarfs to main sequence stars is
\begin{equation}
\frac{N_{wd}}{N_{ms}} = \frac{x_1}{x_2} M_b^{x_2-x_1} \frac{ \left( M_L^{-x_2}-M_{TO}^{-x_2} \right)}
{ 0.56^{-x_1}-0.095^{-x_1}}
\end{equation}
where $x_2$ is the power law in the white dwarf progenitor mass range and $x_1$ is the corresponding lower main
sequence value. The white dwarf progenitors range from the turnoff mass $M_{TO}$ to some mass $M_L$ to be determined,
which is the progenitor mass of the faintest white dwarf we count. This is clearly going to be smaller than 
the canonical $8 M_{\odot}$ limit for white dwarf production, as the white dwarfs that result from the
most massive progenitors will have cooled beyond our
detection limit and will not have been counted.
We count 520 main sequence stars in this
limited range and 222 white dwarfs brighter than V=29. However, the white dwarf numbers are too low because of 
incompleteness (the main sequence stars are somewhat brighter and so not significantly affected by incompleteness).
Once we correct for this, the observed numbers are equivalent to 290 white dwarfs brighter than V=29.
 Thus, with $x_1=-0.9$, we find
\begin{equation}
0.56= - \frac{1.90}{x_2} M_b^{x_2+0.9} \left( M_L^{-x_2} - M_{TO}^{-x_2} \right) \label{Nratio}
\end{equation}
Thus, given a choice of $M_b$, this yields an equation for $T$ in terms of $x$ (since both $M_L$ and $M_{TO}$ are
functions of age $T$). This illustrates, in rough terms, the interrelation between $x$ and $T$ that we found in
previous sections, although it is important to reiterate that the previous solution did not use any information
about the lower main sequence stars.

There is still the matter of choosing $M_b$. Any mass $M_b>0.56 M_{\odot}$ is allowed.
Figure~\ref{Txq} shows the solution for
three different choices of $M_b=0.6,0.8,1.0 M_{\odot}$. They are compared to the (Hess fit) solutions generated for our
default model from the previous section. The covariance of $x$ \& $T$ in the detailed solution
is encouragingly similar to that obtained from this simple calculation with $M_b \sim 0.8 M_{\odot}$. We emphasize
that these values of $x$ come from entirely different lines of argument.

What is a reasonable range for $x$? Clearly a constant mass function
slope $x \sim -0.9$ is not viable if extended into the regime of supernova progenitors or even massive
white dwarfs. If a cluster loses a significant fraction of its mass during the course of stellar evolution,
the cluster will dissolve (Chernoff \& Weinberg 1990), so mass functions that are dominated by high mass stars
are not viable candidates. However, for ages of $\sim 12$~Gyr, $M_L$ at
our completeness limit is
only $\sim 1.3 M_{\odot}$. 
 Clearly, there
is little difference in postulating a break mass at 0.8$M_{\odot}$ or at 1.3$M_{\odot}$. In the latter
case one can adopt the above solution without violating the bounds of significant cluster mass loss.

\subsubsection{The Influence of Mass Segregation}
\label{Segregate}
However, mass segregation is important in M4 (Richer et al 2004a) and so the relation between the
white dwarf number counts and the main sequence number counts is no longer trivial.
Michie-King models of the cluster suggest that the number fraction of white dwarfs relative to
all main sequence stars is approximately constant throughout the cluster. 
 However, the relative
numbers of white dwarfs to turnoff-mass main sequence stars (which more accurately reflect the
spatial distribution of the white dwarfs when they had their progenitor masses) is a factor $\sim 1.3$ larger in
our field than in the cluster as a whole.
 If we repeat the calculation of equation~(\ref{Nratio})
but  extrapolate from $0.6 M_{\odot}$ to $0.8 M_{\odot}$ (using $x_1$=-0.9) and count white
dwarfs relative to only the mass range $0.6-0.8 M_{\odot}$, then the number of main sequence
stars in this range is 205 and thus $N_{wd}/N_{ms}=1.414$ in our field. 
This comparison is also a more robust measure of mass segregation because stars $M>0.6 M_{\odot}$ are
less susceptible to mass-dependent evaporation and tidal stripping. If we perform the calculation
again using these numbers without correcting for mass segregation, we infer $x_2 \sim 0.4$
for a representative case of $T \sim 12$~Gyr (so that $M_L\sim 1.3 M_{\odot}$ and $M_{TO}=0.87 M_{\odot}$)
and $M_b=0.8 M_{\odot}$. However,
if we divide $N_{wd}/N_{ms}$ by 1.3 to convert our field ratio to the global ratio, then we
infer a value $x_2 \sim 1.4$.

This is only an illustrative calculation as it depends on several uncertain factors, but it shows
that the effect of mass segregation is to make our value of $x_2$ appear smaller than the 
true global value. The fact that a star loses mass in becoming a white dwarf means that we see
more white dwarfs in our field than we would if they had retained the masses of their progenitors.
 Thus, we should be very cautious in our interpretation of $x_2$ -- it is
primarily a parameter that describes the probability distribution of white dwarf progenitor
masses in our field. Its relationship to the true global mass function is more uncertain
and the true slope is likely to be somewhat steeper. In fact, King-Michie models for the
full cluster suggest that the global value of $x_2 \sim 1.3$ (Richer et al 2004a).

\subsection{Sensitivity to cooling models}

Using our newly defined default model, we can examine the systematic effects of the various
uncertainties outlined in \S~\ref{Theory}. To keep track of the various effects, we record
the results in Table~\ref{BigTab1}.

\subsubsection{Hydrogen and Helium Layer Masses}
\label{Layers}

If we keep the helium layer mass fixed, and leave the internal core composition unchanged, but
reduce the surface hydrogen layer, we expect faster cooling. If we repeat the default model,
now using a surface hydrogen layer mass $q_H=10^{-6}$, then we obtain the contours of 
Figure~\ref{Tx34}. The best fit solution shifts slightly to $(x,T)$=(-1.6,12.4~Gyr). The
lower age limit drops to 9.8~Gyr. However, this model has a higher $\chi^2$ than the default,
so the lower age limit will be higher than this when we marginalise over the various model
uncertainties.

Similarly, there is an uncertainty in the helium layer mass. If we keep the default model
the same (including $q_H=10^{-4}$) but reduce the helium layer thickness to $q_{He}=10^{-3}$,
we again expect the cooling to be more rapid. The 
solution becomes Figure~\ref{Tx33}. The shift in the $2\sigma$ contours is small, although
the best fit solution does move to $(x,T)=$(-0.5,11.7~Gyr). Again, the $\chi^2$ values
go up across the board, indicating the best fit model at any given point is obtained with
our default parameters.

\subsubsection{Internal Composition}
\label{Core}

In addition to the uncertainty in the surface layer masses, there is also uncertainty in the
composition in the center of the star. The relative amounts of carbon and oxygen vary throughout
the star and are determined by the nuclear burning history of the star on the Horizontal Branch
and AGB stages. The somewhat uncertain $^{12}C(\alpha,\gamma)^{16}O$ nuclear cross-section also plays
an important part. Whereas the different sizes of the surface layers affect the rate of transport
of heat through the star, the internal composition determines the overall thermal energy available
to power the star, through both its influence on the heat capacity and extra sources of heat
such as latent heat or chemical separation.

The most conservative approach would be to simply compare with models of pure carbon and pure oxygen,
but that is too naive, as no stellar evolution models predict pure compositions\footnote{After all,
nobody calculates main sequence models with helium fraction=0!}. Rather, we have
calculated a second set of white dwarf cooling models using a different set of initial conditions.
Our default model was calculated using the chemical compositions 
drawn from the stellar evolution models of Mazzitelli \& D'Antona (1986).
 In addition we have obtained (C.A.Tout, personal communication)
the internal compositional
profiles corresponding to the stellar evolution models of Hurley et al (2000). We will use these
two models as an indication of the variation to be expected due to uncertainties in the 
calculation of interior compositions.

How do the input models differ? Figure~\ref{XX} shows the mass fraction in carbon as a function
of enclosed mass for a $0.6 M_{\odot}$ white dwarf for the case of the Mazzitelli \& D'Antona profile (solid line)
and Hurley et al profile (dashed line). We see that the default model has a higher carbon fraction
in the inner parts 
 and thus a correspondingly higher heat capacity (more ions per unit mass). Thus, we 
expect the Hurley models to cool a little bit faster. The cooling curves for two $0.55 M_{\odot}$ 
models are shown in Figure~\ref{Tout_comp}. The solid line is our default model and the dashed
line is the model with the Hurley et al compositional profile.

Figure~\ref{Tx4} shows the fit for our default set of parameters, but now incorporating cooling
models using the Hurley et al internal profile. The best fit model is now $(x,T)=$(-0.95,12.4~Gyr)
and the minimum $\chi^2_{dof}\sim 0.8$ is even lower than for the default model. The lower age
limit is now 10.4~Gyr and, with the lower $\chi^2$, this limit will hold up after the marginalisation.
The upper age limit moves to beyond $>15$~Gyr (the $2\sigma$ contour extends upwards beyond the left
edge of the plot). Figure~\ref{Hess_Tx4} shows comparison of the new best fit model with the
observations.

\subsection{Initial-Final Mass Relation}
\label{IFMR2}

The extreme range of the variation of white dwarf mass in Figure~\ref{IFMR} is given by our default model
(which yields the lowest white dwarf mass at a given main sequence mass) and the pop~II
relation from Hurley et al (2000), which yields the largest white dwarf mass. If we replace
equation~(\ref{WoodM}) with the latter relation but retain all other features of the default
model, then we obtain the left-hand panel of Figure~\ref{Tx5}.
Of course, a more self-consistent application is to apply the core composition from
Hurley et al with the initial-final mass relation from the same models. The result is
shown in right hand panel of Figure~\ref{Tx5}. There is little significant change in the
overall fits, in either case. 

This is perhaps not too surprising for it is only for
relatively young ages that the massive white dwarfs lie above our detection threshold
and these ages are already excluded from our confidence region. Figure~\ref{MVM} shows
the results of our Monte-Carlo models for two different cases. On the left we show the
white dwarf mass versus V magnitude for our best fit model, using equation~(\ref{WoodM}).
On the right we show the same thing, but for a model of age only 10~Gyr. A much larger
fraction of the white dwarf population lies above our detectability threshold and so
we probe a much larger range of white dwarf masses.

\subsection{Other Group's Models}

As an extension of  section \S~\ref{Core}, we would like to compare with the models of other groups
in the literature. We choose two sets of models from the recent literature -- those of Chabrier
et al (2000) and Salaris et al (2000). These have been chosen specifically because they incorporate
not only up-to-date internal physics but also make an attempt to treat the outer boundary condition
properly using stellar atmosphere models. We know of two other calculations which meet this standard
(Fontaine et al 2001; Prada Moroni \& Straniero 2003) but which did not publish tables that could
be used for this purpose. Figure~\ref{Ltcomp} shows the 0.6~$M_{\odot}$ cooling
curves for the three different groups. The most significant variation occurs at late times
(T$>12$~Gyr) where the differences are of order $\pm 1$~Gyr with respect to our models.

In performing this comparison we encounter the problem that the published models are on a rather coarse grid in
mass and thus not sufficient to properly compare with the full analysis we publish above. To
perform a fair comparison between different models, we will first perform degraded fits
to our models. We will
assume all white dwarfs have mass 0.6~$M_{\odot}$ (since all groups have published models for
this canonical white dwarf mass), but we will keep the $A_V=1.38$. Thus, strictly speaking the
models will not fit the upper part of the main sequence correctly. However, most of the age
discrimination in the models occurs at $I>26$, so this is equivalent to assuming some modest
evolution in the initial-final mass relation such that $M \sim 0.6 M_{\odot}$ at the magnitudes
of interest. Figure~\ref{Tx6} shows the fit using our default cooling models, but  
without enforcing the conditions of \S~\ref{Redden}. Comparing this fit with those below
 affords a direct comparison of the influence
of other cooling models. The character of the solution is the same, with the lower age limit
moving down to 10.4~Gyr. The best fit model has
 $\chi^2_{dof} \sim 1.1$, so the model is still quite acceptable.

The left panel of 
Figure~\ref{Tx61} shows the fit obtained when we use the $0.6 M_{\odot}$ cooling models of 
Chabrier et al (2000). Despite the inconsistency between the mass and the extinction, the
Chabrier et al models also provide an acceptable fit, although with a marginal 
$\chi^2_{dof} \sim 1.7$. The allowed age range is somewhat lower (9.8--12.1~Gyr), not surprising
given that these models are the fastest cooling of those compared in Figure~\ref{Ltcomp}.
The right panel of Figure~\ref{Tx61} shows the results obtained using the models of
Salaris et al. The best fit models show an age range 14.4--16~Gyr, as appropriate to the
slowest cooling models of those discussed. However, the minimum $\chi^2_{dof} \sim 2.1$, indicating
that these models do not provide a good fit. The fits are similarly poor with both the 0.55$M_{\odot}$ and 
$0.6 M_{\odot}$ models.

It is encouraging that the fits obtained with our
models and those of Chabrier et al have similar character and similar lower limits and that although the solutions
obtained with the Salaris models are markedly different they are also a very poor fit -- demonstrating
that the observations are now good enough to rule out some models. Perhaps in the end it
is not surprising that the slowest cooling models are the most susceptible to constraint --
a greater fraction of the entire luminosity function lies above our observational threshold.

\subsection{Helium Atmospheres}
\label{Helium}

We have thus far considered only models with hydrogen atmospheres. However, observations of
the Galactic disk show that a significant fraction of cool white dwarfs possess helium-dominated
atmospheres (Greenstein 1986; Bergeron et al 1990; Bergeron, Ruiz \& Leggett 1997). The
physics that determines this ratio is still somewhat uncertain and it remains
 an issue as to whether the appearance of atmospheric helium is a temporary phase in the life of a white dwarf
-- one in which the dwarf shows a helium atmosphere for a time, but whose cooling history is
still similar to that of a dwarf with a hydrogen atmosphere boundary condition, or whether
the cooling is truly that of a helium atmosphere white dwarf. This is an important distinction
because the low opacity and high photospheric pressures of a cool helium atmosphere lead to
much more rapid cooling. We will examine both cases, although most workers in the field 
favour the former.

In the Galactic disk, the fraction of cool white dwarfs with helium atmospheres lies
somewhere within the range 0.33--0.5.
It is possible that accretion from the ISM may be partially responsible for determining
the balance between hydrogen and helium atmospheres (for a more detailed discussion see appendix~\ref{ChemEvol}.)
If so, the accretion history for white dwarfs in clusters may be significantly different and
so we wish to examine a variety of possibilities.

Thus we will treat the helium fraction as a free parameter. The proper
procedure is then to evaluate the $\chi^2$ at each (x,T) value for a range of helium atmosphere
fractions and then marginalise over the free parameter -- adopting the best fit
helium atmosphere fraction at each value of (x,T). The resulting fit is shown in Figure~\ref{Tx8},
scanning a range of hydrogen fractions from 0.0 to 1.0.
 The left-hand panel shows the resulting
Hess fit contours. The solution looks very similar to the hydrogen case, shifting to only slightly
lower ages. The best fit model is now at $(x,T)$=(-1.2,12.4~Gyr), and is found for a model in
which 60\% of the white dwarfs cool with hydrogen atmosphers and 40\% with helium atmospheres.
This is consistent with the results found in paper~I, in which there was little change in the
model as long as more than 50\% of the white dwarfs have hydrogen atmospheres. This is
primarily because the
 most striking feature of the observed luminosity function is
the sharp rise in the numbers at $I>26$. As we have described in \S~\ref{Jump}, this jump is a 
consequence of the slowing in cooling due to the change in the outer boundary condition that
results when the photospheric opacity drops due to the recombination of atmospheric hydrogen.
When the atmosphere consists of helium instead, the evolution of the outer boundary condition
is different. In particular, recombination occurs at higher effective temperatures. Consequently,
we require a model with a significant fraction of hydrogen atmospheres in order to match the
observations.

If some white dwarfs change their atmospheric appearance as they age, then the cooling ages may still
be more representative of a hydrogen atmosphere boundary condition even if they do appear to show
a helium atmosphere at some point in their lives. In this case, a more accurate reflection of reality is
to simply assume all white dwarfs cool according to a hydrogen atmosphere boundary condition, but
to assume that some fraction show colours more characteristic of helium atmospheres. Figure~\ref{Tx14}
shows the results of a parameter scan if we marginalise over the fraction of stars with helium atmosphere
colours only (i.e. we assume they all still cool with hydrogen atmosphere boundary conditions). Again,
the results do not differ significantly from the default model.

\subsection{Sensitivity to atmosphere models}
\label{AtmosComp}

We have also calculated fits using the atmosphere models of Hansen (1998, 1999). Figure~\ref{Tx10}
shows the fit for our default model. The best fit model moves slightly downwards to
$(x,T)=$(-1.2,12.0~Gyr) and the confidence limits appear to tighten somewhat. However the minimum
$\chi^2_{dof}=2.1$, so that this is not as good a fit as with the Bergeron atmospheres.

\subsection{Binary Fraction}

We inferred in \S~\ref{Features} that our white dwarf sample contains a binary fraction of $\sim 10\%$ or
more. A non-zero binary fraction may also have an influence on our results because it makes stars appear
anomalously bright at fixed colour. We address this by allowing some given fraction of white dwarfs in our
sample to be unresolved binaries. The binaries are assumed to be randomly sampled from the isolated white
dwarf distribution i.e. we do not take into account any influence of binarity on the overall stellar evolution
of the white dwarf. Note, we only consider double degenerate binaries -- even the lowest main sequence star
is bright enough to dominate a white dwarf companion sufficiently to move the system beyond our colour cuts.
The solution for a 10\% binary fraction does not change significantly, with a best fit $(x,T)=$(-1,12.4~Gyr)
and a lower limit of 10.8~Gyr.

\subsection{Proper Motion Cutoff}

Figure~\ref{PMS} shows that our proper motion accuracy gets worse as the stars get fainter. 
Eventually, the scatter of the white dwarfs becomes comparable to the cut we assign to
separate cluster members from background and thus we must examine how sensitive our results
are to the proper motion cut. All the preceding calculations have used the proper motion cut~A
in Figure~\ref{PMS}. In Figure~\ref{TxB} we show the results obtained if we use the more
stringent proper motion cut designated~B for our default model. Note that
we do the comparison self-consistently in that we recompute all the artificial star tests
with the new proper motion limit. The age range does not change
significantly, although the solution is shifted to slightly larger values of $x$. 
Some expansion of the confidence contours is expected as we have reduced the number of 
observed stars by 16\%.

\subsection{Distance Uncertainties}
\label{DeltaD}

The uncertainty in the distance to globular clusters is the principal source of error for the
main sequence fitting method. As outlined in \S~\ref{Redden}, distance errors and extinction
are intimately tied in our analysis method. Keeping the white dwarf mass at the top of the
cooling sequence fixed, a change in the absolute distance modulus is reflected in the extinction. 
A 5\% reduction in the distance for our default model reduces the extinction to $A_V=1.12$.
Recall that the counterintuitive fact that extinction changes in the same sense as the absolute distance
modulus is because of the change in the reddening dominates over the change in the extinction
in matching the observed sequence (\S~\ref{Redden}).

Thus, to investigate the effect of distance uncertainty we allow for a range in distance and
extinction, but require both to meet observational bounds. We consider the true distance modulus
to vary in the range $\mu = 11.18\pm0.18$ (so the distance uncertainty is less for M4 than for most
other clusters, because its proximity to the sun allows astrometric measures to be made) and
consider a loose bound on the extinction of $1.1<A_V<1.5$. We have allowed errors twice as large
as inferred from the subdwarf match, to account for possible differences between white dwarf
and main sequence spectral shapes. We investigate a series of
masses $0.5<M<0.6 M_{\odot}$ and for each mass examine the range of $\mu$ and $A_V$ that
satisfies these constraints. This is shown in Figure~\ref{amu} -- we see that the extinction
limits rule out the lower part of the distance range for 0.5$M_{\odot}$ and the upper part of
the distance range for $0.6 M_{\odot}$. Masses larger than $0.65 M_{\odot}$ are ruled out by this
procedure.

Figure~\ref{TxD2} shows the result of marginalising our model fits over the allowed distance
and extinction range for the full range of upper cooling sequence white dwarf masses 0.5$M_{\odot}$--0.6$M_{\odot}$.
Also shown is the 2$\sigma$ contour for the allowed range of distance and extinction but restricting
the mass to 0.55$M_{\odot}$. The lower age limit thus obtained is 11.1~Gyr. The fact that this is larger
than that obtained for just the 0.55$M_{\odot}$ model is a reflection of the fact that the minimum
$\chi^2$ is found with M=0.6$M_{\odot}$ and $\mu_0=11.1$. The location of the minimum is again
similar to that of prior cases $(x,T)$=(-1.1,12.1~Gyr).

\section{Discussion}
\label{Discuss}

We have presented a rather bewildering array of possible systematic uncertainties which can influence
the model of the cluster white dwarf sequence. To properly assess the global systematic error on our
age determination, we now marginalise over all the model variations. We include all the
variations in physical inputs on our model, including core compositions, atmospheric models and
binary fraction. We do not include the fits obtained using other models from the literature, as
we wish to express the quantifiable model systematics within our model alone. The result is shown
in Figure~\ref{Txbig}. The global minimum is found at (x,T)=(-1.2,12.1~Gyr), with a 2$\sigma$ lower
limit of 10.3~Gyr. A summary of the various model fits is given in Table~\ref{BigTab1}.
Figure~\ref{Hess_best} shows the Hess~fit for our best fit model. Although we have not used the
luminosity function fitting method to locate this minimum, the resulting luminosity functions,
shown in Figure~\ref{LF_best} are also good fits, with $\chi^2=7.9$ for V and $\chi^2=3.8$ for I,
where each has 5 (7-2) degrees of freedom.

The overall character of the solution remains unchanged throughout the variation of the systematic
uncertainties. A strong covariance between age and the parameter $x$ is evident, in the
sense that larger ages require smaller $x$, with ages $\sim 14$~Gyr requiring $x \sim -2$. A shallower
`slope' of $x \sim 0$ implies a best fit age $\sim 11$~Gyr 
 The largest $x$ solutions have $x \sim 0.5$, with an
age $T\sim 11.3$~Gyr. A comparison between the white dwarf and main sequence
number counts yields a similar $x$--$T$ covariance (\S~\ref{numbers}), because larger ages imply a greater
fraction of the total white dwarf population has cooled beyond our detection threshold, requiring a
shallower progenitor mass function to match the observed counts. We should note that the full calculation
does not use any information about the main sequence number counts ($x$ is used only to choose the probability
distribution of progenitor masses in the Monte Carlo model), so that the fact that we obtain the same trend
 from two different lines of calculation is a sign of consistency. We repeat again our cautionary note
from \S~\ref{numbers}, that $x$ should be considered only as a parameter in the probability distribution
of white dwarf progenitors. The relationship between $x$ and the true mass function slope for the 
global cluster population is affected by the dynamical processes in the cluster. As we have shown,
Michie-King models for this cluster indicate that the true mass function has a slope $x>1$.

Our result is thus a best fit age of 12.1~Gyr, with a 95\% lower bound of 10.3~Gyr for the age of M4. This
is remarkably similar (12.6~Gyr best fit, with 95\% lower bound 10.4~Gyr) to that quoted by 
Krauss \& Chaboyer (2003) for the cluster system as a whole. There is
some evidence that there is some dispersion in age amongst the clusters, particularly for the
somewhat more metal-rich group (Stetson, VandenBerg \& Bolte 1996; Rosenberg et al 1999) which includes M4. However, our 
methods are not yet accurate enough to place a strong constraint on this question.
It is also encouraging that solutions of similar character are obtained with the completely
independent models published by Chabrier et al (2000). Even the fact that we are unable to
achieve a good fit with the models of Salaris et al (2000) is encouraging -- it suggests that
our data and methods are now robust enough to begin discriminating between models. This is  a necessary
step in the direction of making the white dwarf cooling method an alternative age discriminator.

Our age determination for M4 using the white dwarf cooling sequence allows us to perform a direct
comparison between the age of the solar neighbourhood and the globular cluster system, using the 
same method (paper~I). Our best fit model has not changed significantly since paper~I,
but the larger error bar due to model uncertainties has brought the lower limit down to 10~Gyr.
This is still larger than the $7.3 \pm 1.5$~Gyr ($2\sigma$ range) we infer for the solar neighbourhood. Although
the best fit models for the two populations support the considerable delay  predicted by
some models for thin disk formation (e.g. Burkert, Hensler \& Truran 1992), our 2$\sigma$ limits
are now consistent with a delay $\sim 1$~Gyr. 
It is also
worth noting the dominant systematics are different in the two cases. In the case of the Galactic
disk, the faintest white dwarfs are dominated by helium atmospheres, which appear to be a minority
constituent in the globular cluster model (\S~\ref{Helium}). Figure~\ref{LDM} shows the sample
of disk white dwarfs from Liebert, Dahn \& Monet (1988), when we apply our preferred value of
extinction and distance to place them on the M4 cooling sequence. The turnover in the disk
luminosity function occurs at $M_I \sim 14.3$ (which corresponds to $I \sim 26.4$ for our default
parameters), well above the faintest white dwarfs in our sample
-- a clear indication of the age difference between the cluster and the Galactic disk.

Our description of Galactic chronology is consistent with other estimates as well.
The ages of solar and super-solar metallicity field stars with Hipparcos parallaxes appear to be
$\sim 7$--8~Gyr (Liu \& Chaboyer 2000; Sandage, Lubin \& VandenBerg 2003) although
some metal poor stars with larger ages ($\sim 10$~Gyr) may be kinematically part
of the thin disk. 

To conclude, in this paper we have performed a detailed analysis fitting models for the
cooling history of the cluster white dwarfs to our observations. We emphasize again that
we fit models to the full luminosity function -- we do not rely on localised features
such as a putative truncation in the luminosity function. We have demonstrated that 
not only is it possible to get excellent fits to the observations, but it is also
possible to rule out significant portions of parameter space. This latter point is
important as it demonstrates that the method has some power to discriminate between
models.
 We stress this because in the past the comparison with white dwarf models
and observations has been rather more qualitative than quantitative. This was in part
due to the poor statistics at the faint end of the disk luminosity function and in
part because the uncertainties injected into the model by an extended star formation 
history. The quality of observations has now reached the level where more detailed
and statistically rigorous theoretical analyses are possible. This is a necessary step in making the white
dwarf cooling sequence a useful tool for age determination.

\acknowledgements
RMR \& MS acknowledge support from proposal GO-8679
 which was provided
by NASA through  grants from the Space Telescope Science Institute, which
is operated by the
 Association of Universities
for Research in Astronomy, Inc., under NASA contract NAS5-26555.
MR also acknowledges the hospitality of the University of British Columbia,
where some of this research was done.
The research of HBR \& GGF is supported in part by 
the  Natural Sciences and Engineering Research Council of Canada.
 HBR extends his appreciation to the Killam Foundation and
the Canada Council through a Canada Council Killam Fellowship. 
BKG acknowledges the support of the Australian Research Council through its
Large Research Program A00105171.

\newpage

\appendix

\section{Chemical Evolution of White Dwarfs}
\label{ChemEvol}

It is an established observational fact that a significant fraction of cool white dwarfs
have helium-rich atmospheres. Indeed, there is much evidence to suggest that the chemical
composition of an individual white dwarf evolves as it cools (Greenstein 1986; Bergeron et al 1990;
Bergeron, Ruiz \& Leggett 1997). The first element is that
the fraction of helium atmosphere dwarfs increases at cooler temperatures. Thus, some
hydrogen atmosphere dwarfs become helium atmosphere dwarfs.
This is thought to be
due to the growth of the surface convection zone, which can dredge up helium from deeper
layers and mix it with a thin surface hydrogen layer. However, it appears that at
lower temperatures the composition reverts back to a hydrogen atmosphere, at least 
temporarily.
The reasons for this evolution are still not well understood, but there is general agreement
that it has to do with the combination of the depth of the convection zone and the amount
of hydrogen on the surface. Indeed, the accretion of hydrogen from the interstellar medium
may play an important role.

If accretion from the ISM is important, then we might reasonably expect differences in the
chemical evolution of population~I and population~II white dwarfs, as the former spend more
time in the high density ISM of the Galactic disk and move will lower velocities. On the other
hand, white dwarfs in clusters may have even higher accretion rates if the intracluster medium
is enhanced as a result of cluster mass loss.

We may estimate the effect of the cluster environment by converting our present-day global
white dwarf birthrate ($\sim 10^{-6} yr^{-1}$ -- see \S~\ref{HotStuff}) into a mass deposition rate
into the cluster ISM ($\sim 3 \times 10^{-7} M_{\odot}.yr^{-1}$). If this mass builds up
in the cluster potential in between disk passages (which most likely result in periodic
stripping of the cluster gas) then we expect roughly $\sim 30 M_{\odot}$ to have built
up in the cluster ISM, leading to a mean density $n \sim 3 cm^{-3}$ (if we use the 
half-mass radius $\sim 3$~pc). This is actually higher than the solar neighbourhood
density. Furthermore, the stellar velocity dispersion is lower in the cluster as 
well, so that the accretion rate is 
\begin{equation}
\dot{M} \sim 4 \times 10^{-14} M_{\odot}.yr^{-1} \left( \frac{n}{3 cm^{-3}} \right) \left(
\frac{V}{5 km.s^{-1}} \right)^{-3},
\end{equation}
approximately 600 times larger than for a corresponding white dwarf in the solar 
neighbourhood. Note that the resulting accretion luminosity in this case is 
$\sim 4 \times 10^{-5} L_{\odot}$ (an issue first raised by Lin, personal communication),
which translates to a magnitude $V \sim 28.8$ -- i.e. right near the limit of our
detection.

However, this is almost certainly an over-estimate. The fact that the clusters do not
exhibit a substantial intracluster medium has been known for many years (Hesser \& Shawl 1977;
Gnapp et al 1996; Hopwood et al 1999). Only recently intracluster gas actually been
conclusively detected (Freire et al 2001). In this case, the cluster is 47~Tuc, and the
density $\sim 0.07 cm^{-3}$ corresponds to only $\sim 0.1 M_{\odot}$ in the inner 2.5~pc
of the cluster. Although this is determined from pulsar dispersion measures and hence only
a measure of the ionized gas, the UV flux resulting from the high stellar density is expected
to ionize most of the cluster gas, so we will adopt a more conservative measure of the M4 intracluster 
density as $0.1 cm^{-3}$. Furthermore, if the gas is predominantly ionized, the thermal velocity
($\sim 9 km.s^{-1}$) is larger than the stellar velocity dispersion, which further reduces the
expected accretion rate, to
  $\sim 2 \times 10^{-16} M_{\odot}.yr^{-1}$
(and reduces the accretion luminosity to well below our detection limit).
This rate is still a factor $\sim 3$ larger than expected for the Galactic disk.

Thus we conclude that, if accretion has a significant influence on the atmospheric chemical evolution
of white dwarfs, then cluster white dwarfs probably accrete more hydrogen than even disk white dwarfs.
Thus, it is quite possible that the temperature range over which disk dwarfs show a dearth of hydrogen
atmospheres (Bergeron, Ruiz \& Leggett 1997) may be smaller for the cluster white dwarfs and that
the overall fraction of dwarfs with hydrogen atmospheres may be larger for the cluster.

This should also introduce a note of caution if we wish to use the cluster white dwarfs to make
any assertions about putative populations of white dwarfs in the Galactic halo. While the two
populations do indeed share similar progenitor histories, their subsequent accretion histories
are likely to be markedly different. Cluster white dwarfs most likely enjoy an abundance of accreted hydrogen
even in the face of the constant removal of gas from the cluster, whereas true halo white dwarfs spend
most of their lives in the very underdense regions of the Galactic halo.

\section{The Gaussian Nature of the Errors}
\label{Errors}

An implicit assumption in the use of the $\chi^2$ statistic to determine confidence
intervals is that the distribution of errors is Gaussian\footnote{The determination of
the best fit model (the minimum of $\chi^2$) is not affected by this, but it is
necessary to assign
statistical significance to a value of $\Delta \chi^2$.}(e.g. Press et al 1992). Since we are dealing with number
counts of stars one might naturally assume this is true (or at least that the errors are
Poissonian, if not Gaussian). However, we note that
much of the spread in the white dwarf cooling sequence is the result of photometric errors.
We wish to examine whether the Poissonian nature of the number counts is still to be expected
when we include the artificial star tests in our calculation.

The variable whose probability distribution we need to examine is $N_i$, the number of stars
counted in a given bin, $i$. Thus, we sample our model distribution a fixed number of times
($N_0$) and then run it through the probability distributions that result from our artificial
star tests. We then examine the distribution of $N_i$. The true value $N_0$ is unknown, so we
fix it such that it yields a prescribed mean value of $N_i$. Figure~\ref{PN} shows the resulting
cumulative distribution (solid histograms) of $N_i$ for two of our faintest bins, $26.5<I<27.0$, $1.45<V-I<1.75$ 
and $26.5<I<27.0$, $1.75<V-I<2.05$, when the input white dwarf was located at the center of the
first bin. The mean values are arbitrary, but of the same order as
the true number counts and specifically chosen so that the two distributions do not coincide
too closely. The dotted lines indicate the cumulative Poisson distributions for the respective mean
values of each bin. We see that the errors are, if anything, slightly better than Poissonian!

\section{An `alternative analysis'}
\label{deMarchi}

Recently, de~Marchi et al (2003) have published a re-analysis of our data which fails to
reproduce our result. We have severe reservations about this work on both observational and
theoretical grounds. 

On the observational side, de~Marchi et al find that they cannot achieve field-cluster
separation at the depths we do. In fact, their limit is almost a magnitude brighter than
ours. The detailed reasons for this difference are discussed in Richer et al (2004b). In
short, de~Marchi et al use image-association stacks generated by the Canadian Astronomy
Data Center and the ESO Space Telescope Coordinating Facility (Micol \& Durand 2002).
The resulting images of faint sources are significantly more diffuse than those determined
by our method, and so the positional accuracy of the de~Marchi et al images is cleary
degraded. However, the principal
fault with the analysis of de~Marchi et al is their refusal to use the recentering
algorithm in ALLSTAR, which severely limits their ability to perform proper motion
separations. They claim that this is prone to measuring noise spikes as cluster members, but
it is easy to demonstrate by a simple experiment that this is an unfounded 
concern. By using the original finding list but randomising the stellar positions,
Richer et al (2004b) show that very few of these spurious `stars' pass the photometric cuts
we apply and those that do are rejected by the cluster proper motion cut. 

Given the flawed data analysis of de~Marchi et al there is little point in a detailed
critique of their theoretical analysis. However, a few points are worth noting.
As we discuss in \S~\ref{Redden}, it is important for internal consistency to ensure that the model white dwarf
sequence actually overlaps the observations for a given choice of mass, distance and
extinction. de Marchi et al quote a single choice for extinction (Harris 1996) which
requires a white dwarf mass $\sim 0.6 M_{\odot}$ at the bright end, yet they quote a range
of white dwarf masses. So, while some of their models are clearly consistent at the
bright end, it isn't clear that they all are. It is also worth noting that some of
the alternative extinction and distances they mention (e.g. Djorgovski et al 1993) can
be ruled out as they would require unphysically low mass ($<0.5 M_{\odot}$) white dwarfs.

The statement that their photometric errors were drawn from a Gaussian raises concern.
As can be seen from Figure~\ref{sigs} the dispersion at fixed magnitude can be
different for stars recovered brighter and fainter than the input magnitude. Furthermore,
the wings of the distribution are often larger than Gaussian -- which is important to
include in the analysis.
 We note here we are talking about the 
distribution of photometric errors, not the distribution of number counts in a given
bin, which we have demonstrated (\S~\ref{Errors}) is indeed well approximated by a Gaussian.

We have demonstrated in this paper that fitting to the full Hess diagram is a much
better method for model fitting than comparing to one-dimensional luminosity functions.
 de Marchi et al state a preference for  the I~LF that may bias their results as our
experience suggests fitting to the I~LF gives consistently lower ages than either
the V~LF or the full Hess fit. It is interesting to note that they furthermore claim
that they require an `unphysical' value of $\alpha=0$ (corresponding to $x=-1$) in
order to match our luminosity function. As we have discussed in \S~\ref{numbers}, the
assumption that $\alpha$ or $x$ is indeed the true mass function slope is a dangerous one, as this cluster clearly shows significant mass segregation.
As a result, the white dwarf population in our field is a product not only
of stellar evolution but cluster dynamical evolution as well. The quantities 
$\alpha$ or $x$ should be viewed as parameters in a probability distribution foremost,
and their interpretation as true physical values should be treated with great caution.

We close with a brief philosophical point. De Marchi et al assert that they would
consider the clear location of the luminosity function peak as a mandatory requirement
for determining the cluster age. If one is going to determine a cluster age, it will
be the result of a model fit and we fail to see why one would believe an age determined
from fitting a model to the cutoff if one does not believe that the model also
accurately describes the luminosity function above the cutoff! 
While the determination of a Galactic disk age has generally been
performed by this method, that was largely because a proper and rigorous statistical treatment 
of the data is 
 hampered in that case by poor statistics and the convolution of white dwarf cooling with an
uncertain star formation rate.
 We have demonstrated in
this paper that the current level of observations in M4 is sufficient to allow us
for the first time to perform proper, statistically rigorous, model comparisons. The demonstration of this
is that we can not only find models that fit the data but furthermore can rule out significant
portions of parameter space. We believe it is time to move beyond the hidebound attitude that only
the faintest white dwarf in a given population can be used for age discrimination.

\newpage
\figcaption[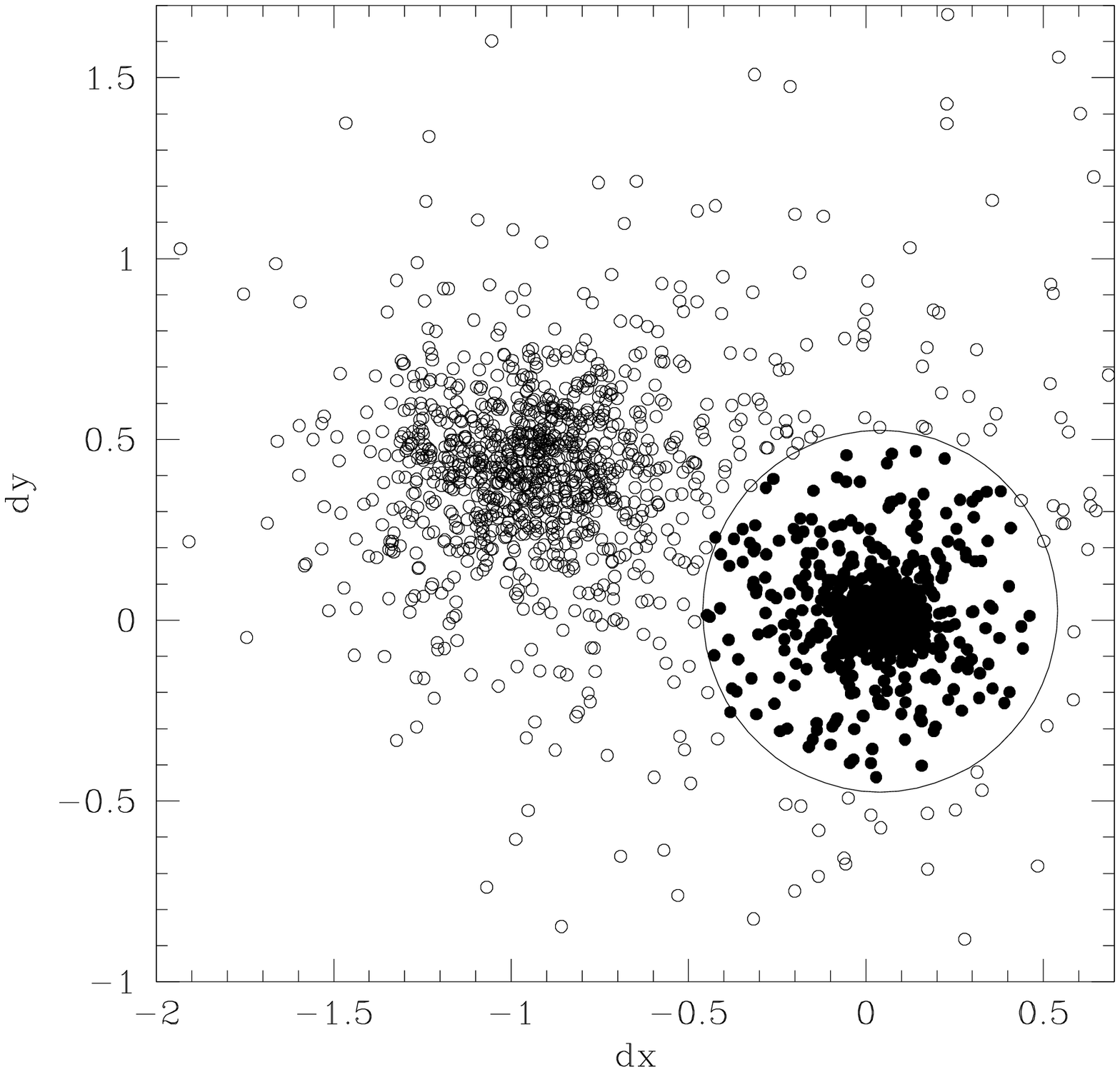]{The proper motion displacements $dx$ and $dy$ (in units of
HST pixels) are shown for the 6 year baseline spanned by our observations. The co-ordinate
system is centered on the bright cluster stars. All objects in the field are shown on this
diagram. A simple proper motion cut of 0.5~pixels separates cluster members (filled circles)
from background (open circles).  \label{plot_pm}}

\figcaption[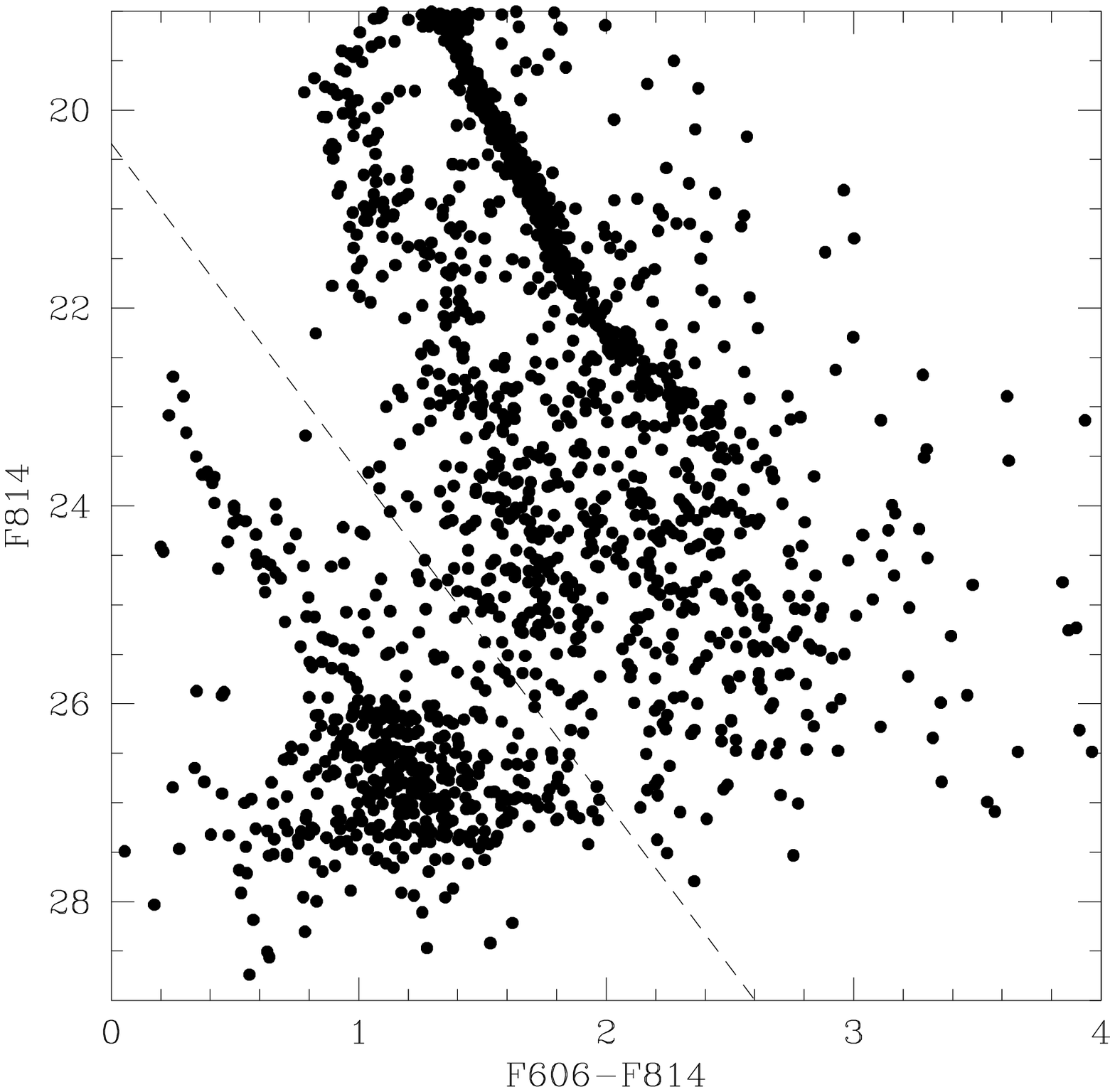]{ The colour magnitude diagram for all detected sources is shown. 
Three principal structures are evident -- the cluster main sequence at the upper right,
the inner halo (i.e. background) main sequence in the middle, and the cluster white dwarf
sequence at the lower left. The dashed line indicates a generous cut to define the region
of the CMD that will contain cluster white dwarf candidates. The proper motion displacements
for {\em all} sources below this line are shown in figure~\ref{PMS}.
 \label{CMD}}

\figcaption[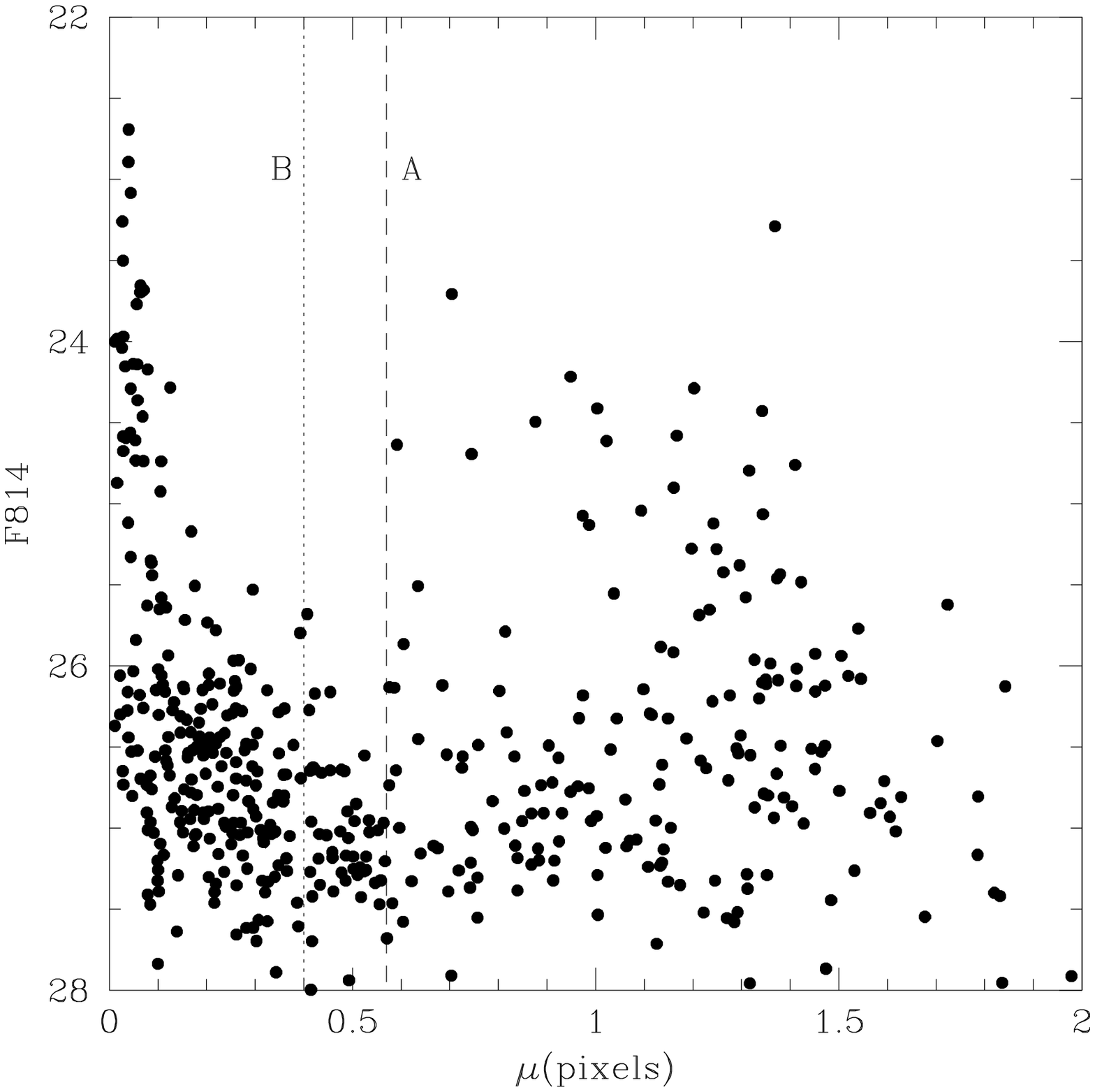]{ Here we show the proper motion displacements of all the 
`cluster white dwarf candidates' of Figure~\ref{CMD}. The horizontal axis $\mu$ is
the total proper motion displacement relative to the frame defined by the bright
cluster main sequence.
 It clearly splits into
two groups, cluster white dwarfs (on the left) and background objects. Even at faint
magnitudes, the concentration of displacements to the left indicates that most objects
are cluster white dwarfs. To investigate the influence on our results of which proper
motion cut we use, we define two different proper motion selection criteria, marked
as A and B. Another point to note is that considerations of completeness will restrict
our attention to white dwarfs with $I<27$.
\label{PMS}}

\figcaption[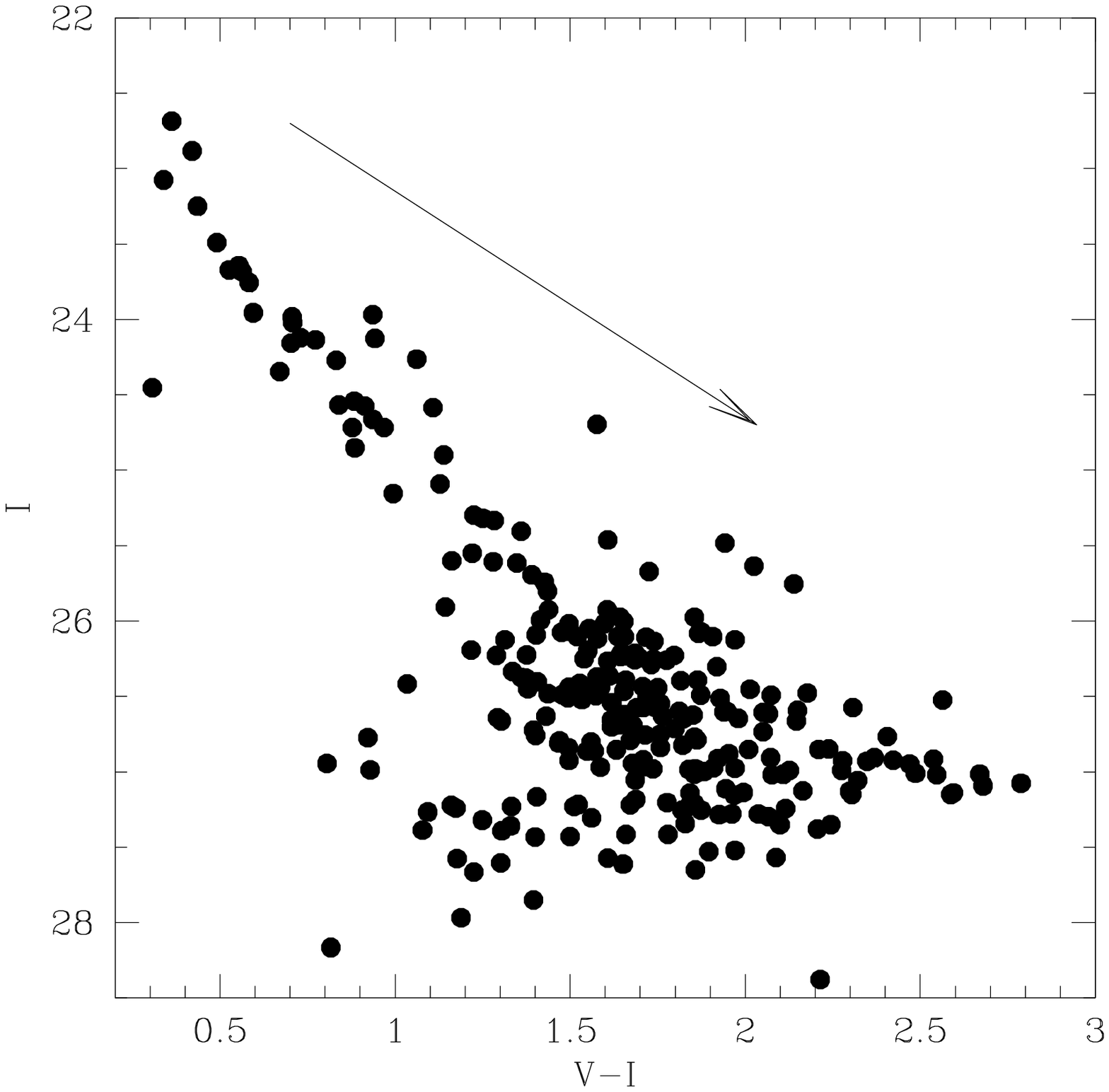]{ The resulting proper motion-selected white dwarf cooling sequence
for Messier~4 is shown. Also shown is the reddening vector -- thus a change in the extinction
moves stars along the sequence much more than across it. \label{Red}}

\figcaption[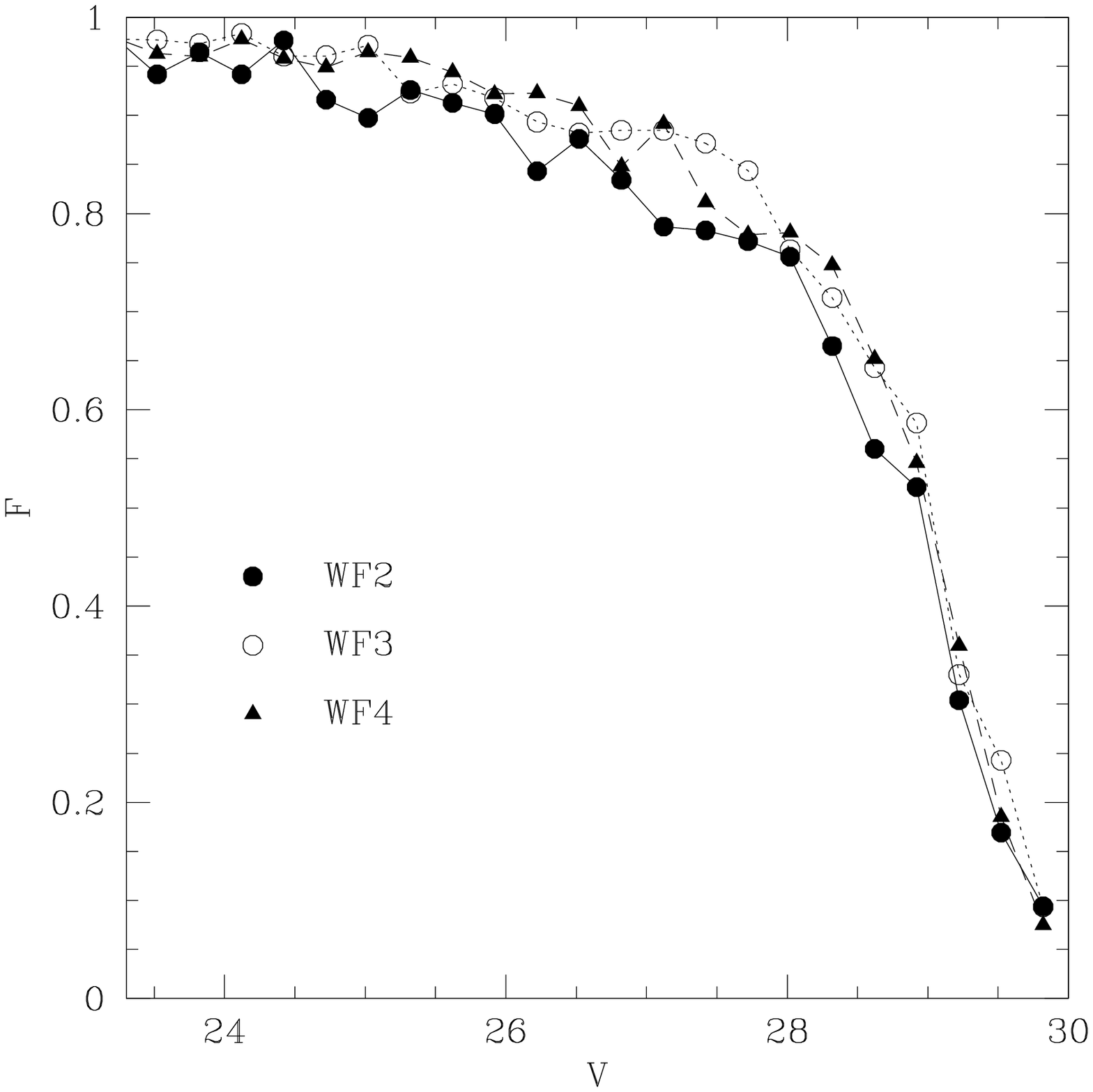]{ The three different curves indicate the fraction of stars recovered
as a function of magnitude for all three WFPC2 chips. This fraction includes stars recovered
at any magnitude but does require that the position be within 0.5 pixels of the true
position i.e. lie within the proper motion cutoff. The 50\% completeness limit is
located at $V \sim 29$ for all three chips.  \label{F0}}

\figcaption[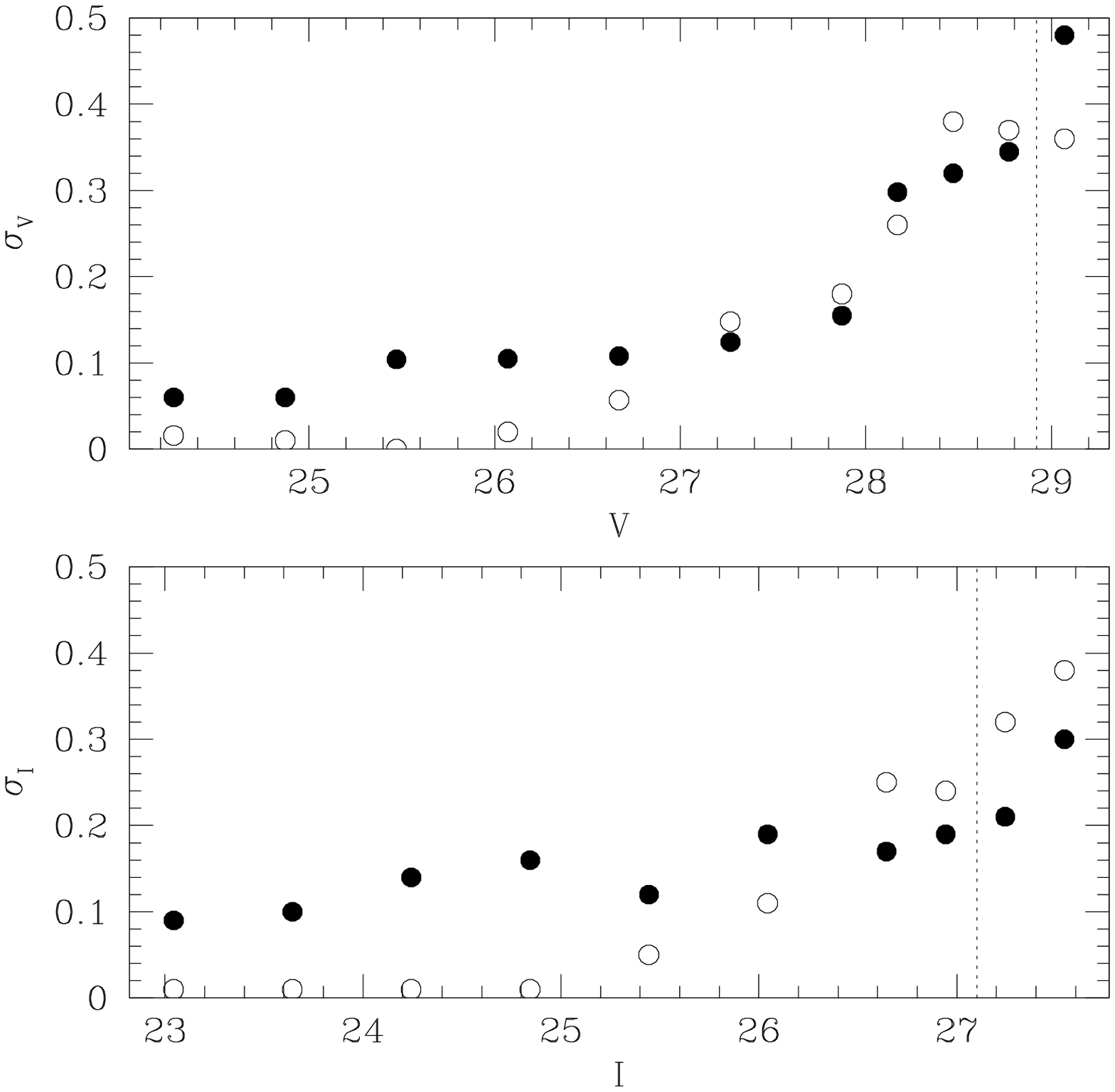]{The points represent the $1\sigma$ bounds on the distribution 
of recovered magnitudes (relative to input magnitude) as a function of input magnitude,
for the V~band (upper panel) and I~band (lower panel). The solid points represent the
tail to brighter magnitudes (so 16\% of artificial stars were recovered at a magnitude
brighter than $-\sigma$ above the input magnitude) and the open circles the tail to
fainter magnitudes (so 16\% lie fainter than mag+$\sigma$). The dotted line indicates
the 50\% completeness limit in both bands. Recall again that all artificial stars counted
also have to pass the proper motion cut. \label{sigs}}

\figcaption[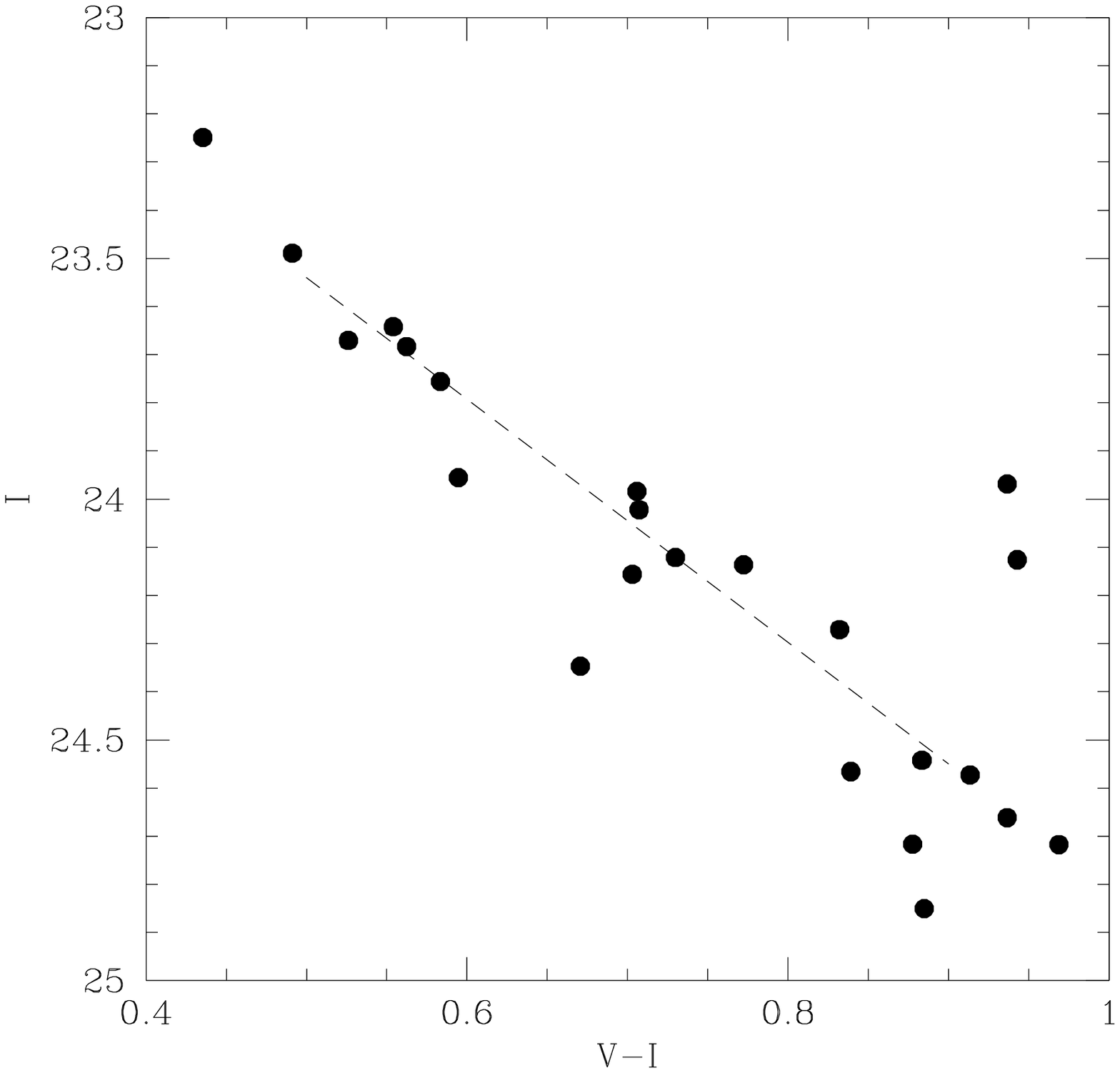]{ The upper part of the cooling sequence (solid points) is shown compared
to the theoretical model fit (dashed line) for the case of a $0.55 M_{\odot}$ model. The line
can be adjusted not only vertically (by distance and change in mass) but also with a horizontal
component (due to the changes in reddening associated with changes in extinction).
\label{CMDfit}}

\figcaption[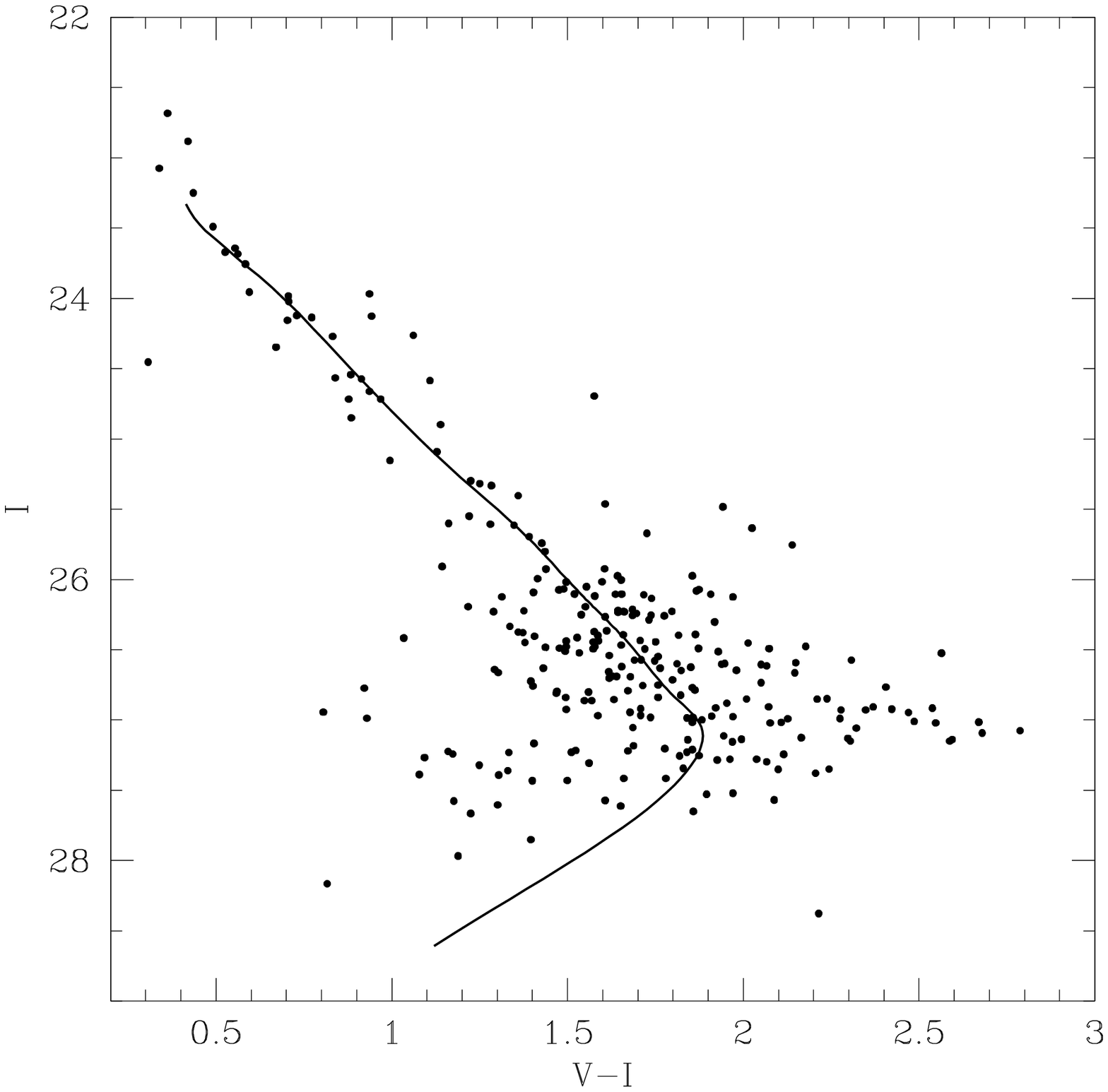]{ The solid curve shown is the cooling sequence for a white dwarf of
fixed radius. The value chosen in this case was appropriate to a cool $0.55 M_{\odot}$ dwarf and
thus the extinction is inferred to be $A_V=1.39$ to fit the upper part of the cooling sequence. \label{CMD_comp}}

\figcaption[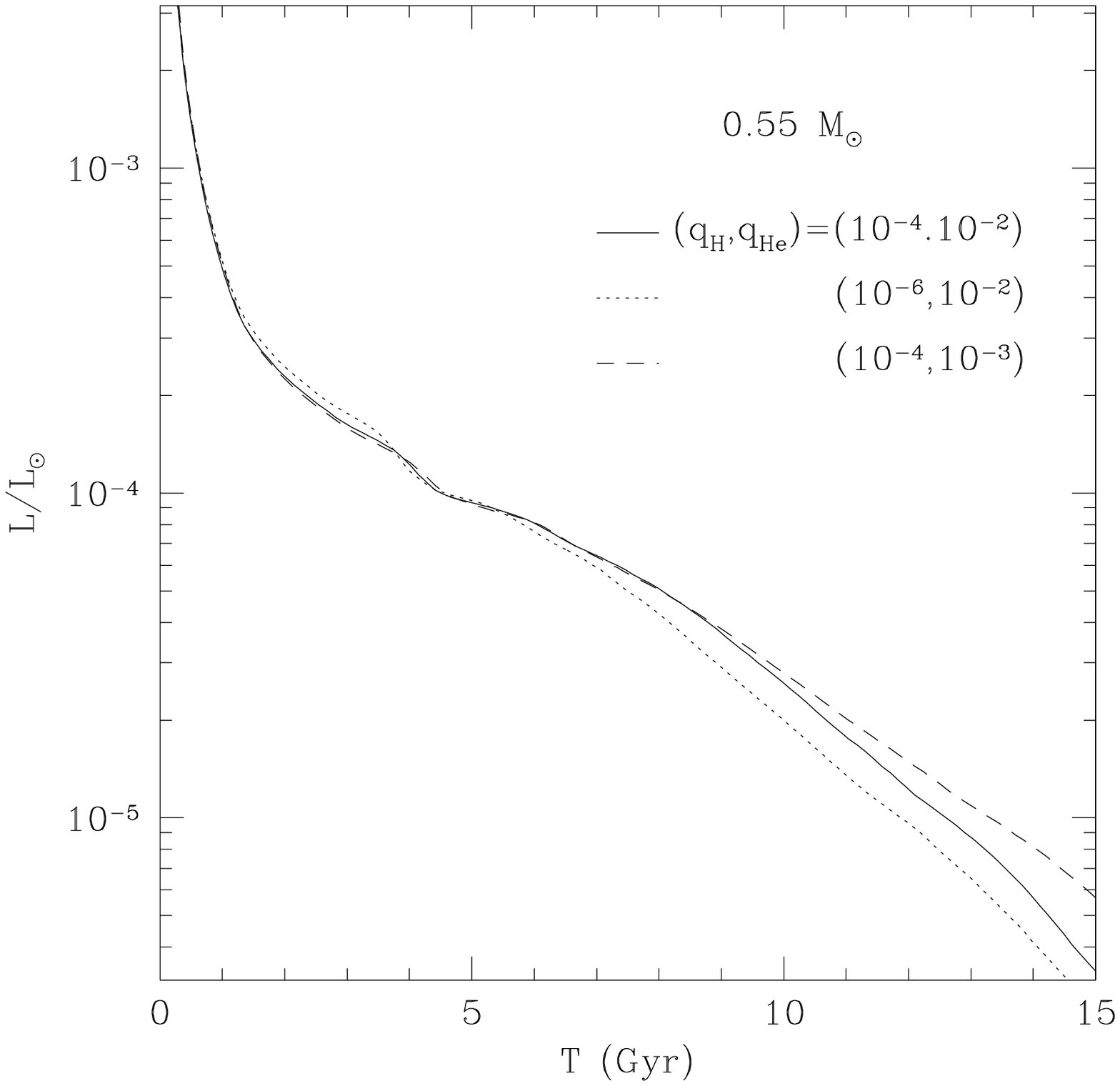]{The three models shown are identical 0.55$M_{\odot}$ cooling models except for
the different hydrogen and helium layer masses. Our default values are for mass fractions
$q_H=10^{-4}$ and $q_{He}=10^{-2}$. These are the most commonly used values and also at the `thick' end
of the expected spectrum. More massive hydrogen layers are unlikely (it would result in significant
nuclear burning i.e. the star would become a red giant again) but they could potentially be thinner.
We show the cooling curve for a $q_H=10^{-6}$ as well (dotted curve). The effects of a thinner helium
layer are also shown (dashed curve). \label{Cool}}

\figcaption[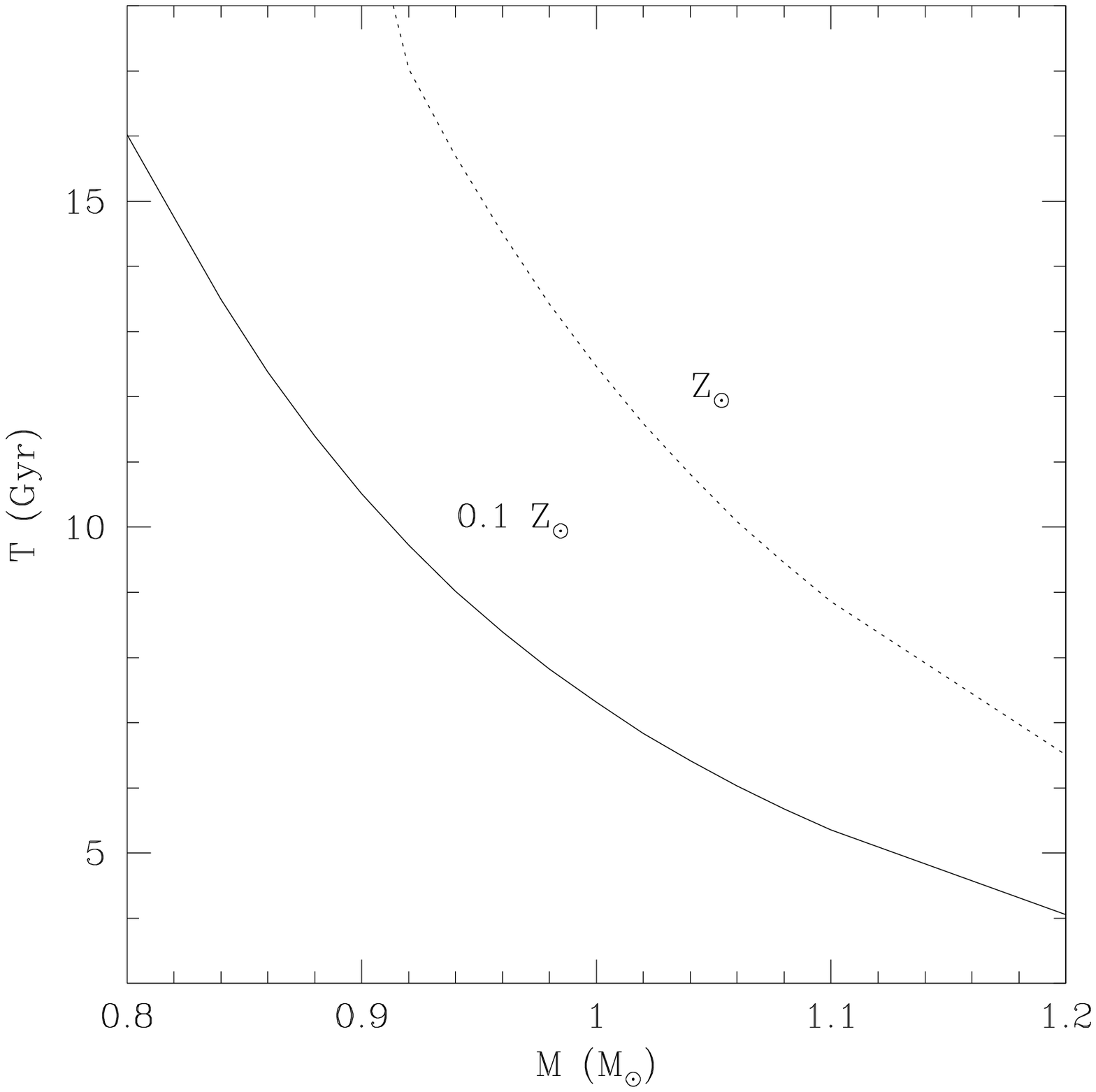]{ The solid line represents the duration of the pre-white dwarf evolutionary phases as
a function of Zero-age main sequence mass for stars of low metallicity appropriate to M4. The dotted
line indicates the corresponding value for a solar metallcity sequence. Both sets of models are taken
from the results of Hurley et al (2000).
 \label{tms}}

\figcaption[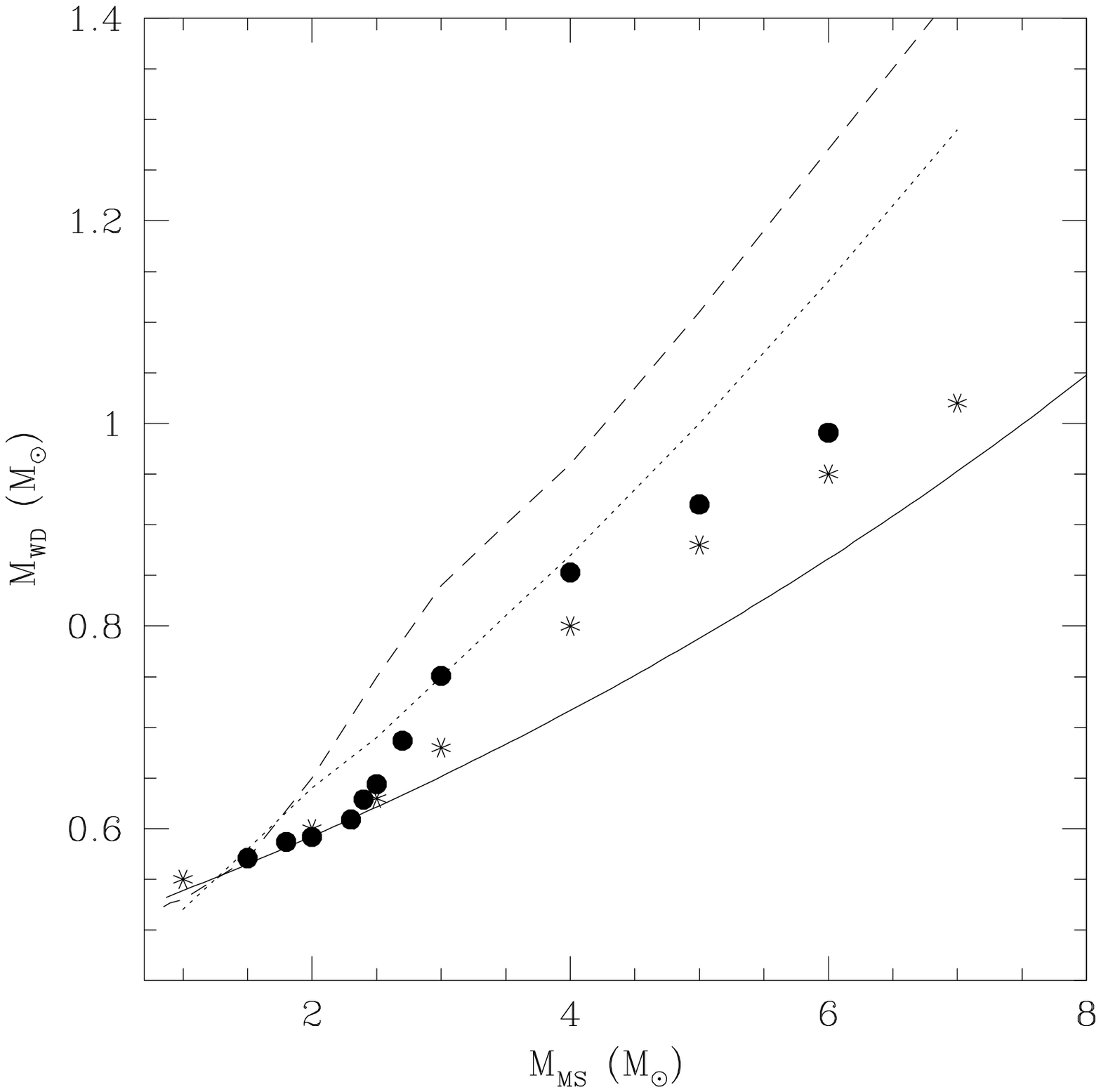]{The solid line is model A from Wood (1992), which is the model for default relation~(\ref{WoodM}).
The dotted ($Z_{\odot}$) and dashed ($0.1 Z_{\odot}$) are relations from Hurley et al (2000) which indicate the
range of variation expected from changes in metallicity between population~I and population~II white dwarfs. The solid
points are the $Z=0.05 Z_{\odot}$ models of Dominguez et al (2000), while the asterisks represent the
empirical relation from Weidemann (2000). \label{IFMR}}

\figcaption[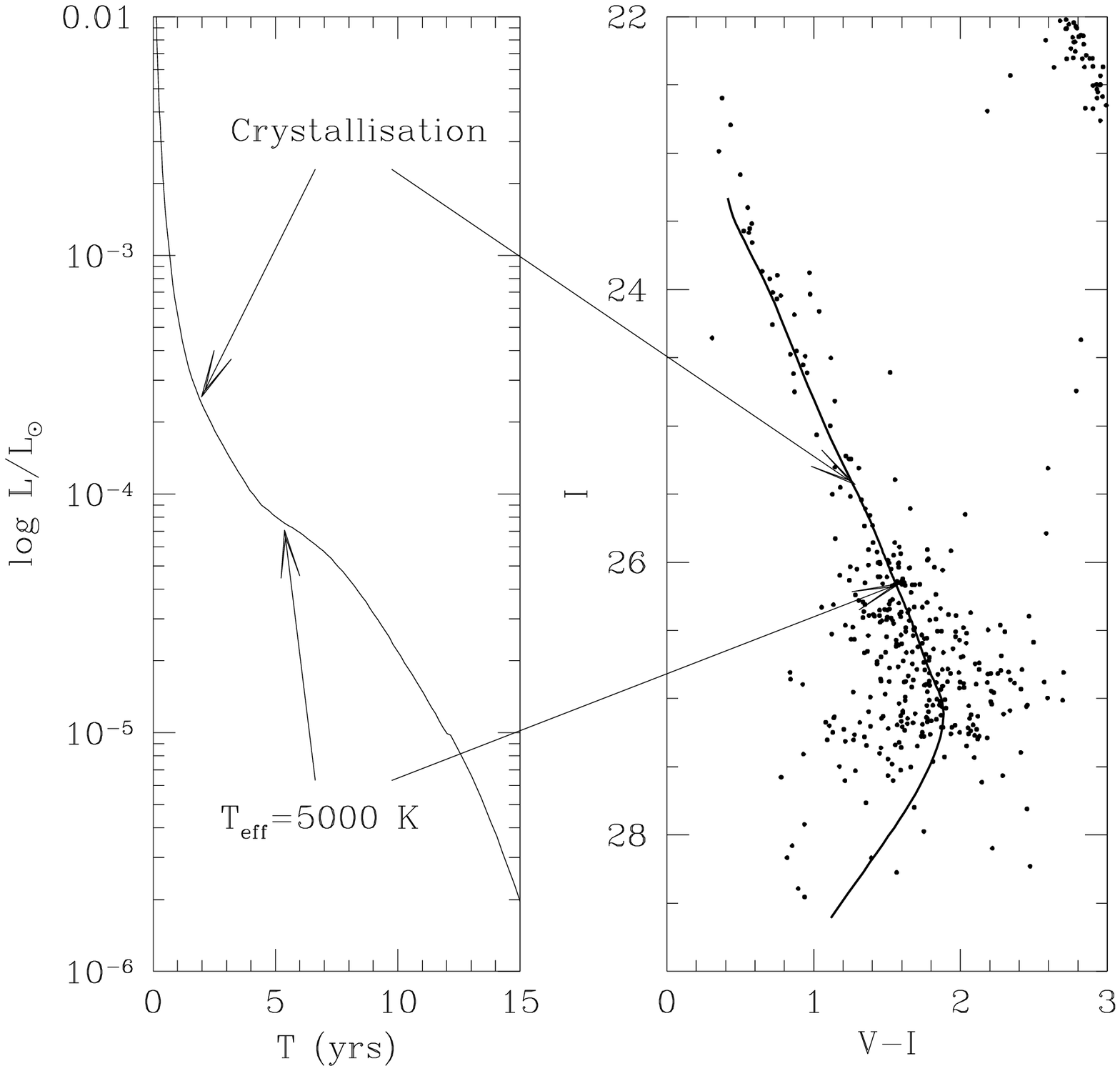]{The left-hand panel shows the cooling curve for a $0.55 M_{\odot}$ model. The
right hand panel shows the same cooling sequence in the colour-magnitude diagram, compared to the observed
cooling sequence. The arrows indicate the corresponding positions of two important epochs. The first
is the onset of core crystallisation, which contributes latent heat. The second is the point at
which the effective temperature reaches 5000~K. This corresponds to a drop in photospheric hydrogen
opacity and a change in the boundary condition of the cooling models. The result is a flattening of
the cooling curve which causes a corresponding jump in the observed luminosity function.  \label{Lt}}

\figcaption[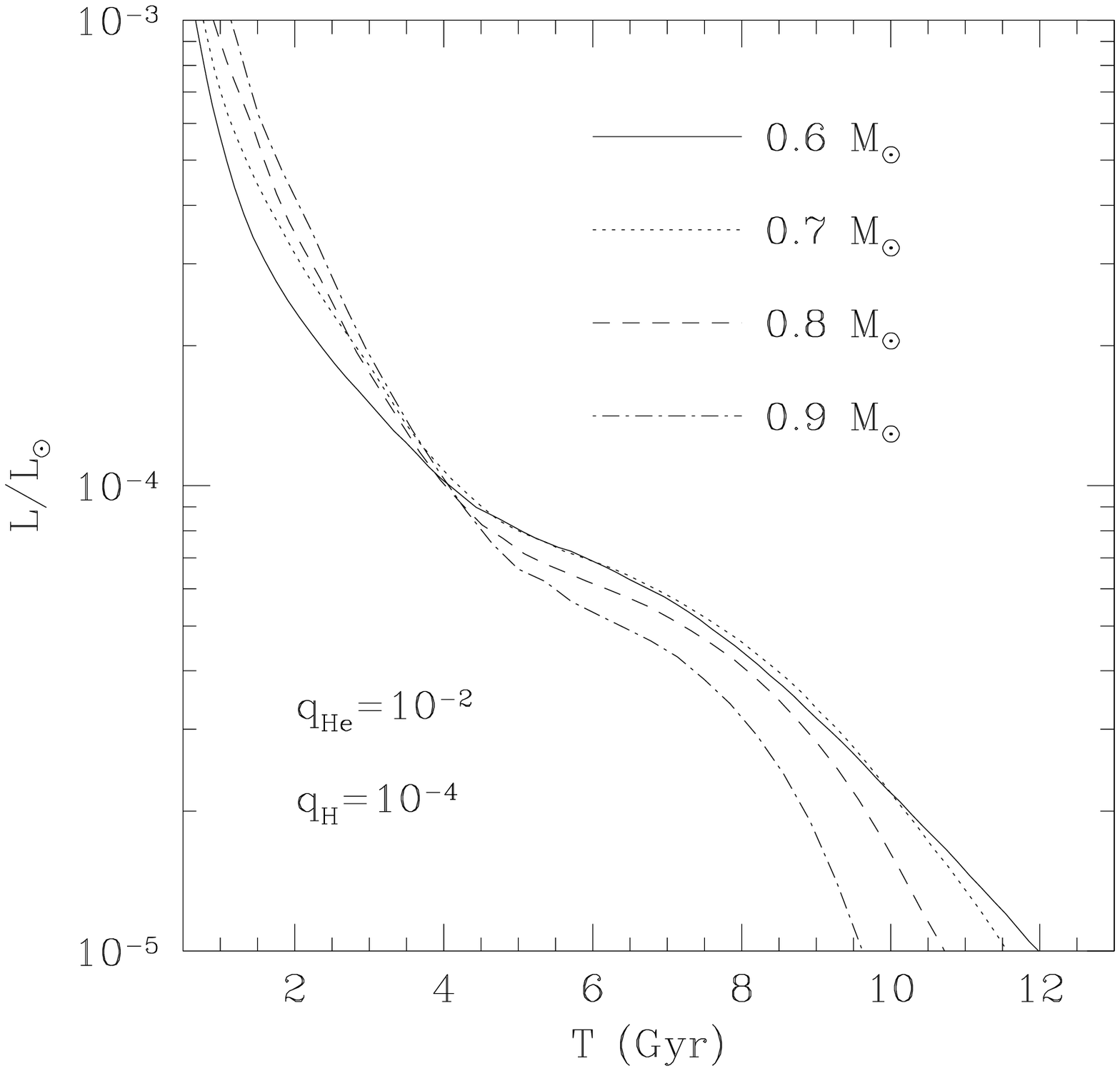]{The different curves indicate the cooling curves for white dwarfs of different
mass. At early times more massive white dwarfs are brighter because their larger mass corresponds to
a larger heat capacity. However, the higher central densities means that more massive white
dwarfs crystallize sooner and at late times cool faster because they enter the Debye cooling regime
earlier.  \label{Mcool}}

\figcaption[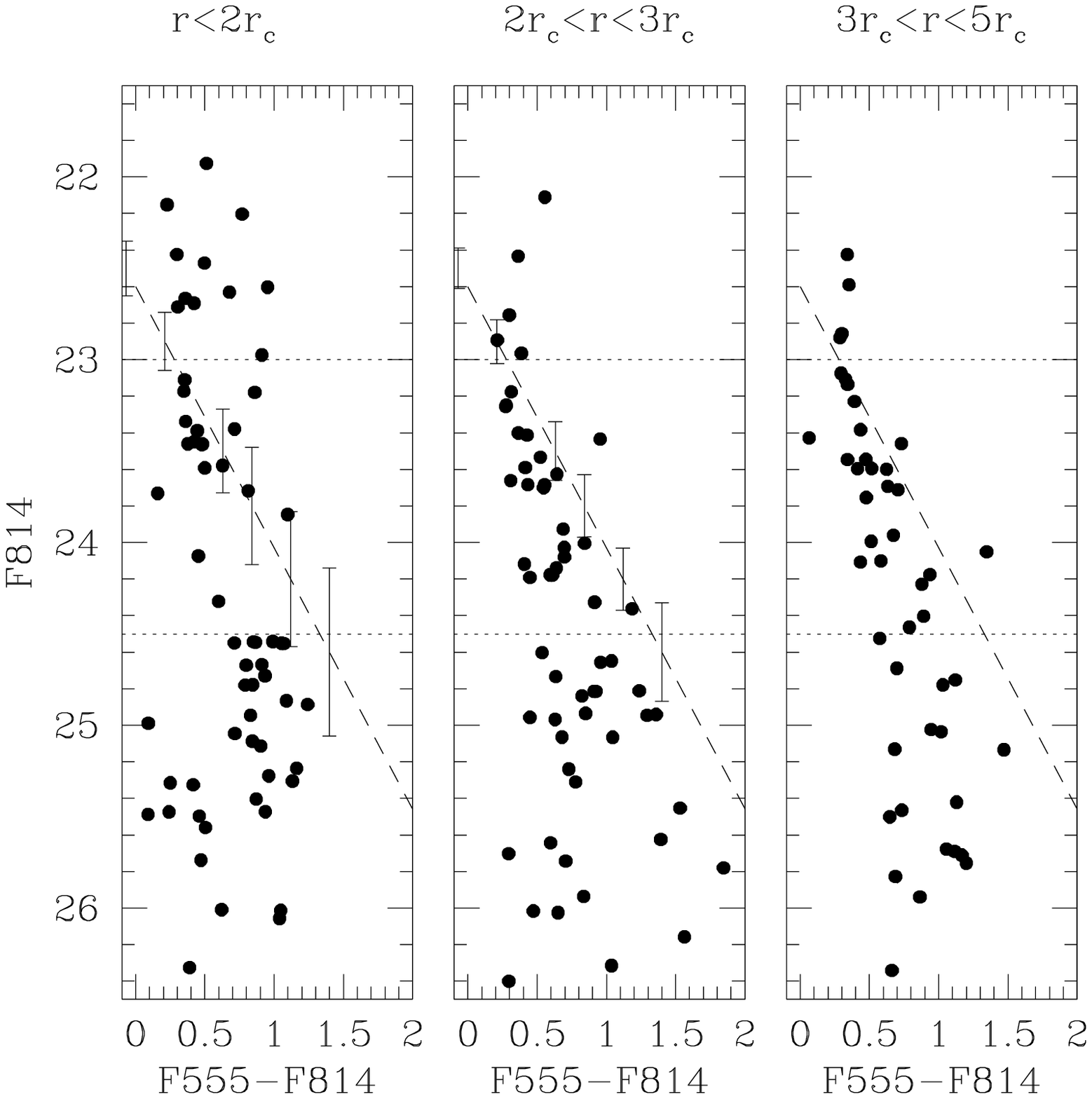] {The left-hand panel shows the upper cooling sequence for the annulus fro 1--2 core radii of
the cluster M4. The middle and right-hand panels show the same for the ranges 2--3 and 3--5 core radii.
The horizontal dotted lines bracket the region in which we determine the binary fraction. The dashed line
is the designated upper envelope of single white dwarfs. Our binary fraction is determined by counting what
fraction of the white dwarfs in this magnitude range lie to the right of the dashed line. The
error bars in the left and middle panels indicate the photometric uncertainties at given
places along the cooling sequence. \label{binfrac}}

\figcaption[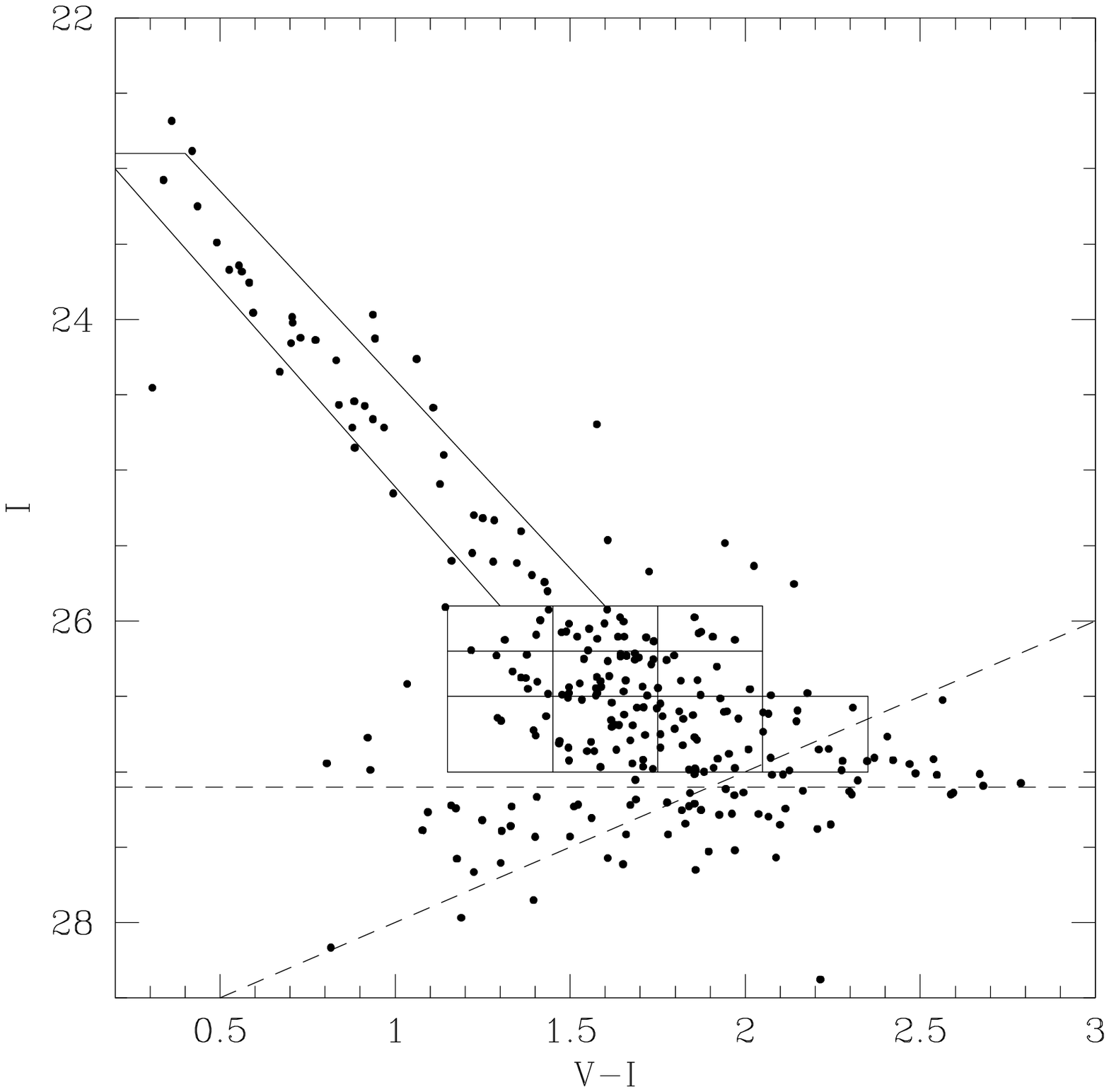]{ The boxes delineate the grid elements for the 
$\chi^2$ fits to the full colour distribution (Hess fits). The upper cooling
sequence is defined as a single grid point since there is little age information in
the detailed distribution at these bright magnitudes. The dashed lines indicate the 50\% completeness limits
for V and I. \label{Grid2}}

\figcaption[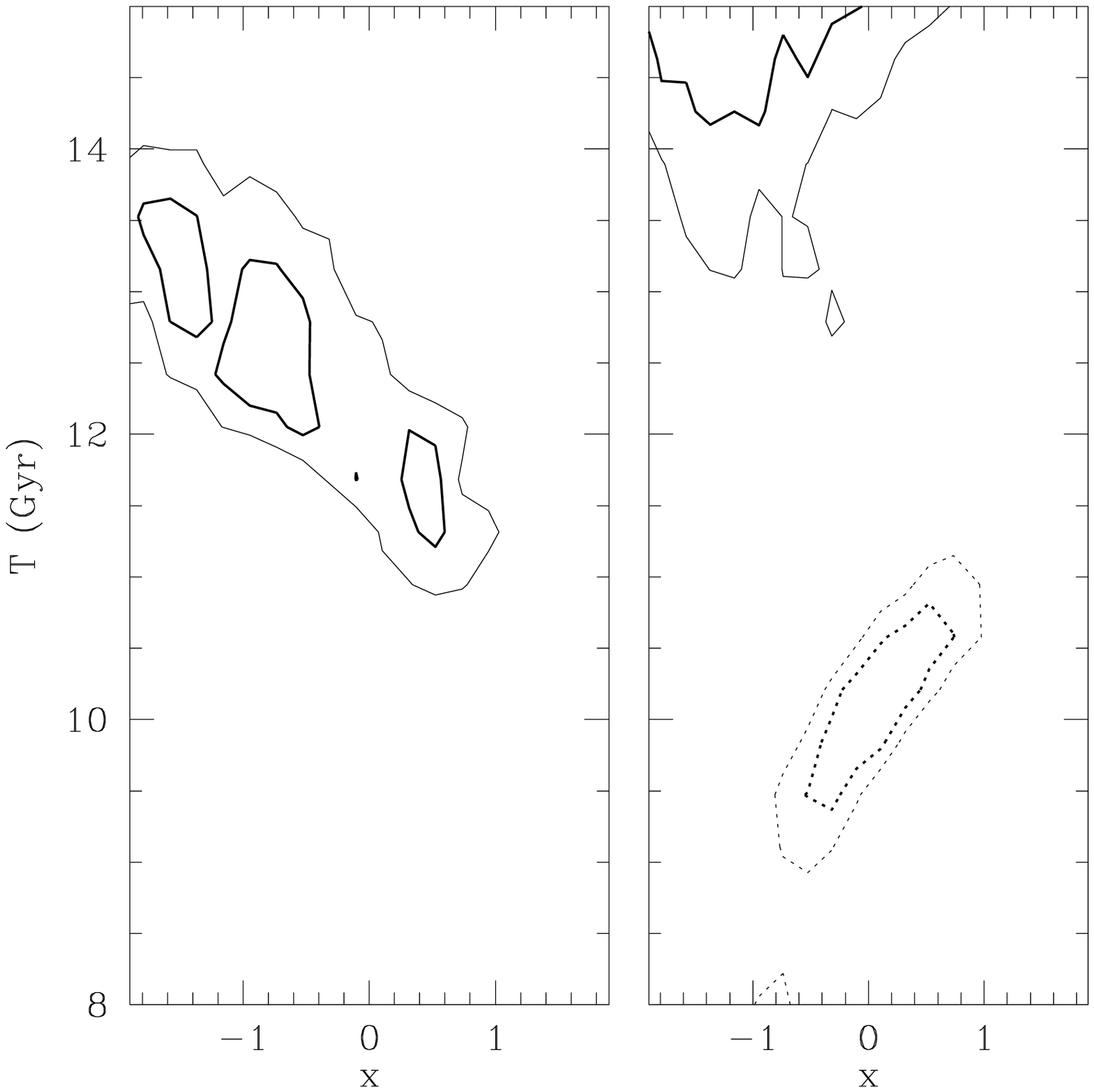]{The left-hand panel shows the $68\%$ (heavy solid contour) and $95\%$ (thin solid
contour) confidence intervals that result from the Hess diagram fits to our default model. The right
hand panel shows the corresponding results if we fit only to the V~band luminosity function (solid
contour) or the I~band luminosity function (dotted contour).
\label{Tx3}}

\figcaption[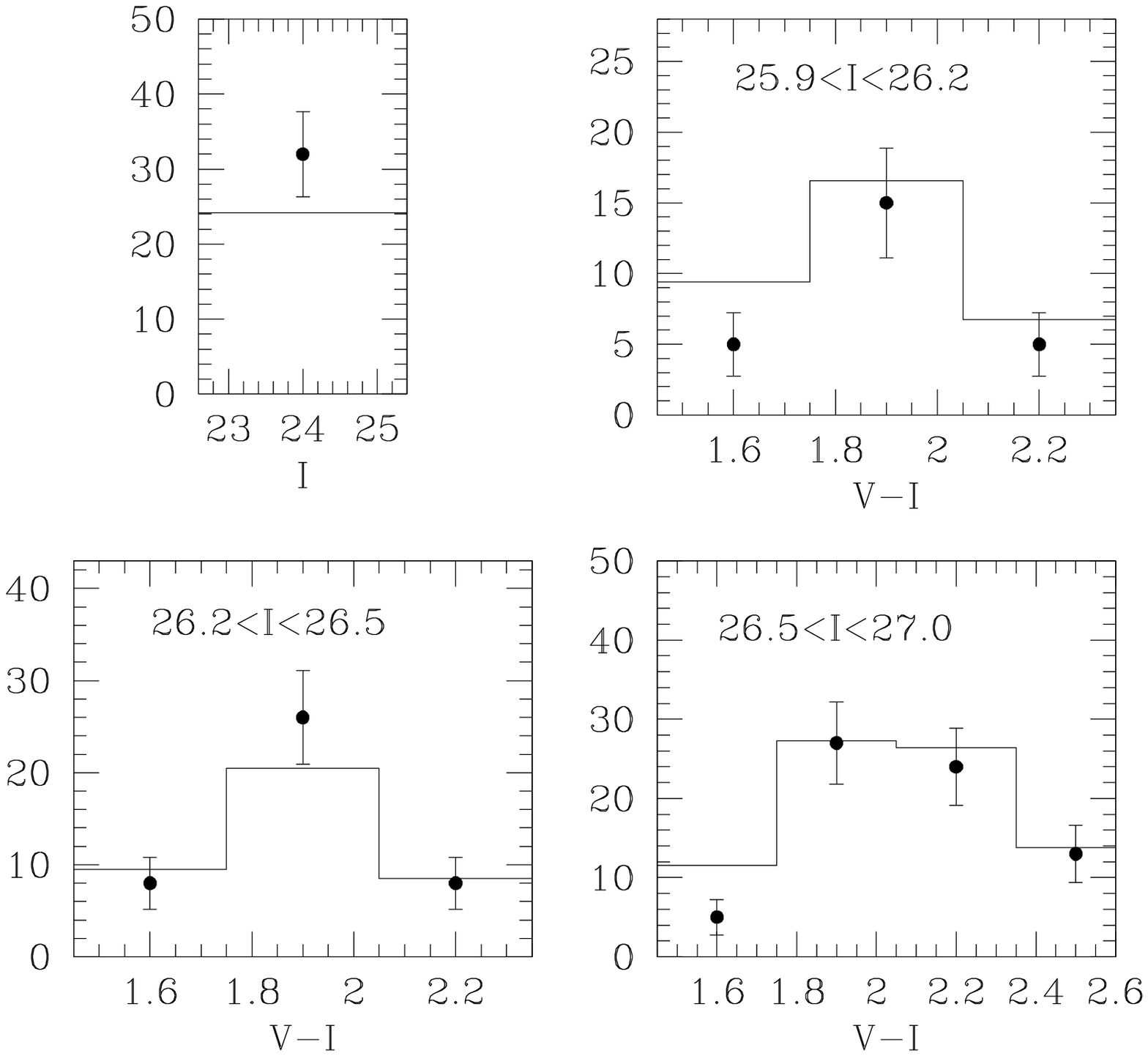]{ The upper left-hand panel shows the number counts in the large,
bright white dwarf bin. The other three bins compare the colour distribution in the
labelled magnitude ranges with the observations.
 The models match both the ratio of bright and faint white 
dwarfs as well as the colour distribution of the faint white dwarfs. \label{Hess_Tx3}}

\figcaption[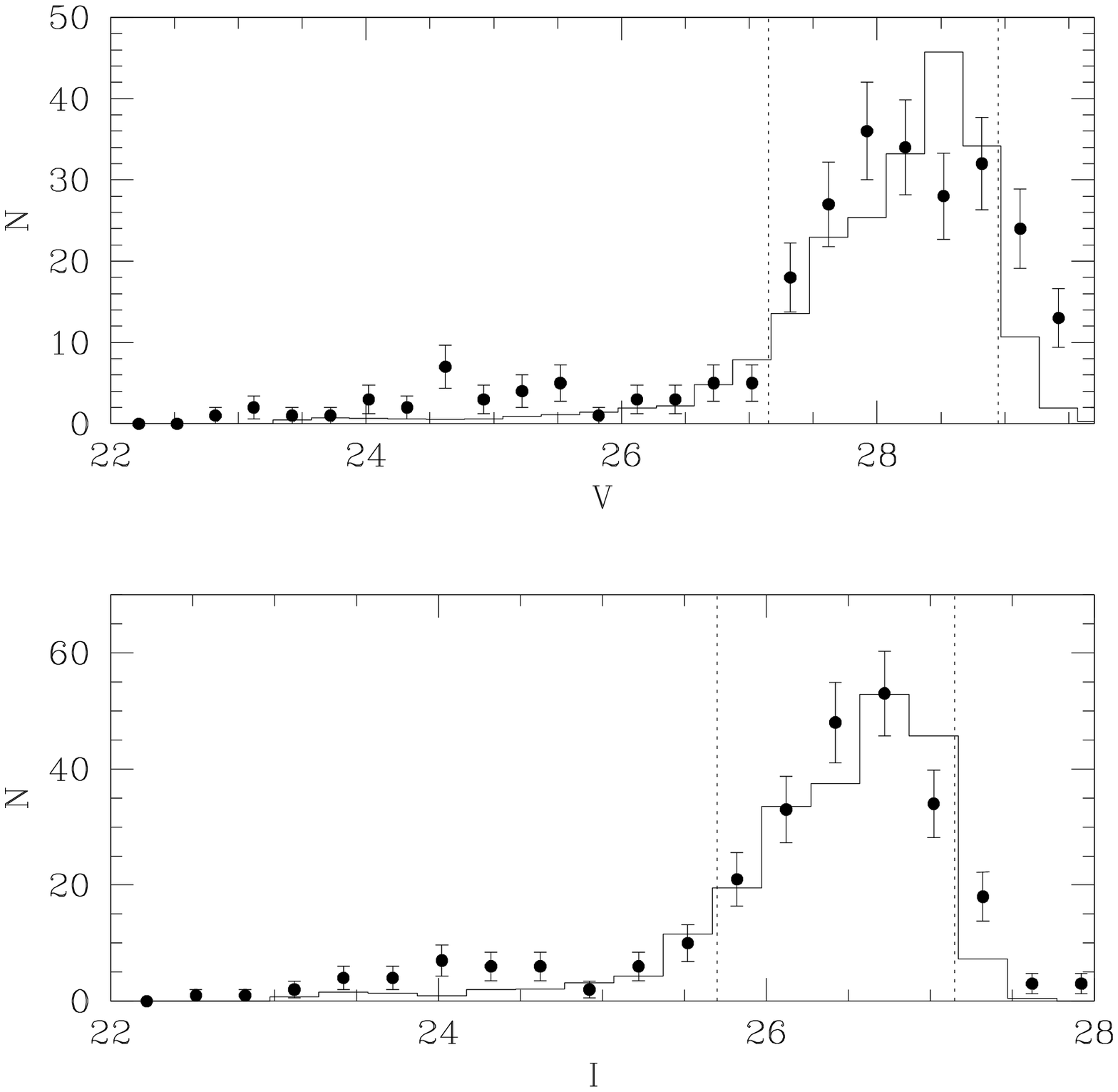] { The upper panel shows the V-band luminosity function and the bottom
panel shows the corresponding I~LF. The dashed lines in each case enclose the region where
the $\chi^2$ fit was performed. Fainter bins were ignored because of poor completeness, while
bins brighter than this region were summed together and included as a single bin in
the fit. \label{LF_Tx3}}

\figcaption[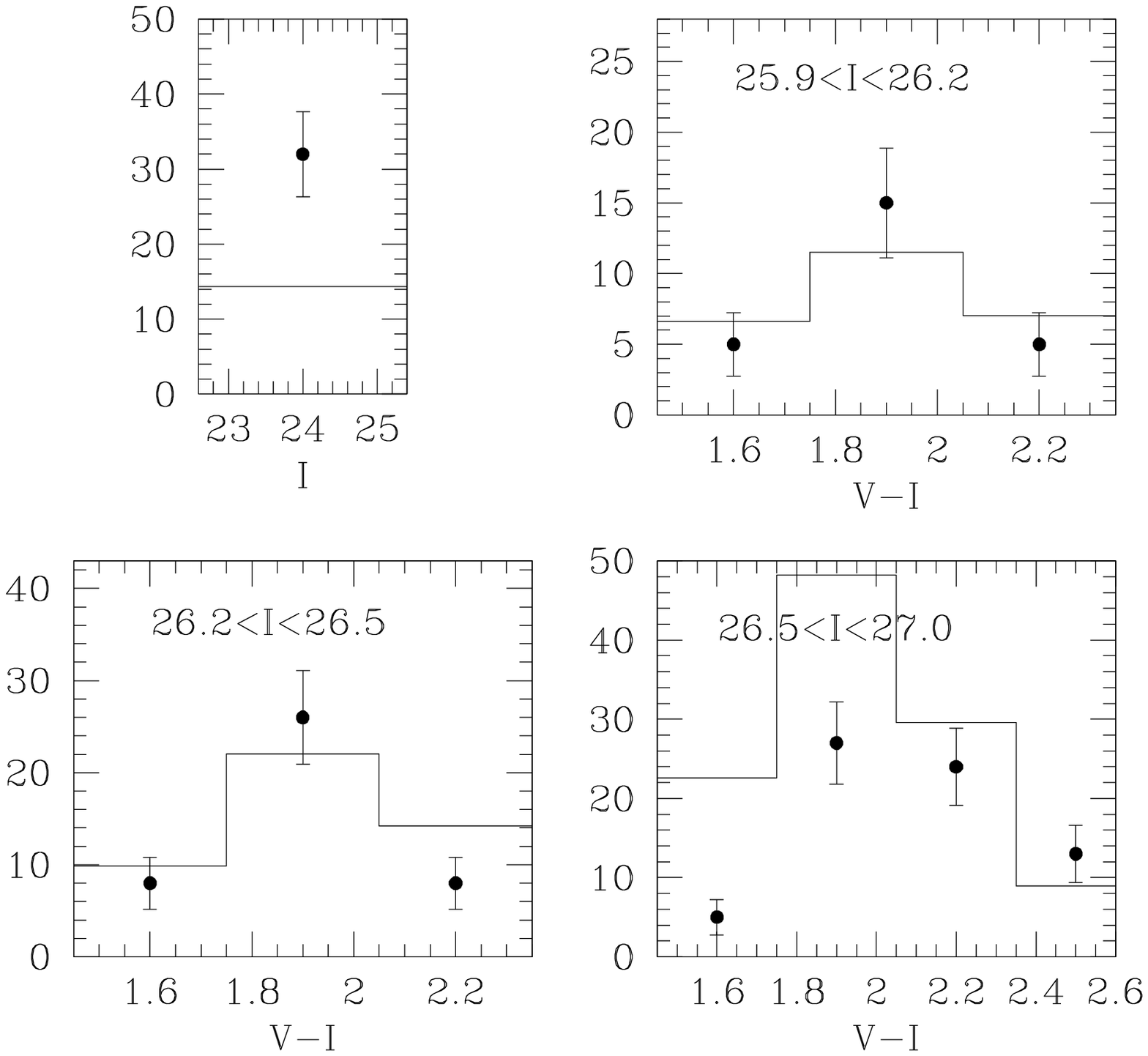] { This fit is for a young (10~Gyr) solution and clearly the model 
contains too many faint white dwarfs relative to the bright end. Furthermore, the colour distribution
is too skewed towards the blue at the faint end. \label{Hess_bad1}}

\figcaption[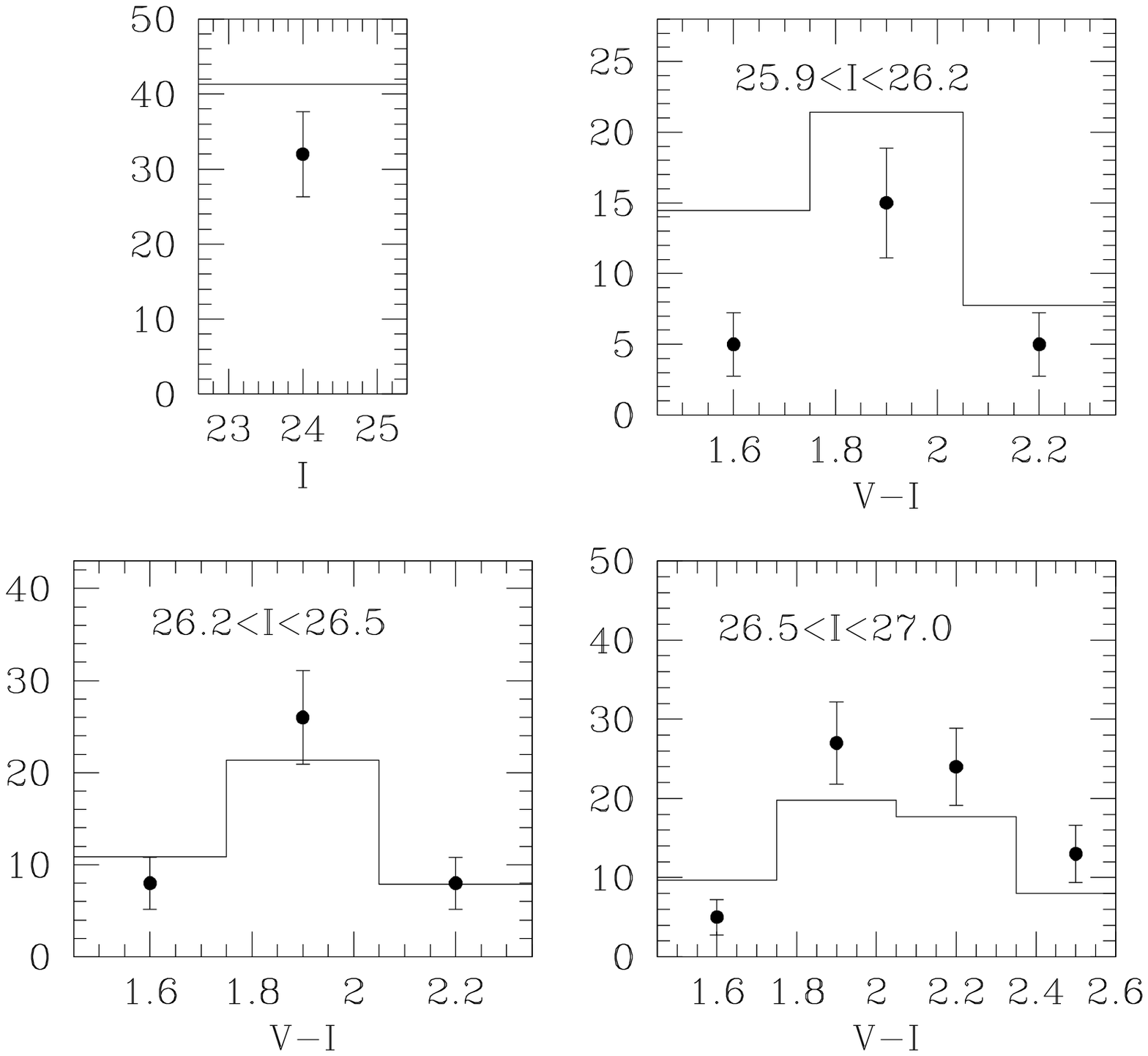] {In this case we have the opposite problem to that in Figure~\ref{Hess_bad1} --
namely that there are too many bright white dwarfs relative to the faint white dwarfs. The colour distribution
is also flatter than that shown by the observations. \label{Hess_bad2}}

\figcaption[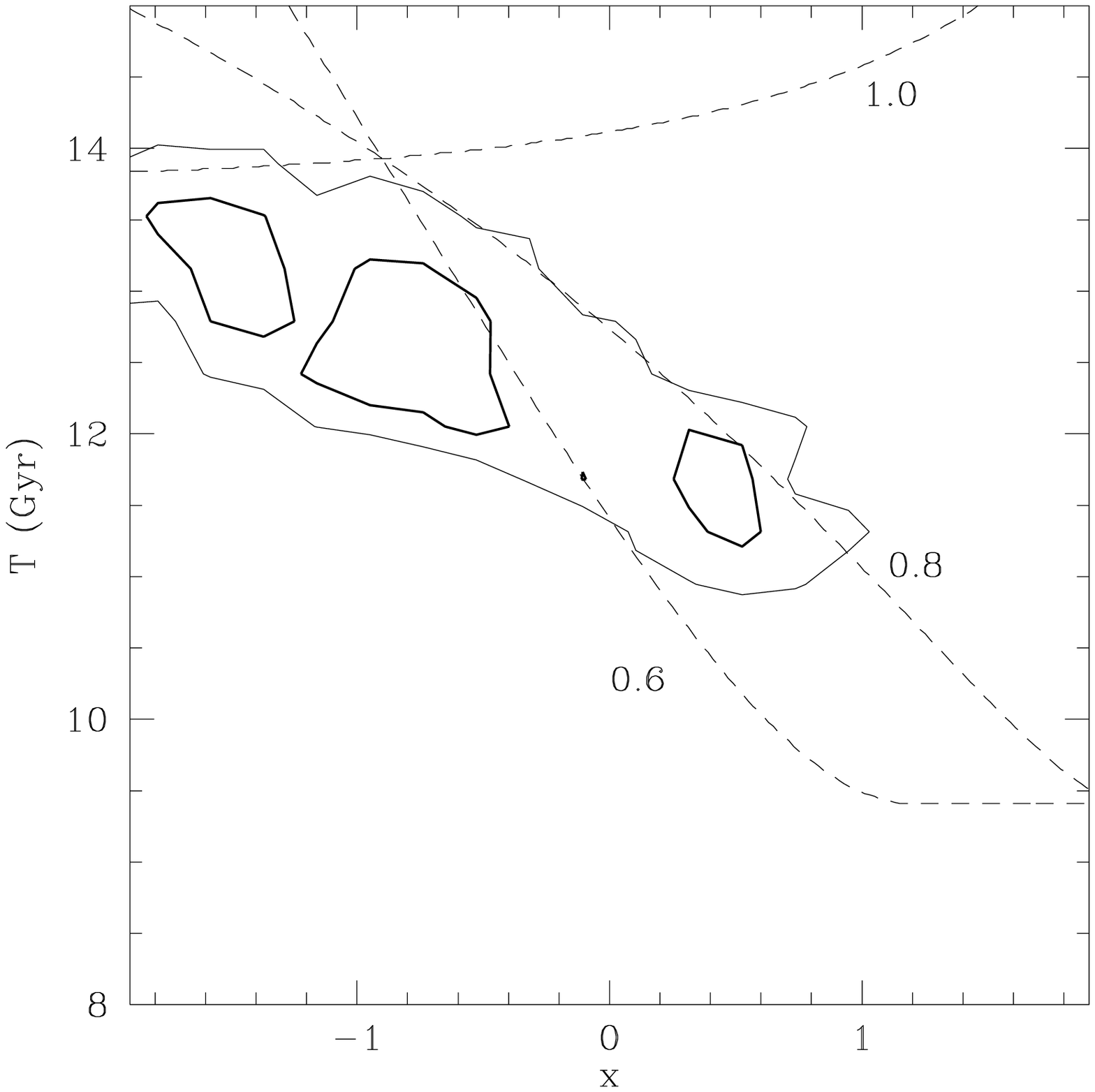]{ The solid contours show the 1 and 2~$\sigma$ bounds from the Hess fits 
for our default model. The dashed lines indicate the relation between x and T we obtain if
we match the white dwarf and main sequence number counts, as described in \S~\ref{numbers},
for three different assumed break masses. \label{Txq}}

\figcaption[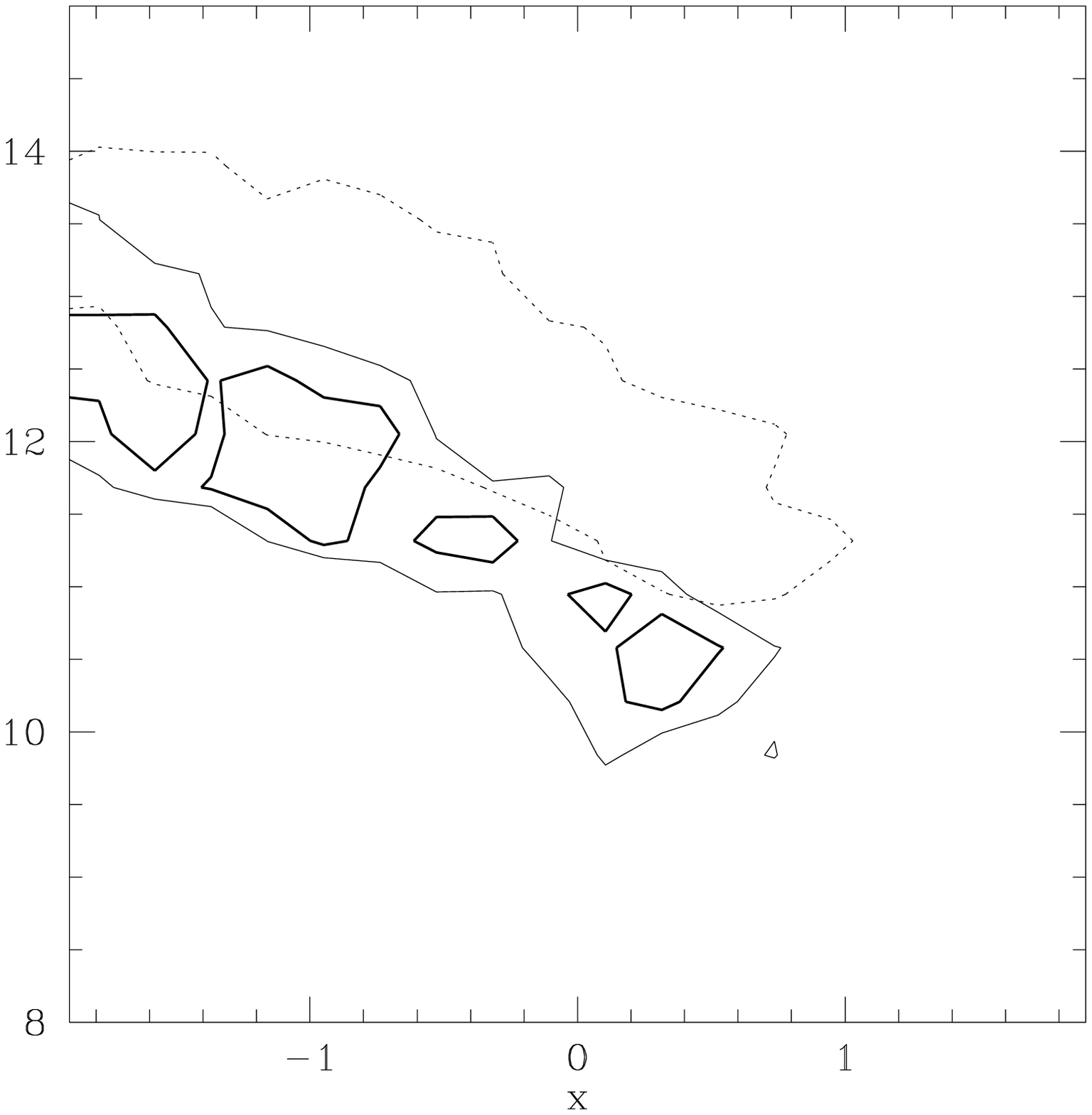]{The model for a thin hydrogen envelope gives a similar best fit age but
the lower age limit decreases. The solid curves show the $1\sigma$ and $2\sigma$ contours
for this fit, compared with the $2\sigma$ contour from Figure~\ref{Tx3} (dotted line).
 \label{Tx34}}

\figcaption[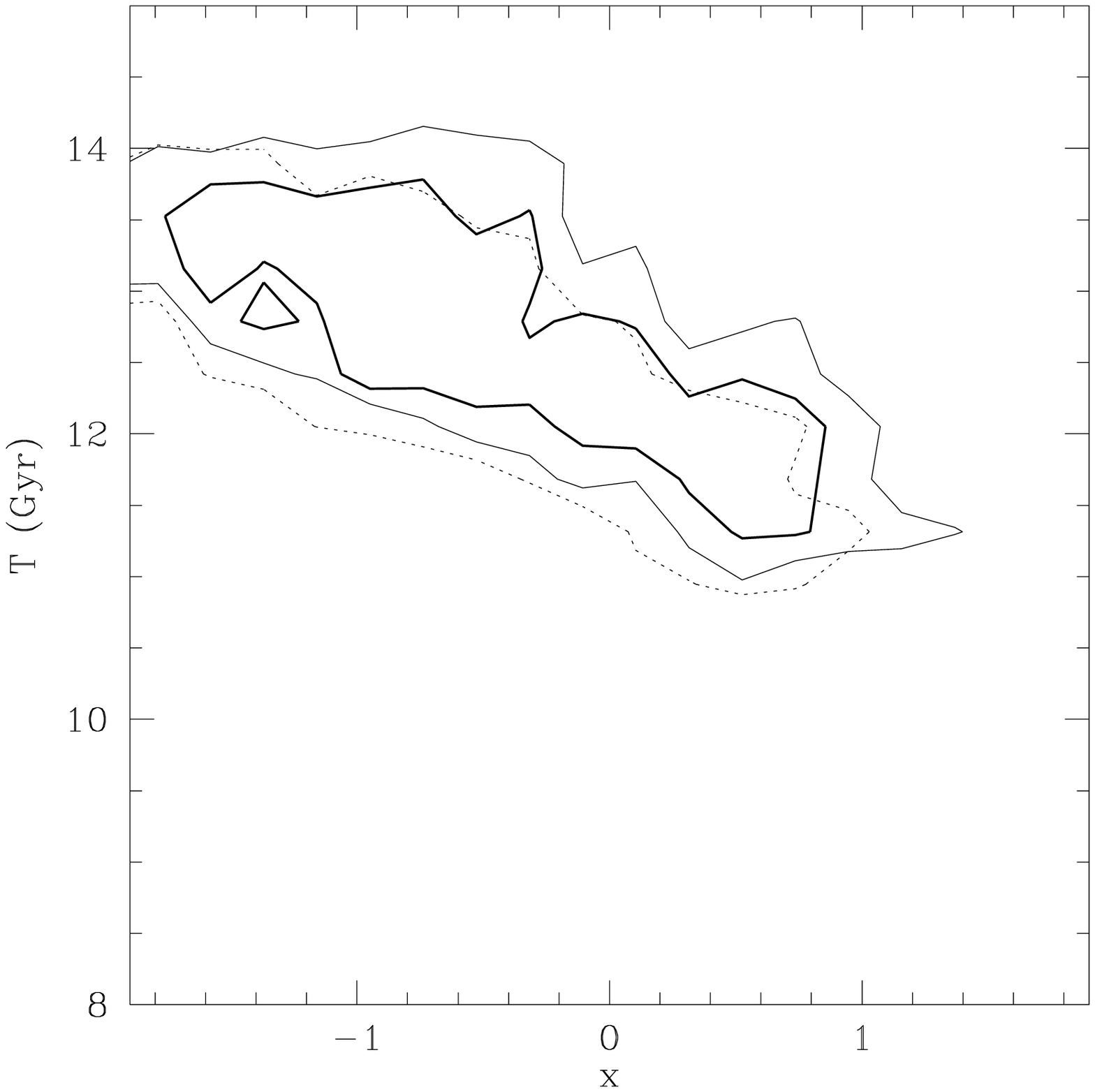]{A thinner helium layer increases the age bounds slightly. Once again,
the solid contours indicate the 1 and 2~$\sigma$ intervals for the model under discussion
and the dotted contour shows the 2$\sigma$ interval for our default model, included as 
a point of reference.
  \label{Tx33}}

\figcaption[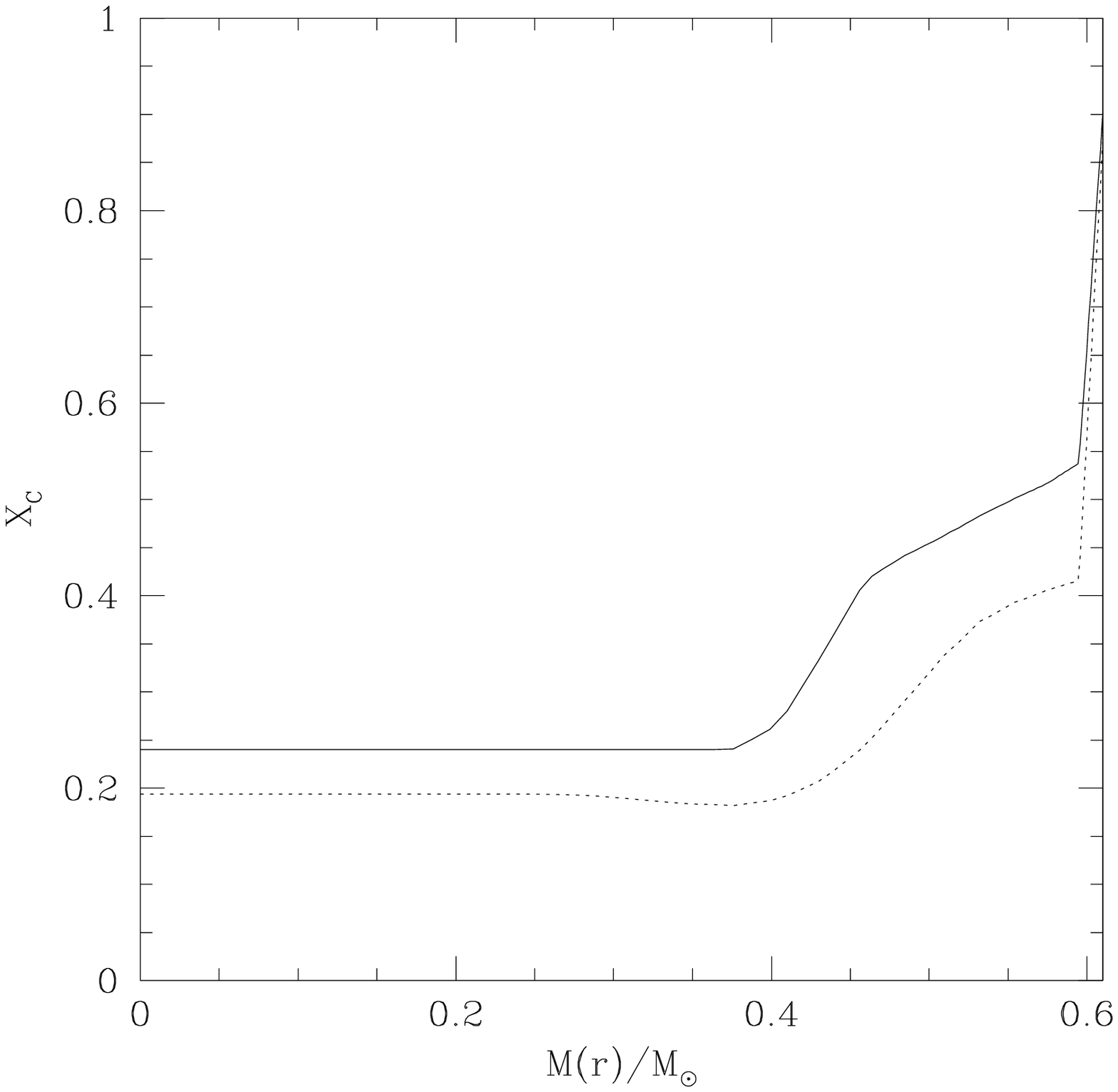]{ The solid line shows the profile of carbon fraction $X_C$ as a function
of mass enclosed for our default 0.6$M_{\odot}$ model. The dashed line shows the profile
we obtain from the Z=0.001 Hurley et al models.   \label{XX}}

\figcaption[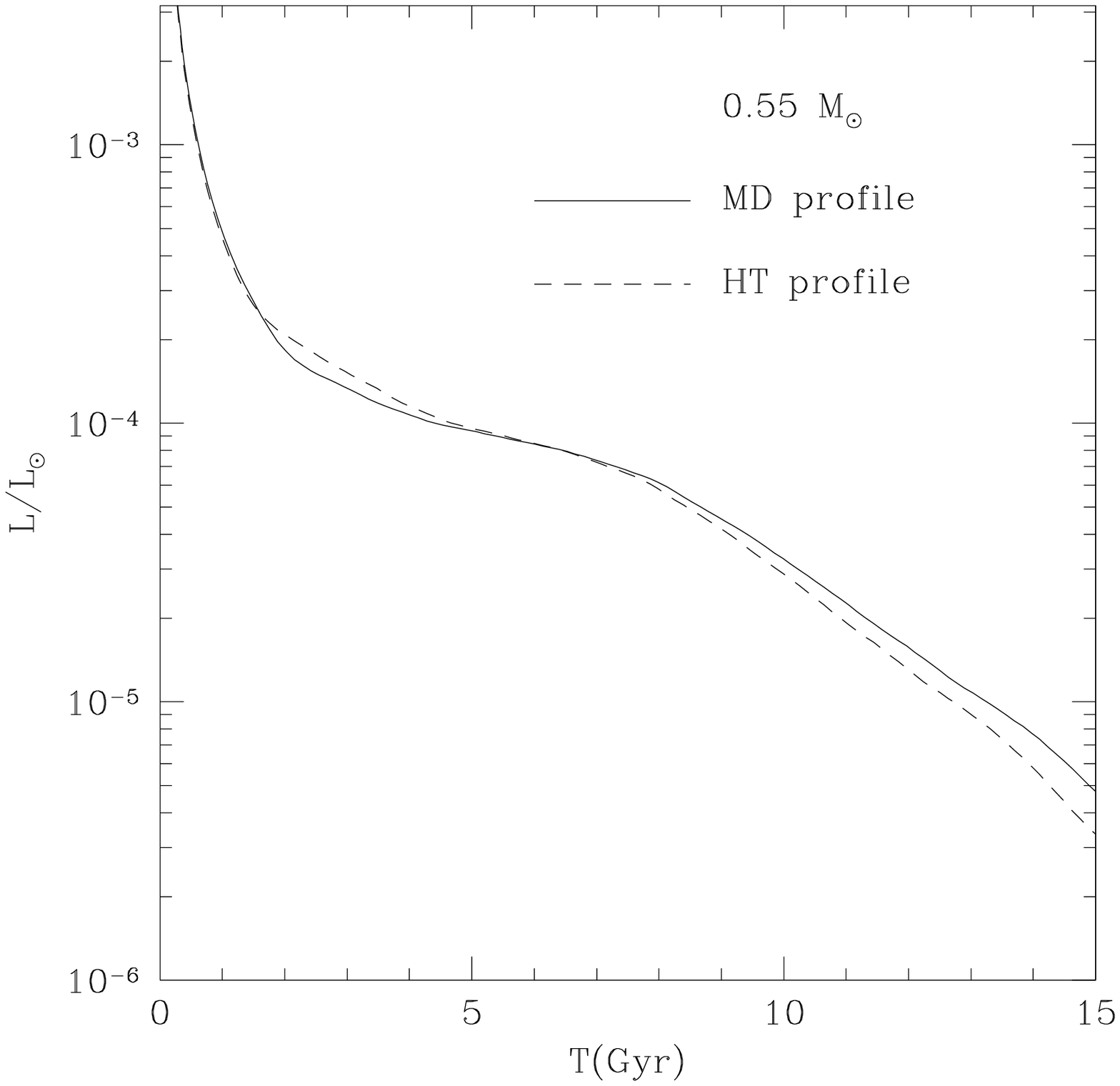]{ As in Figure~\ref{XX}, the solid line indicates our standard
model, while the dashed line is for the model using the Hurley et al (2000) inputs.
The greater oxygen content of the latter model results in earlier crystallisation and
thus faster cooling at late times. \label{Tout_comp}}

\figcaption[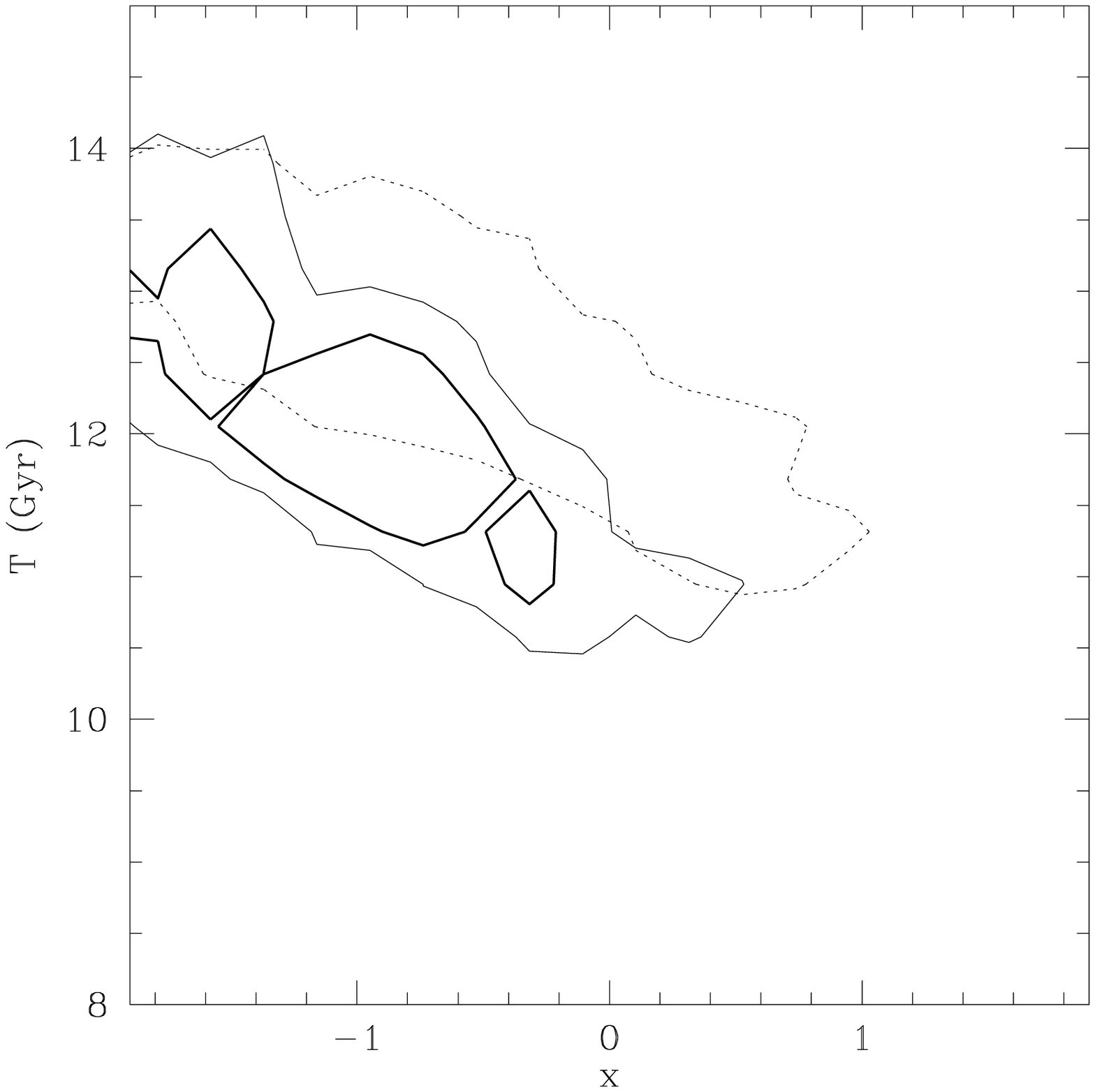]{ The Hurley et al core composition shifts the $2\sigma$ confidence
region (thin solid contour) to slightly lower values than the equivalent region for the default models
(dotted contour).
\label{Tx4}}

\figcaption[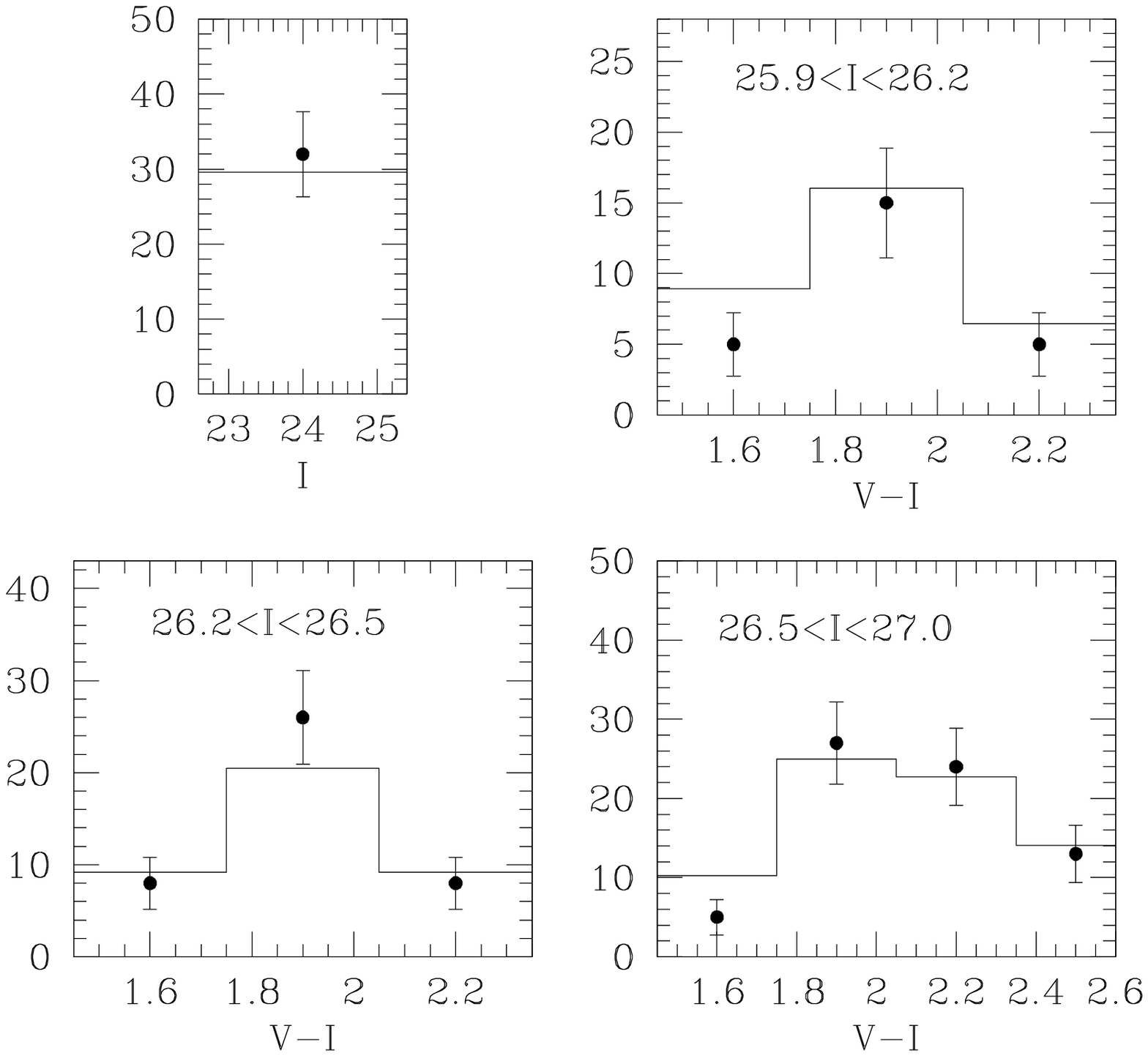]{ Here we compare the model for the case $(x,T)$=(-0.95,12.1~Gyr) with
the observations. The colour distributions and the relative normalisations of the different
bins all provide excellent fits. \label{Hess_Tx4}}

\figcaption[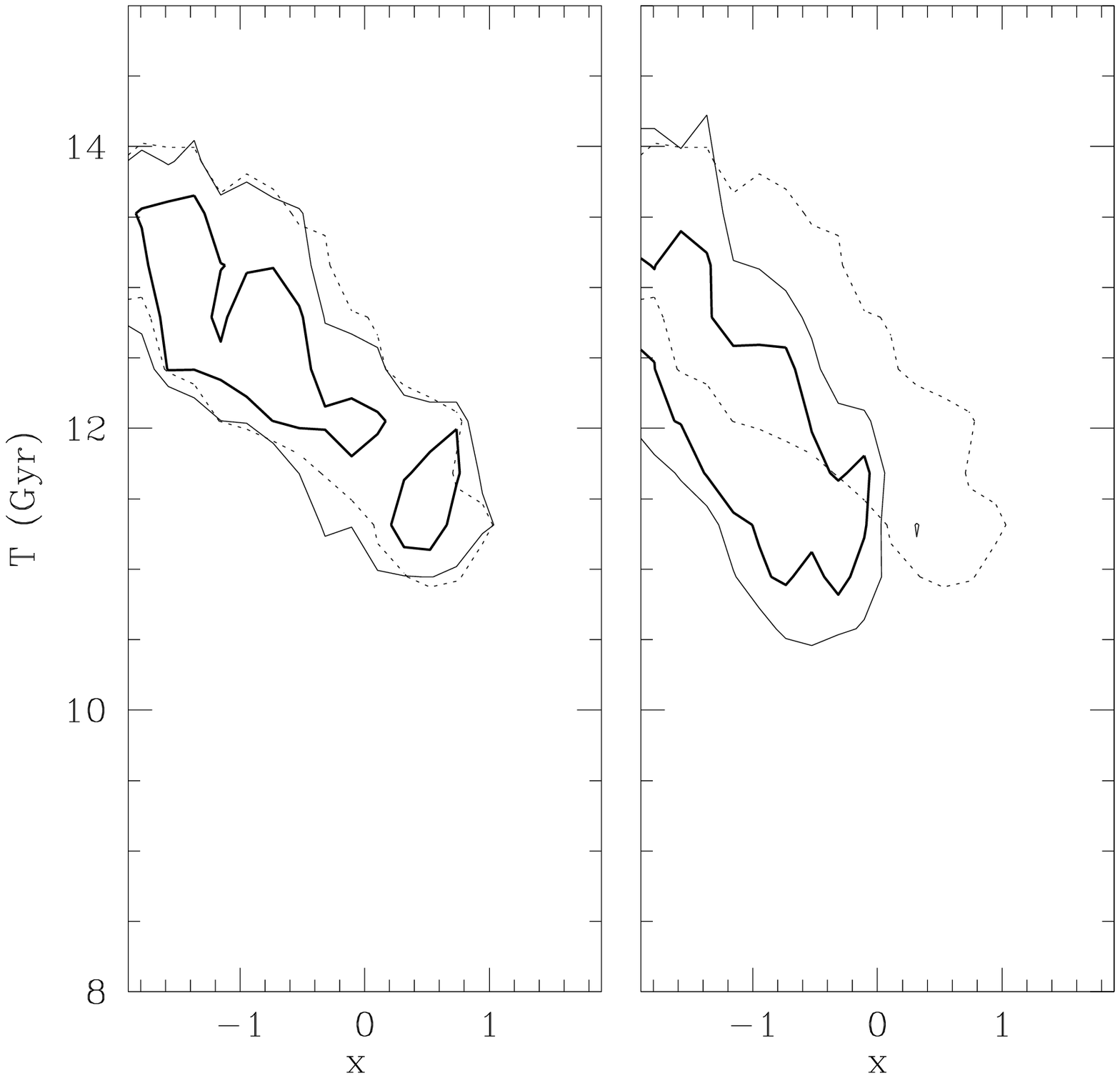] {The left-hand panel shows the results of combining the Hurley et al initial-final
mass relation with our default model. The dotted line indicates the $2\sigma$ contours of the default
model, demonstrating that there is very little difference. The right hand panel shows the same
thing but now using the core composition from Hurley et al as well.
 \label{Tx5}}

\figcaption[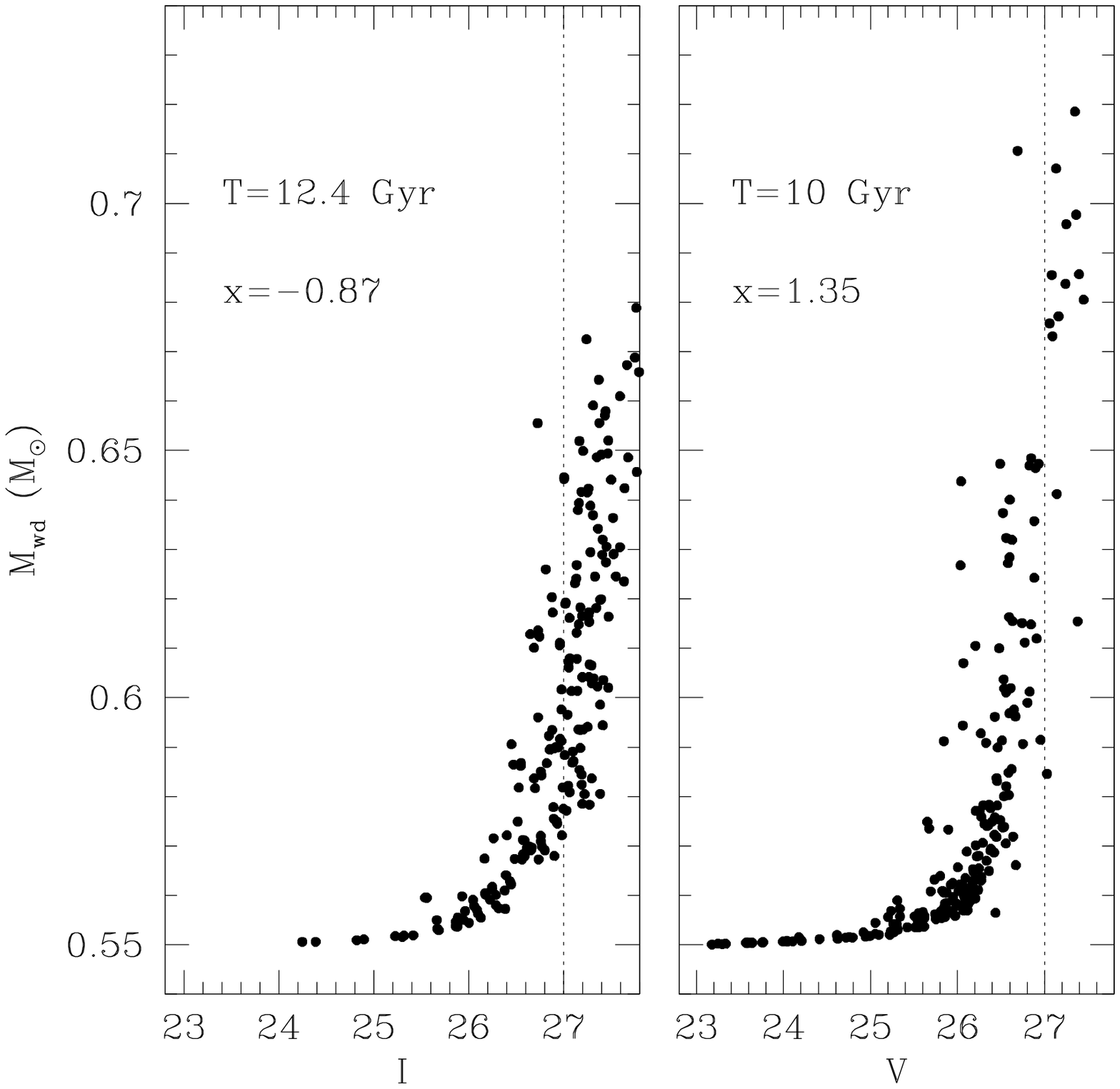]{ The left hand panel shows the I magnitude versus white dwarf mass for
a model of age $12.4$~Gyr. The right hand panel shows the same for a model of 10~Gyr.
In both cases, the main sequence--white dwarf mass relation was defined as in equation~(\ref{WoodM}).
Each of the two panels contains 300 model stars. The scatter in the diagrams is the result
of photometric errors derived from our artificial star tests. The dotted line indicates the
50\% completeness limit used in our Hess fits. These diagrams illustrate that, in
 the case of older
solutions, the more massive white dwarfs have cooled beyond the detection limit and we
probe only a limited range of white dwarf masses. Consequently our results
are not very sensitive to the exact form of the main sequence--white dwarf
relation.   \label{MVM}}

\figcaption[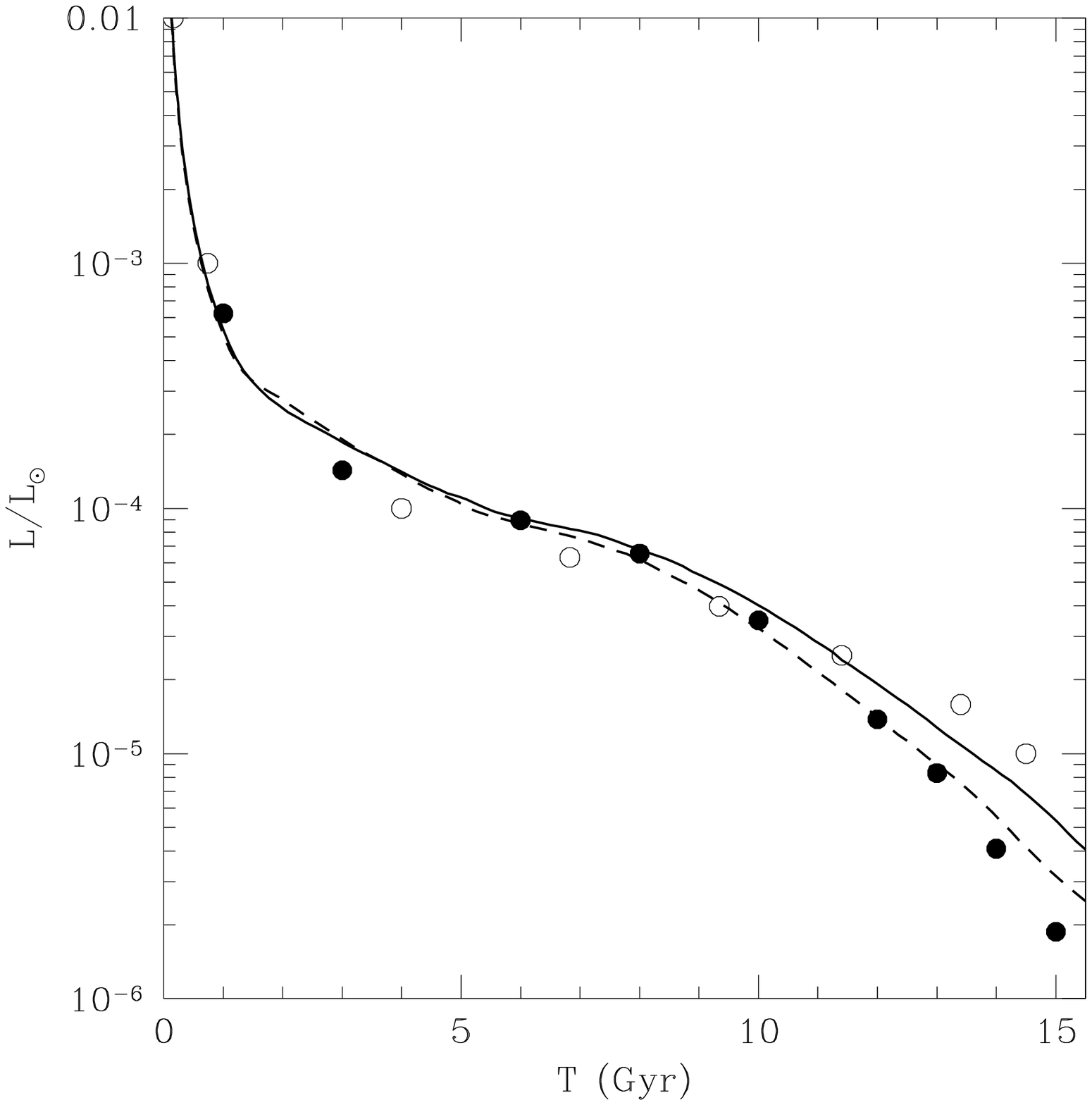] {The solid line is our default cooling model and the
dashed line is the same but with the core composition of Hurley et al (2000). The the
filled points are from Chabrier et al (2000) and the open points are from
Salaris et al (2000). At late times, these other two solutions bracket ours, with
the Chabrier models cooling the fastest and the Salaris models the slowest. The
latter would cool even more slowly if we included their calculated delay due to
phase separation. \label{Ltcomp}}

\figcaption[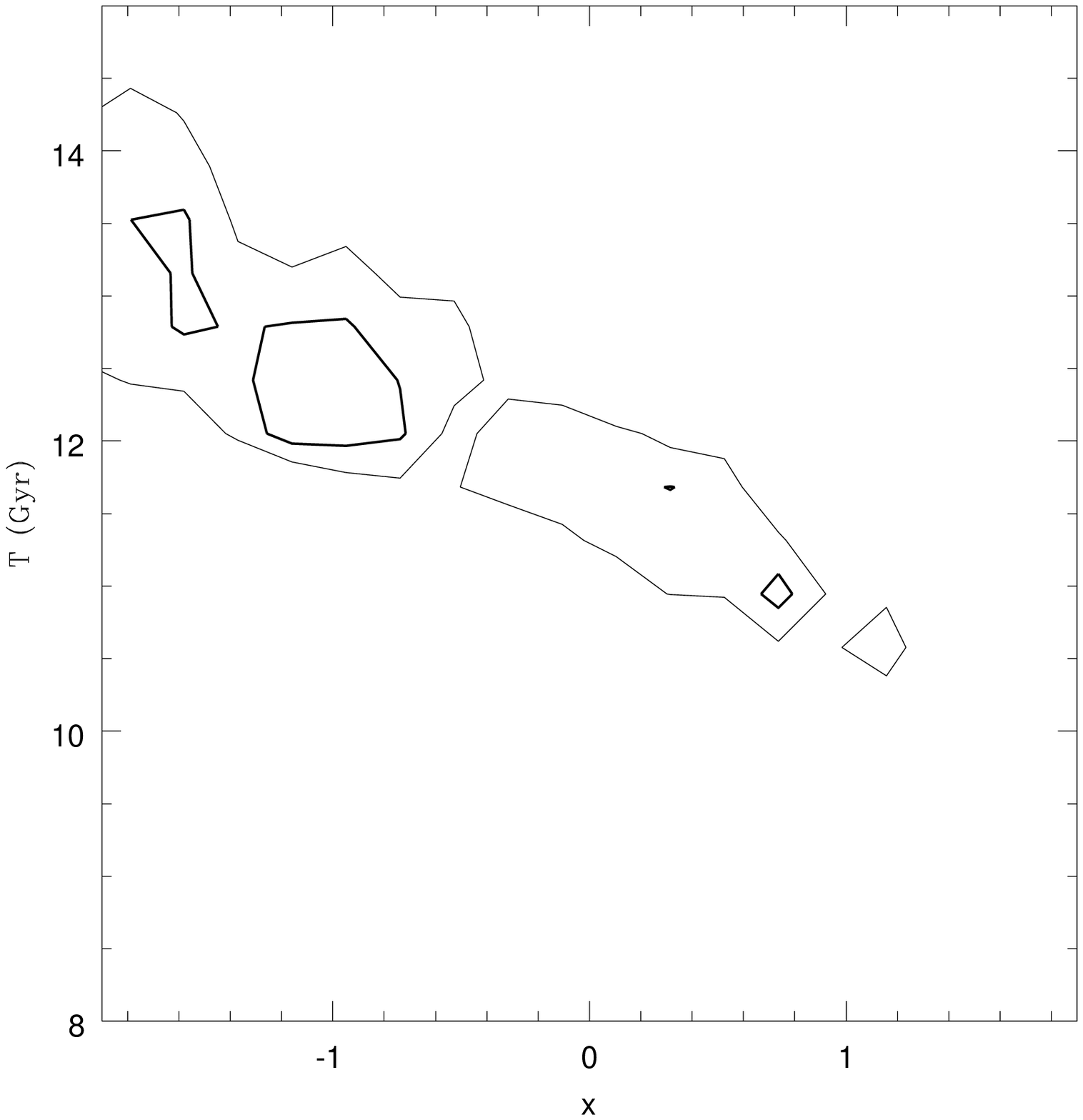]{This shows the solution obtained, using our models but
under the approximation we have to adopt to calculate the contours shown in
Figure~\ref{Tx61}. In particular the assumption of $0.6 M_{\odot}$ but
using our extinction of $A_V=1.39$ means that the upper part of the cooling
sequence is not accurately matched. The contours are similar to before, but
the best-fit $\chi^2$ is slightly worse. \label{Tx6}}

\figcaption[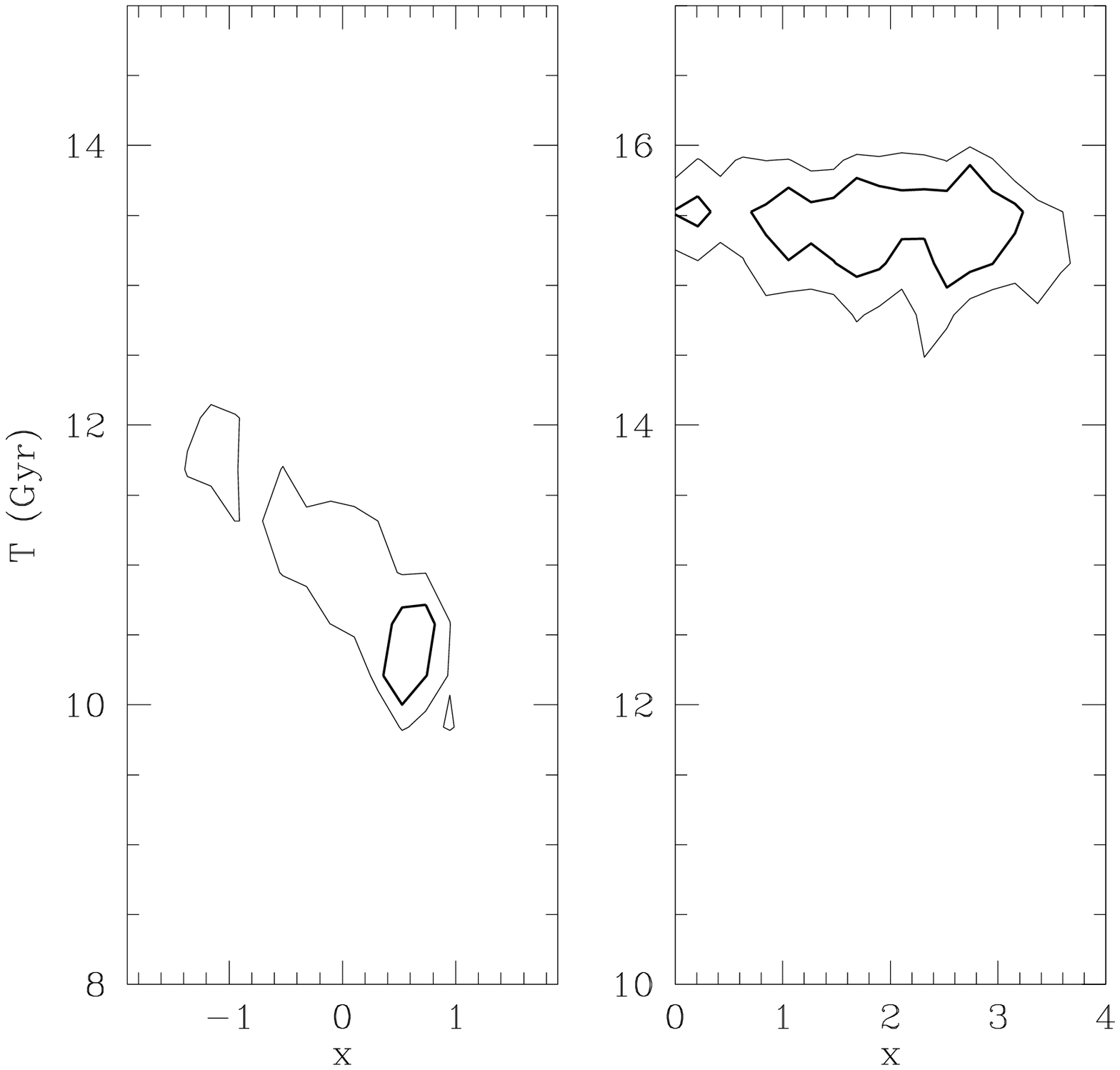]{ The left hand panel shows the fit for the Chabrier et al (2000) models.
The fit is acceptable, despite the necessary
approximation of $0.6 M_{\odot}$ with $A_V=1.39$. Again the solution shows the
familiar covariance between x and T. The $2\sigma$ limit in this case extends down
to ages of 9.8~Gyr. 
The right hand panel shows the corresponding fit for the Salaris models. The
$\chi^2$ for these fits does not correspond to a good model.
 \label{Tx61}}

\figcaption[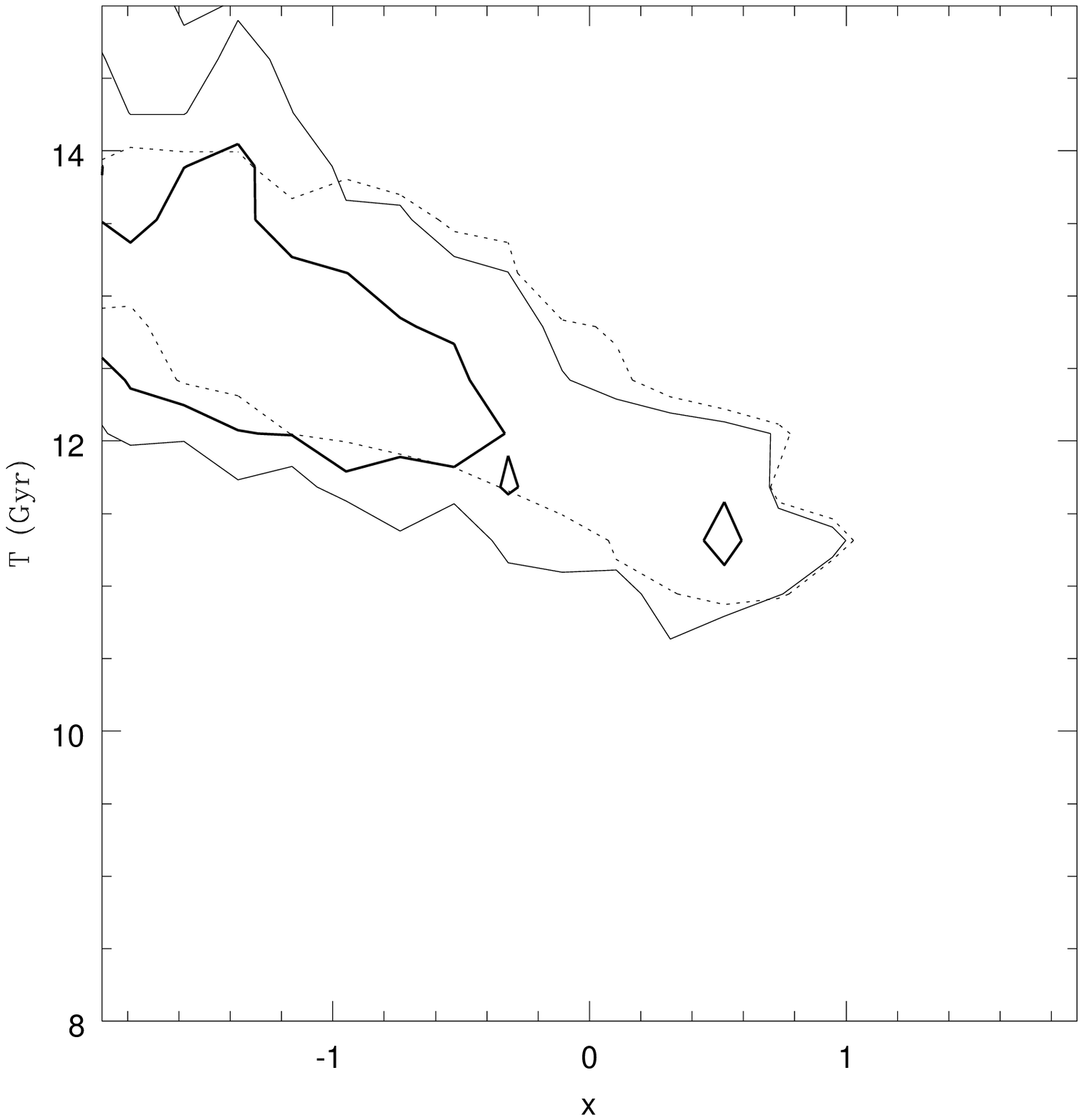] {The solid contours shows the usual confidence regions based on
the fit to the observed Hess diagram, now marginalised over Helium atmosphere fraction at 
each point.
 The dotted contour shows the 2$\sigma$ contour
from the default model. The small difference indicates that the uncertainty in the
helium atmosphere fraction does not have a big effect on the determination of the
best model.
\label{Tx8}}

\figcaption[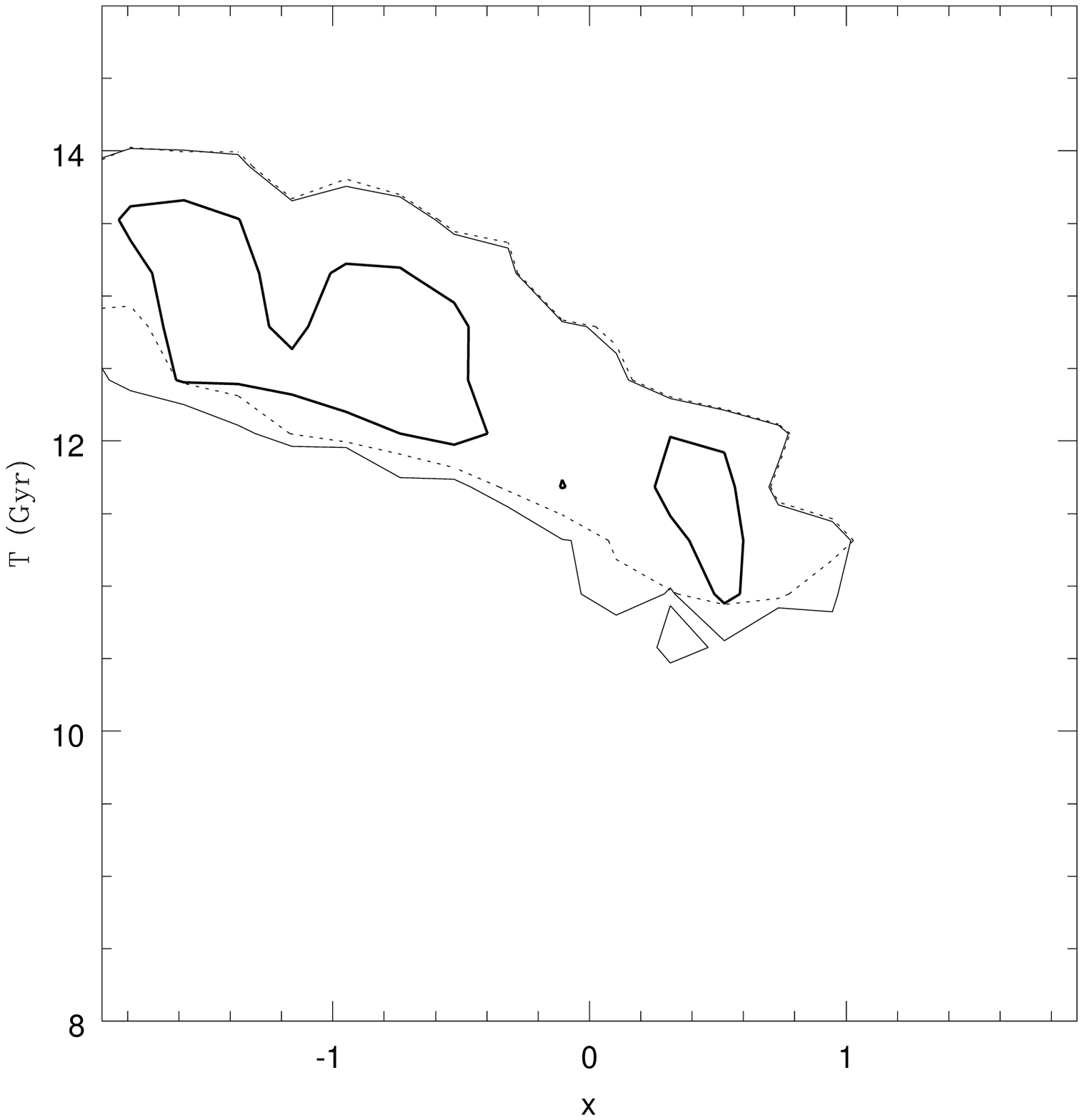] {This figure is the same as Figure~\ref{Tx8}, except for the
fact that now the helium models still cool with outer boundary conditions determined
by hydrogen atmospheres, although their colours are calculated with helium atmospheres.
This is designed to mimic the case when the helium atmosphere appearance is only
a temporary phenomenon in the life of a white dwarf. Again, the influence of the
atmospheric uncertainties is small.
 \label{Tx14}}

\figcaption[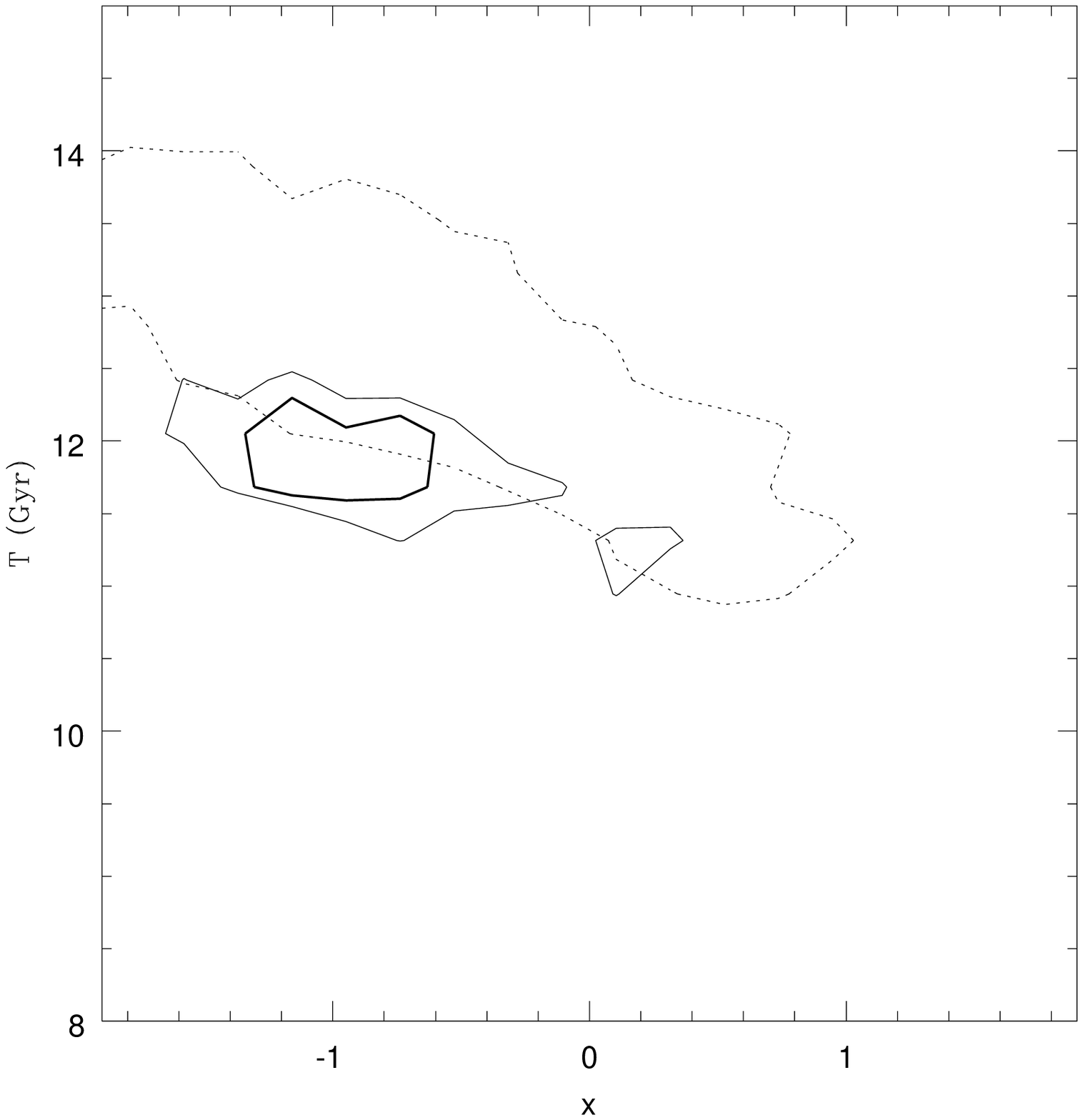] {Using the atmosphere models of Hansen (1998) we find a 
similar solution to before, but with a worse $\chi^2$. The higher minimum $chi^2$ is
responsible for the narrower confidence intervals, since the latter correspond to
fixed $\Delta \chi^2$. \label{Tx10}}

\figcaption[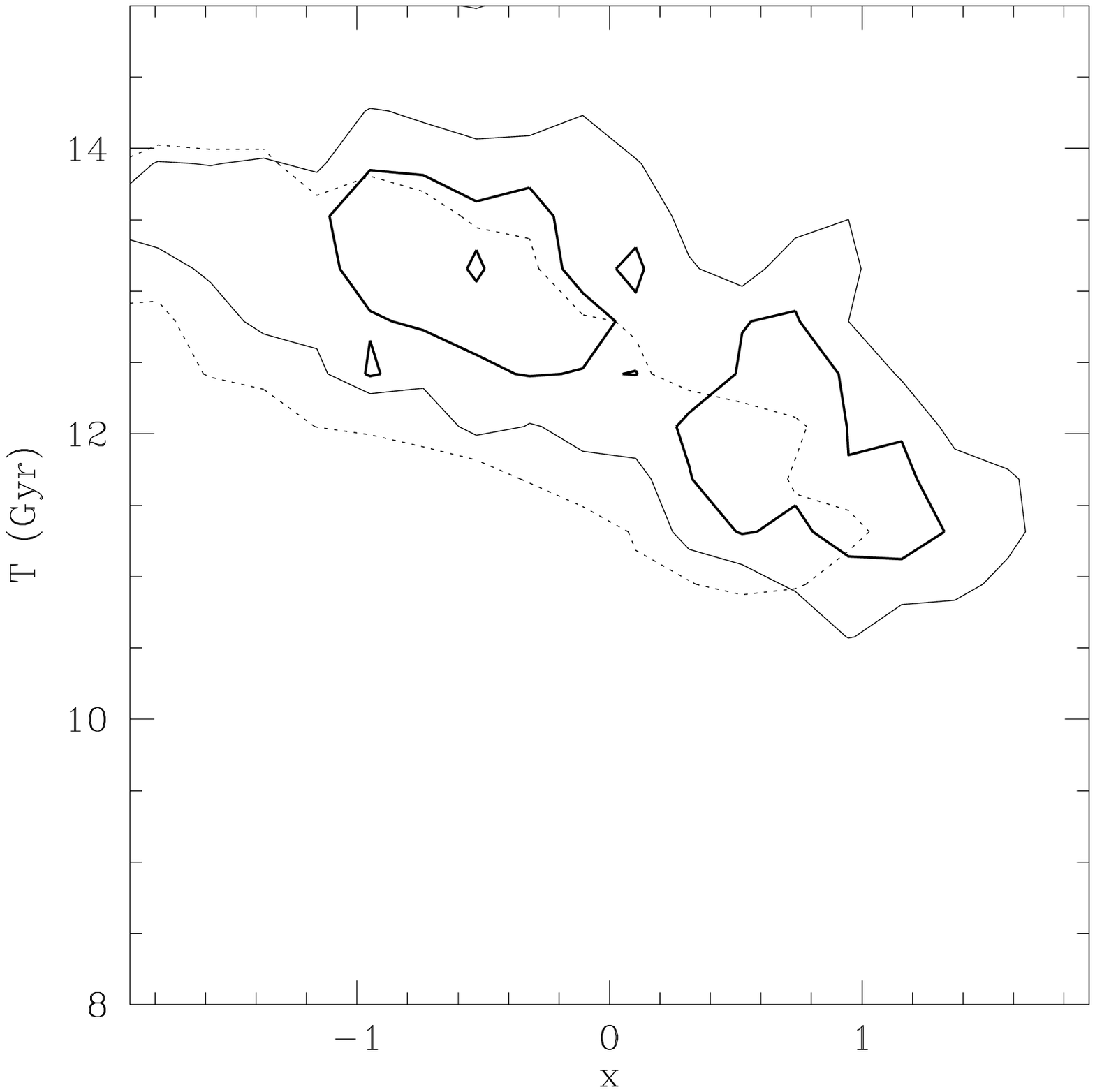]{ The solid curves indicate the confidence intervals if we evaluate
our default model but now using the proper motion cut labelled B in Figure~\ref{PMS}.
The dotted curve is once again the $2\sigma$ limit for the default model using proper motion
cut~A. The age range is not significantly affected, but the range of $x$ values does now
extend to larger values. \label{TxB}}

\figcaption[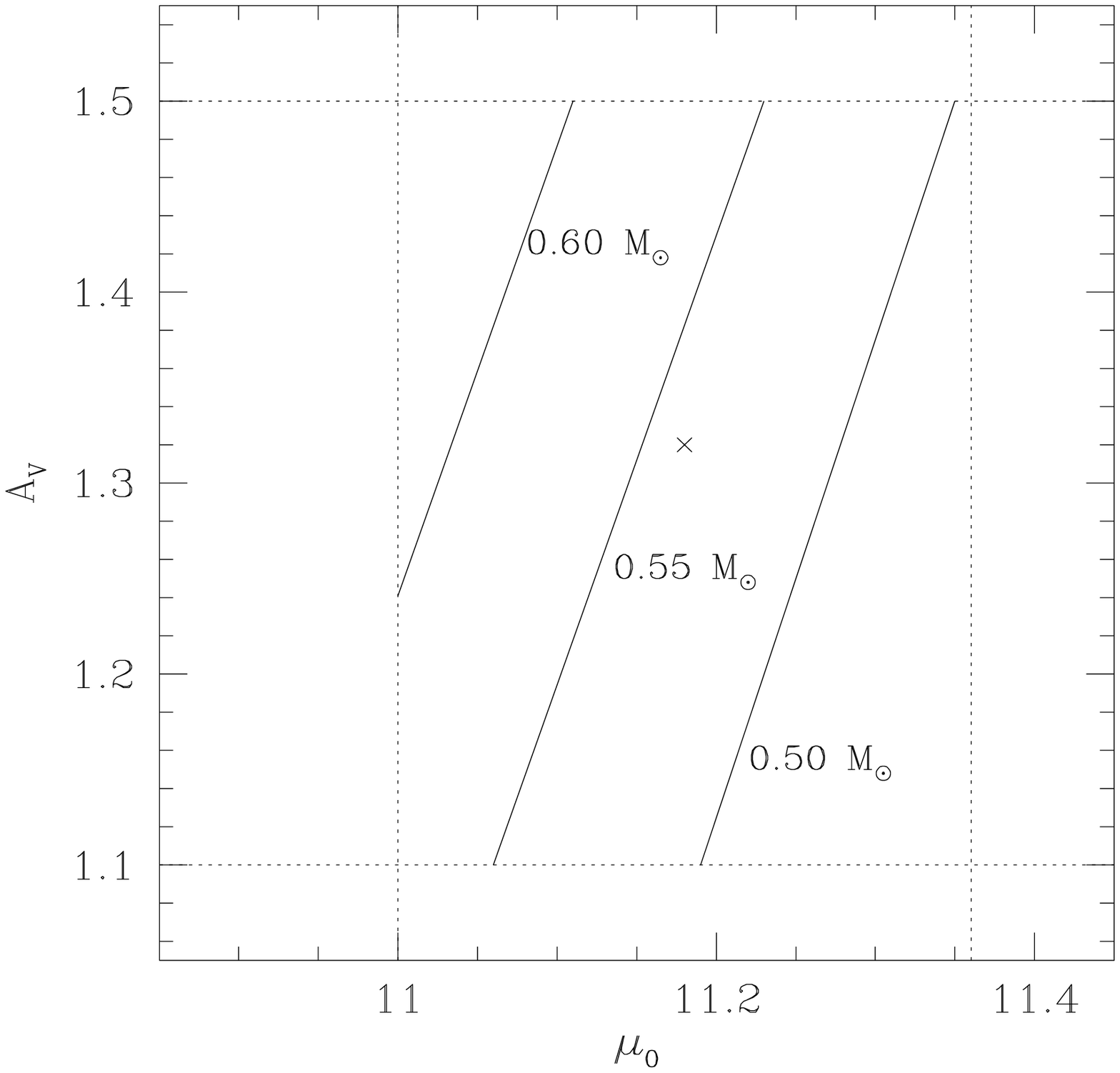] {The dotted lines delineate the bounds on the distance and extinction
to M4. The solid lines indicate the relation between $\mu$ and $A_V$ that is required to fit
the upper part of the cooling sequence, for each stated white dwarf mass. The cross indicates
the location of the values preferred by Richer et al (1997). This lies close to our current
default model ($\mu=11.18$, $A_V=1.39$). \label{amu}}

\figcaption[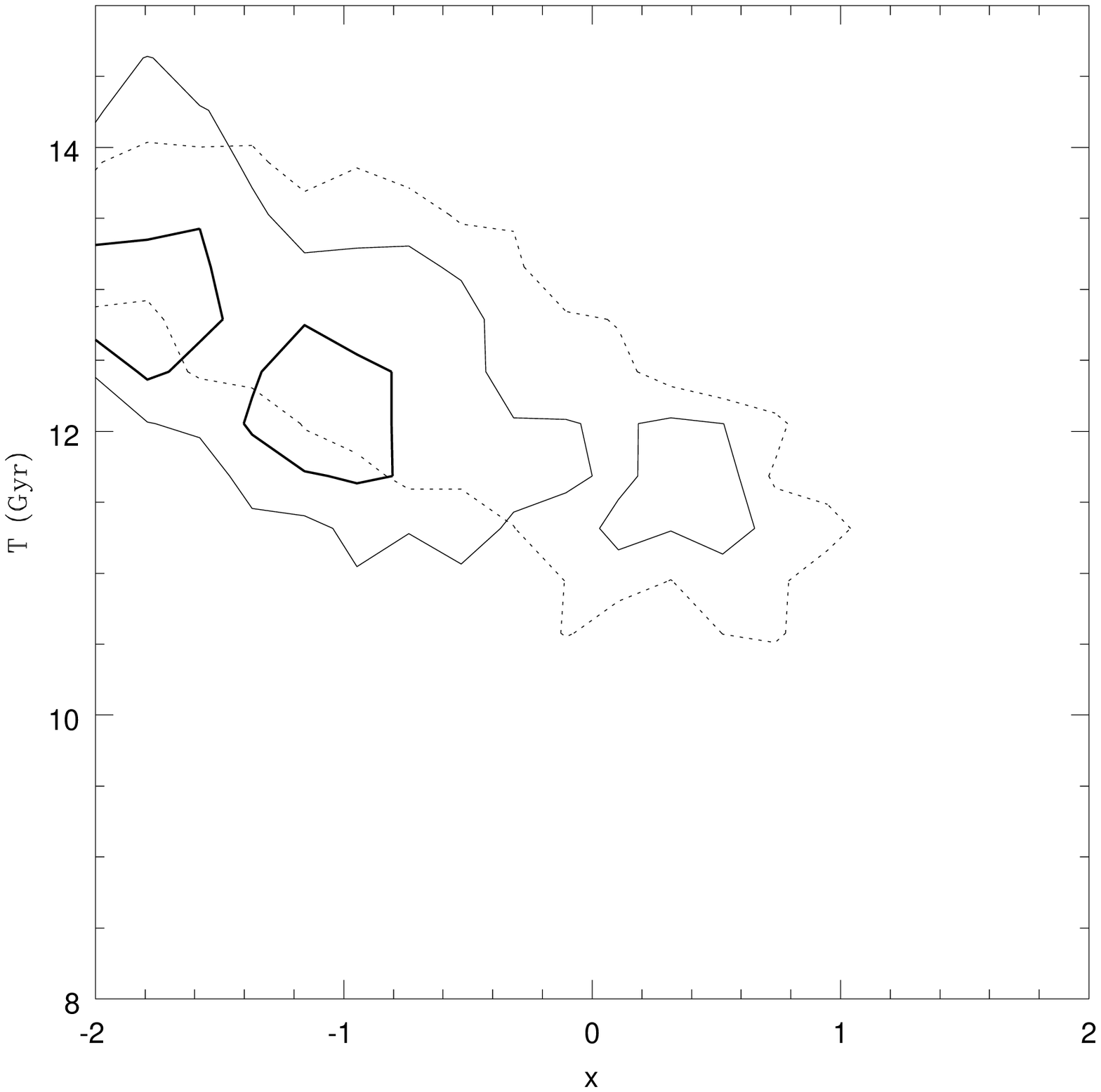]{The solid contours indicate the usual 1 and $2\sigma$ bounds obtained
by marginalising over the range of masses $0.5 M_{\odot}$--$0.6 M_{\odot}$ and the range of
distance and extinction outlined in Figure~\ref{amu}. The dotted contour is different from
our usual reference model, and shows the
$2\sigma$ range if we restrict our attention to 0.55$M_{\odot}$ but still allow the
distance/extinction to vary.
  \label{TxD2}}

\figcaption[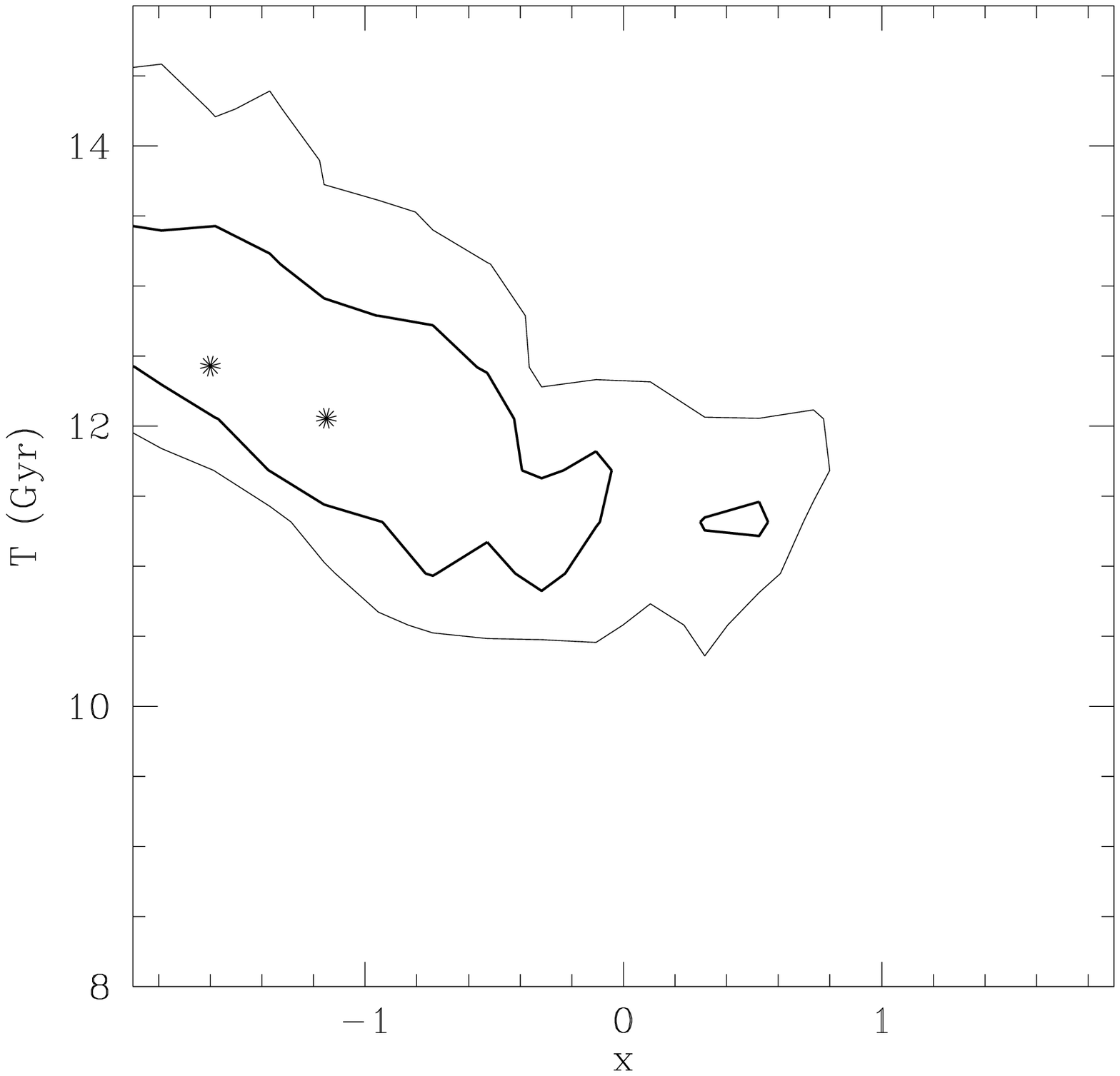]{The solid lines indicate the 1$\sigma$ (heavy line) and 2$\sigma$ (thin line)
confidence intervals when we marginalise over our various systematic model uncertainties.
The two asterisks indicate the locations of the two almost equal minima in the $\chi^2$ surface.
 \label{Txbig}}

\figcaption[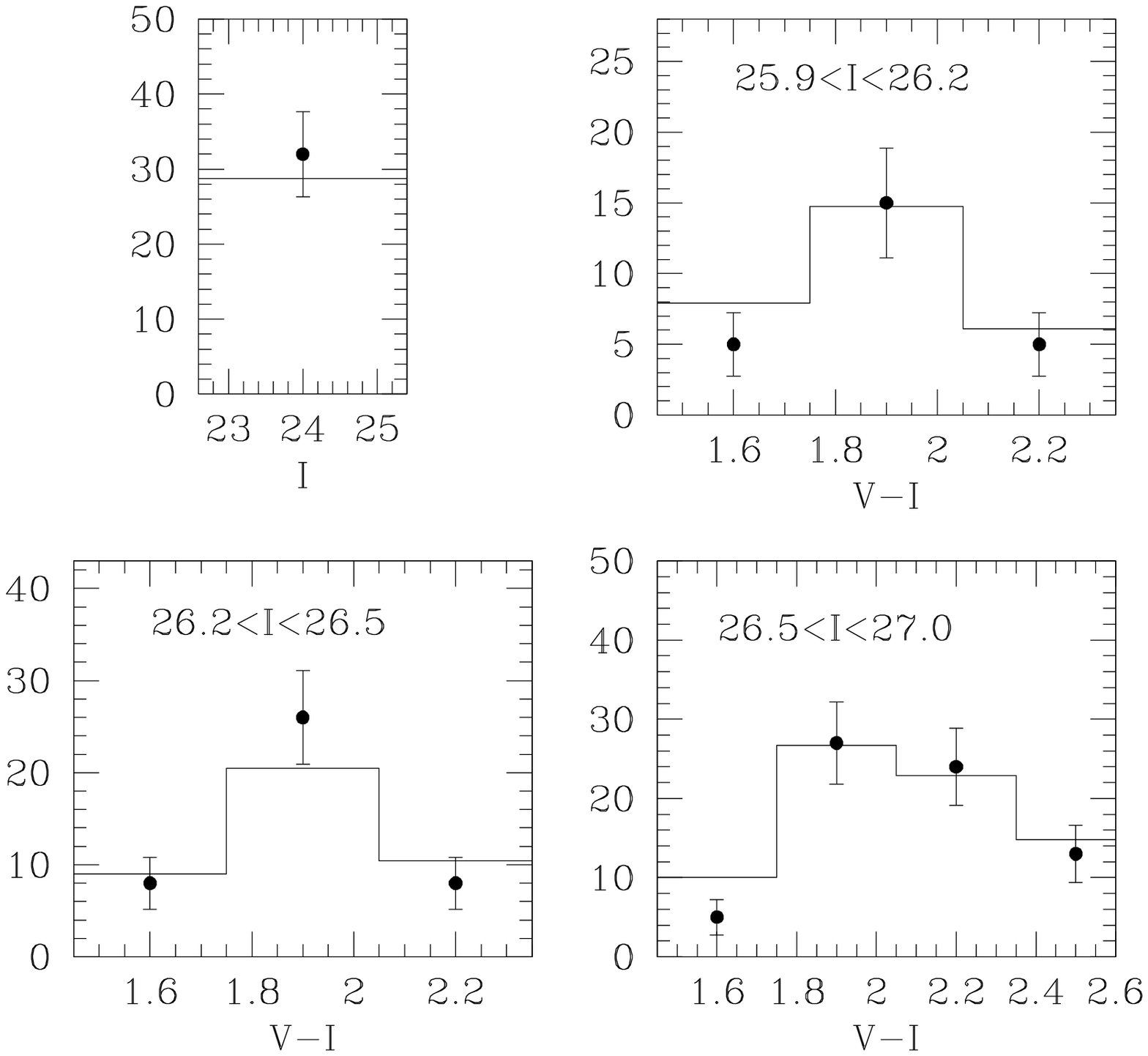]{Our best fit model clearly meets all the various requirements imposed
by the observations. The overall ratio of bright white dwarfs to faint white dwarfs is
matched while the colour distribution of the faint white dwarfs is also well described. \label{Hess_best}}

\figcaption[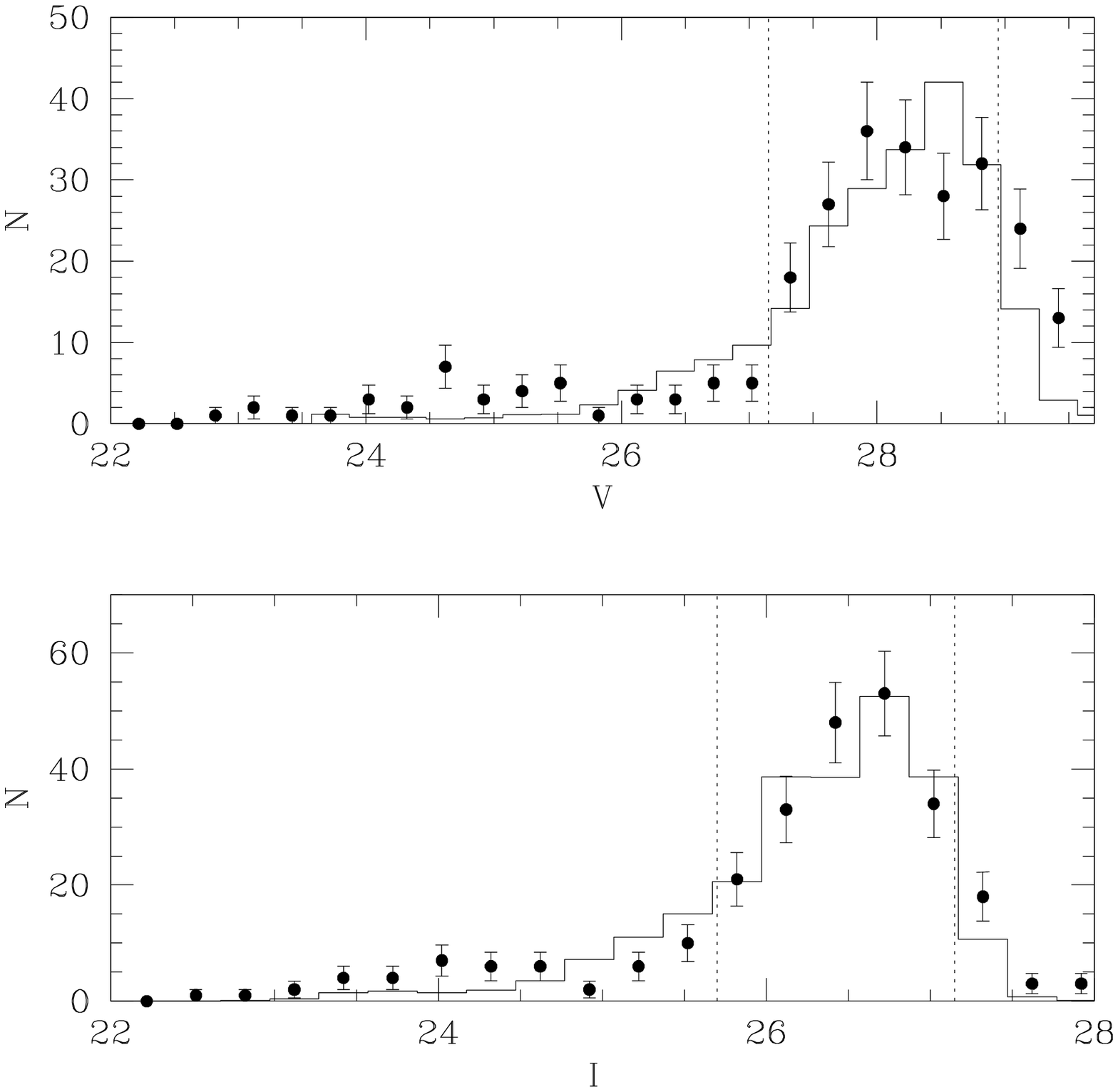]{The upper panel shows the V~LF while the lower panel shows the I~LF. 
For both measures the rise in the luminosity function is well matched as is the plateau before
the number counts drop off due to the incompleteness. Thus, although we have used the Hess diagram
to find our best model, we clearly get a good fit to both of the luminosity functions as well.  \label{LF_best}}

\figcaption[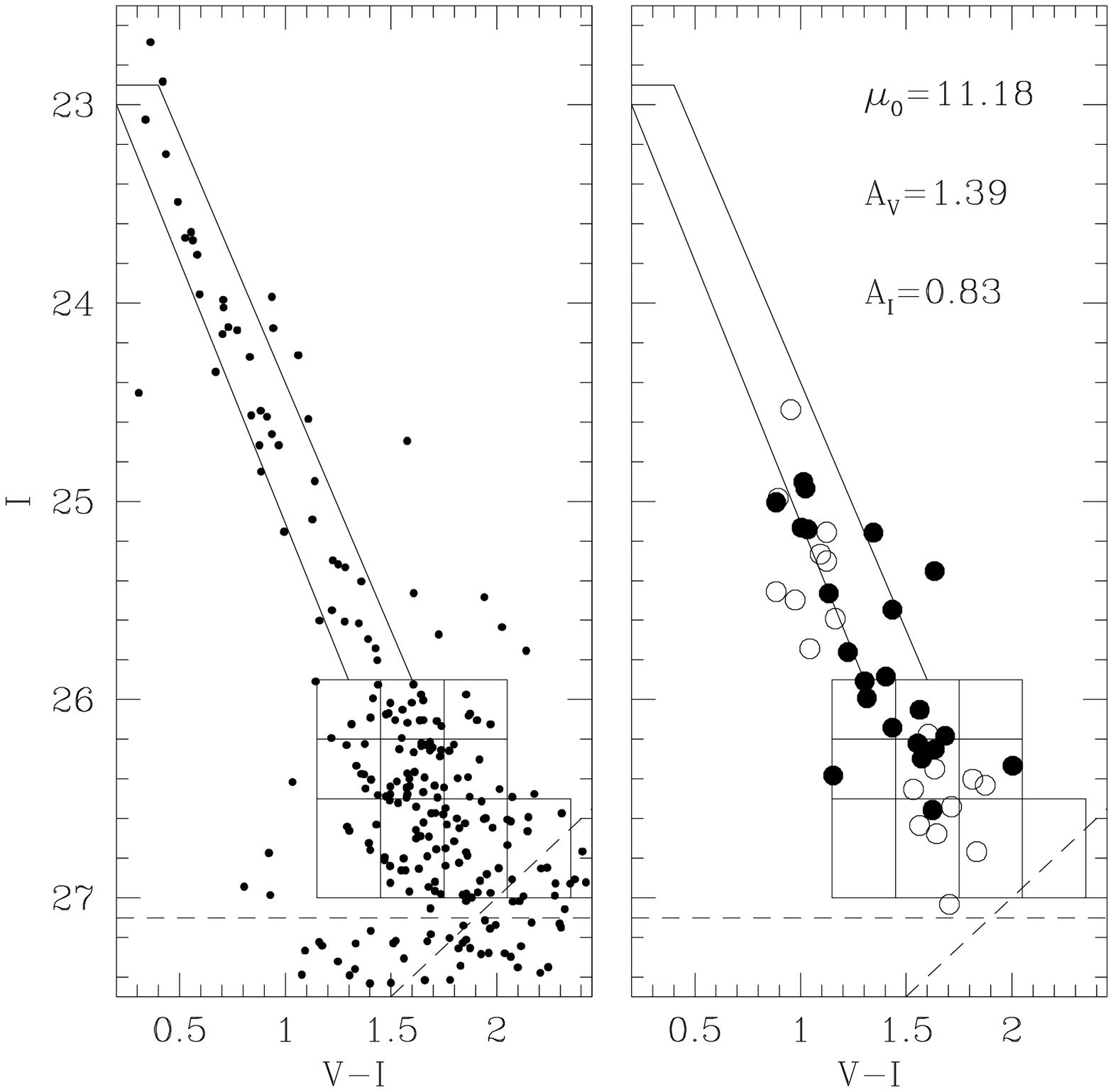]{The left hand panel shows the M4 white dwarf sequence, with our analysis
grid superimposed. The right hand panel shows the same grid but with the solar neighbourhood
white dwarfs (Liebert et al 1988) plotted at the location of M4. We have used the photometry
of Leggett, Ruiz \& Bergeron (1998) and an extinction of $A_V=1.39$. The filled circles indicate
dwarfs with hydrogen atmospheres and the open circles indicate those with helium atmospheres.
Recall that the faintest 8 white dwarfs fall beyond the peak of the disk luminosity function.
The fact that the brighter of the disk white dwarfs lie at the edge of our upper bin is
an indication of the mass difference between disk white dwarfs and the brightest cluster
white dwarfs.
\label{LDM}}

\figcaption[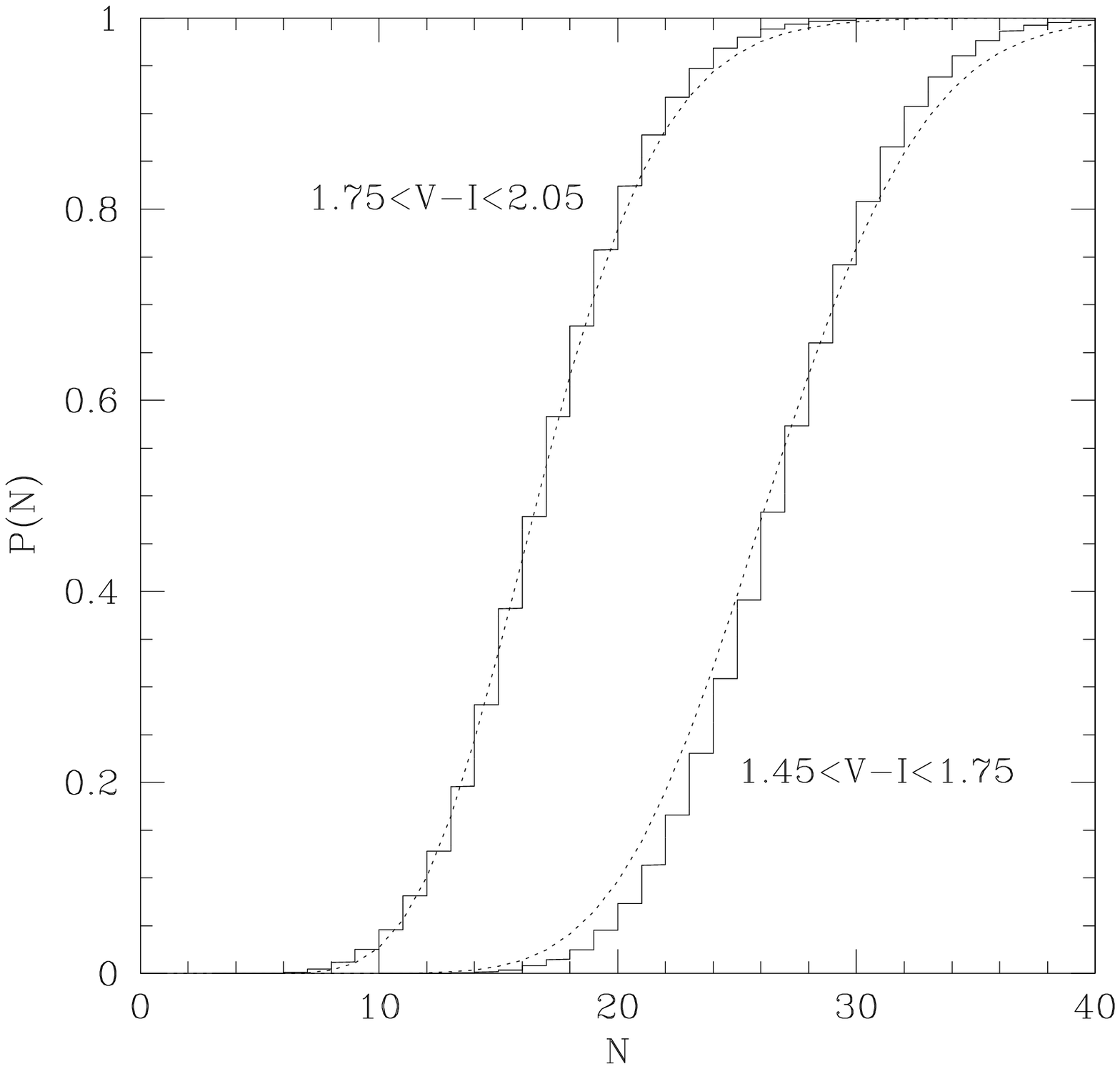]{ The solid histograms indicate the probability distribution of number counts
for a fixed number of input stars. The stars were input with I=26.7, V-I=1.75. The two bins
cover the colours as indicated and $26.5<I<27.0$. The dotted lines are the incomplete
$\Gamma$ functions Q(N,17) and Q(N,26), corresponding to the appropriate cumulative Poisson
distributions. \label{PN}}

\clearpage
\plotone{fig1.ps}
FIGURE~1
\clearpage
\plotone{fig2.ps}
FIGURE~2
\clearpage
\plotone{fig3.ps}
FIGURE~3
\clearpage
\plotone{fig4.ps}
FIGURE~4
\clearpage
\plotone{fig5.ps}
FIGURE~5
\clearpage
\plotone{fig6.ps}
FIGURE~6
\clearpage
\plotone{fig7.ps}
FIGURE~7
\clearpage
\plotone{fig8.ps}
FIGURE~8
\clearpage
\plotone{fig9.ps}
FIGURE~9
\clearpage
\plotone{fig10.ps}
FIGURE~10
\clearpage
\plotone{fig11.ps}
FIGURE~11
\clearpage
\plotone{fig12.ps}
FIGURE~12
\clearpage
\plotone{fig13.ps}
FIGURE~13
\clearpage
\plotone{fig14.ps}
FIGURE~14
\clearpage
\plotone{fig15.ps}
FIGURE~15
\clearpage
\plotone{fig16.ps}
FIGURE~16
\clearpage
\plotone{fig17.ps}
FIGURE~17
\clearpage
\plotone{fig18.ps}
FIGURE~18
\clearpage
\plotone{fig19.ps}
FIGURE~19
\clearpage
\plotone{fig20.ps}
FIGURE~20
\clearpage
\plotone{fig21.ps}
FIGURE~21
\clearpage
\plotone{fig22.ps}
FIGURE~22
\clearpage
\plotone{fig23.ps}
FIGURE~23
\clearpage
\plotone{fig24.ps}
FIGURE~24
\clearpage
\plotone{fig25.ps}
FIGURE~25
\clearpage
\plotone{fig26.ps}
FIGURE~26
\clearpage
\plotone{fig27.ps}
FIGURE~27
\clearpage
\plotone{fig28.ps}
FIGURE~28
\clearpage
\plotone{fig29.ps}
FIGURE~29
\clearpage
\plotone{fig30.ps}
FIGURE~30
\clearpage
\plotone{fig31.ps}
FIGURE~31
\clearpage
\plotone{fig32.ps}
FIGURE~32
\clearpage
\plotone{fig33.ps}
FIGURE~33
\clearpage
\plotone{fig34.ps}
FIGURE~34
\clearpage
\plotone{fig35.ps}
FIGURE~35
\clearpage
\plotone{fig36.ps}
FIGURE~36
\clearpage
\plotone{fig37.ps}
FIGURE~37
\clearpage
\plotone{fig38.ps}
FIGURE~38
\clearpage
\plotone{fig39.ps}
FIGURE~39
\clearpage
\plotone{fig40.ps}
FIGURE~40
\clearpage
\plotone{fig41.ps}
FIGURE~41
\clearpage
\plotone{fig42.ps}
FIGURE~42
\clearpage
\plotone{fig43.ps}
FIGURE~43
\clearpage

\begin{deluxetable}{lllllll} 
\tablecolumns{7} 
\tablewidth{0pc} 
\tablecaption{M4 Parameter Scan \label{BigTab1}}
\tablecomments{ The notation `$\cdots$' indicates where we have
marginalised over the parameter in question.} 
\tablehead{ 
\colhead{$M_0/A_V$}  & \colhead{Cooling} & \colhead{$q_H/q_{He}$}   &  \colhead{Atmosphere}
& \colhead{Best Fit} & \colhead{ $\chi^2_{dof}$} & \colhead{Age (95\%)}  \\
 &  &  &  & \colhead{(Gyr)} & \colhead{min} & \colhead{Gyr}
}
\startdata 
$0.55 M_{\odot}/1.385$ & Hansen (default) & $10^{-4}/10^{-2}$ & Bergeron H & 12.4 & 0.97 & 10.9--14.0 \\
$0.55 M_{\odot}/1.385$ & Hansen (default) & $10^{-6}/10^{-2}$ & Bergeron H & 12.4 & 1.14  & 9.8--13.6 \\
$0.55 M_{\odot}/1.385$ & Hansen (default) & $10^{-4}/10^{-3}$ &  Bergeron H & 11.7 & 1.13 &  11.0--14.2 \\
$0.55 M_{\odot}/1.385$ & Hansen (Hurley) & $10^{-4}/10^{-2}$ & Bergeron H & 12.1 & 0.77 & $>10.5$ \\
$0.60 M_{\odot}/1.385$ & Hansen (default) & $10^{-4}/10^{-2}$ & Bergeron H & 12.4 & 1.09 & 10.4--14.5 \\
$0.60 M_{\odot}/1.385$ & Chabrier & $10^{-4}/10^{-2}$  & Bergeron H  & 10.2 & 1.7 & 9.8--12.1 \\
$0.60 M_{\odot}/1.385$ & Salaris & $10^{-4}/10^{-2}$  & Bergeron H & 15.5 & 2.1 & 15.0-16.3 \\
$0.55 M_{\odot}/1.385$ & Salaris & $10^{-4}/10^{-2}$  & Bergeron H & 15.5 & 2.1 & 14.5-16.0 \\
$0.55 M_{\odot}/1.385$ & Hansen (default) & $f_H$ $10^{-4}/10^{-2}$ & Bergeron $f_H$ H  & 12.4 & 0.98 & 10.5--14.0 \\
                       &                  & 1-$f_H$ $0/10^{-2}$       &  \& $1-f_H$ He & & \\
$0.55 M_{\odot}/1.385$ & Hansen (default) & $10^{-4}/10^{-2}$ & Bergeron $f_H$ H  & 14.6 & 0.9 & $>11.6$ \\
                       &                  &        &  \& $1-f_H$ He & & \\
$0.55 M_{\odot}/1.385$ & Hansen (default) & $10^{-4}/10^{-2}$ & Bergeron H & 12.5 & 0.92 & 11.0--14.1 \\
  & Hurley IFMR & \\
$0.55 M_{\odot}/1.385$ & Hansen (Hurley) & $10^{-4}/10^{-2}$ & Bergeron H & 12.4 & 0.74 & $>10.5$ \\
                       &  Hurley IFMR   &                   &  &     &     & \\
$0.55 M_{\odot}/1.385$ & Hansen (default) & $10^{-4}/10^{-2}$ & Hansen H & 12.1 & 2.1 & 10.9--12.5 \\
$0.55 M_{\odot}/\cdots$ & Hansen (default) & $10^{-4}/10^{-2}$ & Bergeron H & 12.4 & 0.98 & 10.5--14.1 \\
$\cdots/\cdots$          & Hansen (default) & $10^{-4}/10^{-2}$ & Bergeron H & 12.1 & 0.76 & 11.1--14.6 \\
\hline
$\cdots/\cdots$          & $\cdots$           & $\cdots$            & $\cdots$     & 12.1 & 0.74 & $>10.3$ \\
\enddata 
\end{deluxetable} 

\end{document}